\documentclass[
  aps,
  prx,
  reprint,
  groupedaddress,
  longbibliography,
  amsmath,
  amssymb,
  nofootinbib
]{revtex4-2}

\usepackage{graphicx}
\usepackage{tikz}
\usepackage{tikz-cd}
\usetikzlibrary{arrows.meta,decorations.markings}
\usepackage[colorlinks=true,allcolors=blue]{hyperref}
\usepackage{array,colortbl}

\newif\ifAppendixContents
\newcommand{\AppendixContentsStart}{\AppendixContentstrue}
\makeatletter
\renewcommand{\tableofcontents}{%
  \begingroup
    \AppendixContentsfalse
    \let\FullContentsLine\contentsline
    \renewcommand{\contentsline}[4]{%
      \ifAppendixContents\else\FullContentsLine{##1}{##2}{##3}{##4}\fi}
    \let\l@title\@gobbletwo
    \let\l@abstract\@gobbletwo
    \let\l@subsubsection\@gobbletwo
    \let\addcontentsline\@gobblethree
    \def\tocmax@section{2.5em}%
    \@fileswfalse
    \print@toc{toc}%
  \endgroup
}
\makeatother

\newcommand{\paragraf}[1]{\par\medskip\noindent{\bf #1}}

\tikzset{
  topological arrow/.style={
    postaction={decorate},
    decoration={markings,mark=at position #1 with {
      \arrow{Stealth[length=3.6pt,width=3pt]}
    }}
  },
  topological arrow/.default=0.55,
  topological arrows/.style args={#1/#2}{
    postaction={decorate},
    decoration={markings,
      mark=at position #1 with {\arrow{Stealth[length=3.6pt,width=3pt]}},
      mark=at position #2 with {\arrow{Stealth[length=3.6pt,width=3pt]}}
    }
  },
  reversed topological arrow/.style={
    postaction={decorate},
    decoration={markings,mark=at position #1 with {
      \arrowreversed{Stealth[length=3.6pt,width=3pt]}
    }}
  },
  reversed topological arrow/.default=0.55,
  reversed topological arrows/.style args={#1/#2}{
    postaction={decorate},
    decoration={markings,
      mark=at position #1 with {\arrowreversed{Stealth[length=3.6pt,width=3pt]}},
      mark=at position #2 with {\arrowreversed{Stealth[length=3.6pt,width=3pt]}}
    }
  }
}

\definecolor{tleModule}{HTML}{000000}
\definecolor{tleRho}{HTML}{355C7D}
\definecolor{tleJunction}{HTML}{000000}
\definecolor{tleInk}{HTML}{000000}
\tikzset{
  tle diagram/.style={x=1cm,y=1cm,line cap=round,line join=round,
    font=\footnotesize,text=black},
  tle module/.style={draw=tleModule,line width=1.05pt,reversed topological arrow},
  tle rho/.style={draw=tleRho,line width=1.05pt,reversed topological arrow},
  tle auxiliary/.style={draw=tleJunction,line width=1.05pt,topological arrow=0.80},
  tle site/.style={text=black,font=\scriptsize,above=5pt,inner sep=0pt},
  tle leaf/.style={below=3pt,inner sep=0pt},
  tle grade/.style={font=\scriptsize,inner sep=0pt},
  tle object/.style={font=\scriptsize,inner sep=0pt},
  tle move/.style={draw=tleInk,-{Stealth[length=4.5pt]},line width=0.75pt}
}

\newcommand{\TwoLegEvaluationOverview}[1][]{%
  \begin{tikzpicture}[tle diagram,#1]
    \begin{scope}
      \foreach \xa/\xb in {-0.2/0.4,0.4/2,2/2.6} {
        \draw[tle module] (\xb,0)--(\xa,0);
      }
      \node[tle site] at (0,0) {$g_{i-1}$};
      \node[tle site] at (1.2,0) {$g_i$};
      \node[tle site] at (2.4,0) {$g_{i+1}$};
      \foreach \x in {0.4,2.0} {
        \draw[tle rho] (\x,0)--(\x,-1.9);
        \node[tle leaf,text=black] at (\x,-1.9) {$\rho$};
      }
      \node[tle grade,text=black,left=3pt] at (0.4,-1.55) {$q_-$};
      \node[tle grade,text=black,right=3pt] at (2,-1.55) {$q_+$};
      \node[tle object] at (1.2,-2.85)
        {$(g_{i-1}\triangleleft\rho)\triangleleft\rho$};
    \end{scope}

    \begin{scope}[xshift=3.6cm]
      \foreach \xa/\xb in {-0.2/0.4,0.4/2,2/2.6} {
        \draw[tle module] (\xb,0)--(\xa,0);
      }
      \node[tle site] at (0,0) {$g_{i-1}$};
      \node[tle site] at (1.2,0) {$g_i$};
      \node[tle site] at (2.4,0) {$g_{i+1}$};
      \draw[tle rho] (0.4,0)--(0.4,-1.9);
      \draw[tle rho] (2,0)--(2,-1.9);
      \draw[tle auxiliary] (0.4,-0.75)--(1.2,-1.25)--(1.2,-1.9);
      \node[tle leaf,text=black] at (0.4,-1.9) {$\rho$};
      \node[tle leaf,text=black] at (1.2,-1.9) {$X$};
      \node[tle leaf,text=black] at (2,-1.9) {$\rho$};
      \node[tle grade,text=black,left=3pt] at (0.4,-1.55) {$q_-s$};
      \node[tle grade,text=black,right=3pt] at (1.2,-1.55) {$s^{-1}$};
      \node[tle grade,text=black,right=3pt] at (2,-1.55) {$q_+$};
      \node[tle object] at (1.2,-2.85)
        {$\bigl(g_{i-1}\triangleleft(\rho\otimes X)\bigr)\triangleleft\rho$};
    \end{scope}

    \begin{scope}[xshift=7.2cm]
      \foreach \xa/\xb in {-0.2/0.4,0.4/1.2,1.2/2,2/2.6} {
        \draw[tle module] (\xb,0)--(\xa,0);
      }
      \node[tle site] at (0,0) {$g_{i-1}$};
      \node[tle site] at (0.8,0) {$g_i s$};
      \node[tle site] at (1.6,0) {$g_i$};
      \node[tle site] at (2.4,0) {$g_{i+1}$};
      \draw[tle rho] (0.4,0)--(0.4,-1.9);
      \draw[tle auxiliary] (1.2,0)--(1.2,-1.9);
      \draw[tle rho] (2,0)--(2,-1.9);
      \node[tle leaf,text=black] at (0.4,-1.9) {$\rho$};
      \node[tle leaf,text=black] at (1.2,-1.9) {$X$};
      \node[tle leaf,text=black] at (2,-1.9) {$\rho$};
      \node[tle grade,text=black,left=3pt] at (0.4,-1.55) {$q_-s$};
      \node[tle grade,text=black,right=3pt] at (1.2,-1.55) {$s^{-1}$};
      \node[tle grade,text=black,right=3pt] at (2,-1.55) {$q_+$};
      \node[tle object] at (1.2,-2.85)
        {$\bigl((g_{i-1}\triangleleft\rho)\triangleleft X\bigr)\triangleleft\rho$};
    \end{scope}

    \begin{scope}[xshift=10.8cm]
      \foreach \xa/\xb in {-0.2/0.4,0.4/2,2/2.6} {
        \draw[tle module] (\xb,0)--(\xa,0);
      }
      \node[tle site] at (0,0) {$g_{i-1}$};
      \node[tle site] at (1.2,0) {$g_i s$};
      \node[tle site] at (2.4,0) {$g_{i+1}$};
      \draw[tle rho] (0.4,0)--(0.4,-1.9);
      \draw[tle rho] (2,0)--(2,-1.9);
      \draw[tle auxiliary] (2,-0.75)--(1.2,-1.25)--(1.2,-1.9);
      \node[tle leaf,text=black] at (0.4,-1.9) {$\rho$};
      \node[tle leaf,text=black] at (1.2,-1.9) {$X$};
      \node[tle leaf,text=black] at (2,-1.9) {$\rho$};
      \node[tle grade,text=black,left=3pt] at (0.4,-1.55) {$q_-s$};
      \node[tle grade,text=black,right=3pt] at (1.2,-1.55) {$s^{-1}$};
      \node[tle grade,text=black,right=3pt] at (2,-1.55) {$q_+$};
      \node[tle object] at (1.2,-2.85)
        {$(g_{i-1}\triangleleft\rho)\triangleleft(X\otimes\rho)$};
    \end{scope}

    \begin{scope}[xshift=14.4cm]
      \foreach \xa/\xb in {-0.2/0.4,0.4/2,2/2.6} {
        \draw[tle module] (\xb,0)--(\xa,0);
      }
      \node[tle site] at (0,0) {$g_{i-1}$};
      \node[tle site] at (1.2,0) {$g_i s$};
      \node[tle site] at (2.4,0) {$g_{i+1}$};
      \foreach \x in {0.4,2.0} {
        \draw[tle rho] (\x,0)--(\x,-1.9);
        \node[tle leaf,text=black] at (\x,-1.9) {$\rho$};
      }
      \node[tle grade,text=black,left=3pt] at (0.4,-1.55) {$q_-s$};
      \node[tle grade,text=black,right=3pt] at (2,-1.55) {$s^{-1}q_+$};
      \node[tle object] at (1.2,-2.85)
        {$(g_{i-1}\triangleleft\rho)\triangleleft\rho$};
    \end{scope}

    \foreach \x/\n/\lab in {2.7/1/\mathcal I_f,
                             6.3/2/(\alpha^{\mathcal M})^{-1},
                             9.9/3/\alpha^{\mathcal M},
                             13.5/4/\mathcal J_g^\dagger} {
      \draw[tle move] (\x,-0.95)--(\x+0.6,-0.95);
      \node[above=4pt] at (\x+0.3,-0.95) {$\lab$};
      \node[below=4pt] at (\x+0.3,-0.95) {$(\n)$};
    }
  \end{tikzpicture}%
}

\definecolor{anyonModule}{HTML}{000000}
\definecolor{anyonRho}{HTML}{355C7D}
\definecolor{anyonJunction}{HTML}{A46A24}
\definecolor{anyonInk}{HTML}{000000}
\definecolor{anyonMuted}{HTML}{000000}

\tikzset{
  anyon diagram/.style={
    x=0.98cm,
    y=1cm,
    line cap=round,
    line join=round,
    font=\small,
    text=black
  },
  anyon module/.style={draw=anyonModule,line width=1.05pt},
  anyon leg/.style={draw=anyonRho,line width=1.05pt},
  anyon auxiliary/.style={draw=black,line width=1.05pt},
  anyon junction/.style={
    circle,
    fill=black,
    draw=white,
    line width=0.45pt,
    inner sep=0pt,
    minimum size=4.8pt
  },
  anyon module label/.style={text=black,above=5pt,inner sep=1pt},
  anyon rho label/.style={text=black,below=4pt,inner sep=1pt},
  anyon junction label/.style={
    text=black,
    font=\footnotesize,
    anchor=north west,
    inner sep=0pt,
    xshift=5pt,
    yshift=-6pt
  },
  anyon continuation/.style={text=black,inner sep=0pt},
  anyon operator/.style={
    draw=black,
    fill=white,
    text=black,
    line width=0.8pt,
    rounded corners=2pt,
    minimum height=0.65cm,
    inner sep=5pt
  }
}

\newcommand{\AnyonLegTerm}[1][]{%
  \begin{tikzpicture}[anyon diagram,#1]
    \foreach \x in {0,1.5,4.5,6} {
      \draw[anyon leg,reversed topological arrows={0.17/0.83}]
        (\x,0) -- (\x,-1.8);
      \node[anyon rho label] at (\x,-1.8) {$\rho$};
      \node[text=black,above=4pt,inner sep=1pt] at (\x,0) {$\rho$};
    }
    \node[anyon continuation] at (3,-0.15) {$\cdots$};
    \node[anyon continuation] at (3,-1.65) {$\cdots$};
    \node[anyon operator,minimum width=6.6cm] at (3,-0.9) {$h^{(n)}$};
  \end{tikzpicture}%
}

\newcommand{\AnyonChain}[1][]{%
  \begin{tikzpicture}[anyon diagram,#1]
    \foreach \xa/\xb in {-0.5/0,0/2,2/4,4/6,6/6.5} {
      \draw[anyon module,reversed topological arrow] (\xb,0)--(\xa,0);
    }
    \node[anyon continuation] at (-0.82,0) {$\cdots$};
    \node[anyon continuation] at (6.82,0) {$\cdots$};

    \foreach \x in {0,2,4,6} {
      \draw[anyon leg,reversed topological arrow] (\x,0) -- (\x,-1.05);
      \node[anyon rho label] at (\x,-1.05) {$\rho$};
      \node[anyon junction] (junction-\x) at (\x,0) {};
    }

    \node[anyon module label] (module-left) at (1,0) {$m_{j-1}$};
    \node[anyon module label] (module-middle) at (3,0) {$m_j$};
    \node[anyon module label] (module-right) at (5,0) {$m_{j+1}$};
    \node[anyon junction label] at (junction-2) {$x_{j,j-1}$};
    \node[anyon junction label] at (junction-4) {$x_{j+1,j}$};

    \draw[draw=black,line width=0.8pt] (-0.5,-2.6) -- (6.5,-2.6);
    \node[anyon continuation] at (-0.82,-2.6) {$\cdots$};
    \node[anyon continuation] at (6.82,-2.6) {$\cdots$};
    \foreach \x/\j in {1/{j-1},3/j,5/{j+1}} {
      \node[circle,draw=black,fill=white,line width=0.8pt,
        inner sep=0pt,minimum size=6pt] at (\x,-2.6) {};
      \node[text=black,above=5pt,inner sep=1pt,font=\footnotesize]
        at (\x,-2.6) {$m_{\j}$};
    }
    \node[text=black,below=5pt,inner sep=1pt,font=\footnotesize]
      at (2,-2.6) {$x_{j,j-1}$};
    \node[text=black,below=5pt,inner sep=1pt,font=\footnotesize]
      at (4,-2.6) {$x_{j+1,j}$};
  \end{tikzpicture}%
}

\newcommand{\AnyonBoxTerm}[3][]{%
  \begin{tikzpicture}[anyon diagram,baseline=-0.99cm,#1]
    \pgfmathsetmacro{\anyonBoxCenter}{0.9*(#2-1)}
    \pgfmathsetlengthmacro{\anyonBoxWidth}{(1.8*(#2-1)+0.85)*1cm}
    \foreach \j in {1,...,#2} {
      \draw[anyon leg,reversed topological arrows={0.17/0.83}]
        ({1.8*(\j-1)},0) -- ({1.8*(\j-1)},-1.8);
      \node[text=black,above=4pt,inner sep=1pt]
        at ({1.8*(\j-1)},0) {$\rho$};
      \node[anyon rho label] at ({1.8*(\j-1)},-1.8) {$\rho$};
    }
    \node[anyon operator,minimum width=\anyonBoxWidth]
      at (\anyonBoxCenter,-0.9) {$#3$};
  \end{tikzpicture}%
}

\newcommand{\VillainAnyonChain}[1][]{%
  \begin{tikzpicture}[anyon diagram,#1]
    \node[font=\footnotesize,anchor=south west,inner sep=0pt] at (-0.82,0.85)
      {$\cC=\left[\Vec_{U(1)_m}\boxtimes\Vec_{U(1)_w}\right]^\omega$};
    \node[text=black,font=\footnotesize,anchor=south east,inner sep=0pt]
      at (6.82,0.85)
      {$\cM=\Vec_{U(1)}$};
    \foreach \xa/\xb in {-0.5/0,0/2,2/4,4/6,6/6.5} {
      \draw[anyon module,reversed topological arrow] (\xb,0)--(\xa,0);
    }
    \node[anyon continuation] at (-0.82,0) {$\cdots$};
    \node[anyon continuation] at (6.82,0) {$\cdots$};
    \foreach \x in {0,2,4,6} {
      \draw[anyon leg,reversed topological arrow] (\x,0) -- (\x,-0.95);
      \node[anyon junction] (villain-junction-\x) at (\x,0) {};
    }
    \foreach \x in {0,4,6} {
      \node[anyon rho label] at (\x,-0.95) {$\rho$};
    }
    \node[anyon rho label] at (2,-0.95)
      {$\rho=\int_{\bR}^{\oplus}t$};
    \node[anyon module label] at (1,0) {$\phi_{i-1}$};
    \node[anyon module label] at (3,0) {$\phi_i\in\bR/2\pi\bZ$};
    \node[anyon module label] at (5,0) {$\phi_{i+1}$};
    \node[anyon junction label] at (villain-junction-2) {$n_{i,i-1}$};
    \node[anyon junction label] at (villain-junction-4) {$n_{i+1,i}\in\bZ$};
    \node[font=\footnotesize,anchor=north west,inner sep=0pt] at (-0.82,-1.65)
      {$\cC^*_{\cM}=\Vec_{\bR}$};
  \end{tikzpicture}%
}

\newcommand{\TwoLegCoordinateGuide}[1][]{%
  \begin{tikzpicture}[x=1cm,y=0.90cm,font=\small,
    line cap=round,line join=round,#1]
    \foreach \xa/\xb in {-0.2/1.5,1.5/4.5,4.5/6.2} {
      \draw[anyon module,reversed topological arrow] (\xb,0)--(\xa,0);
    }
    \node[above=12pt] at (0.25,0) {$g_{i-1}$};
    \node[above=18pt] at (3,0) {$g_i\xrightarrow{\ s\ }g_i s$};
    \node[above=12pt] at (5.75,0) {$g_{i+1}$};
    \foreach \x/\lab in {1.5/-,4.5/+} {
      \draw[anyon leg,reversed topological arrow] (\x,0)--(\x,-3.35);
      \node[anyon junction] at (\x,0) {};
      \node[below=4pt] at (\x,-3.35) {$\rho$};
    }
    \node[font=\footnotesize,fill=white,inner sep=2pt]
      at (1.5,-0.40) {$(\bde_{q_-},x_-)$};
    \node[font=\footnotesize,fill=white,inner sep=2pt]
      at (4.5,-0.40) {$(\bde_{q_+},x_+)$};
    \node[font=\footnotesize,fill=white,inner sep=2pt]
      at (1.5,-2.95) {$(\bde_{q_-s},x_-)$};
    \node[font=\footnotesize,fill=white,inner sep=2pt]
      at (4.5,-2.95) {$(\bde_{s^{-1}q_+},x_+)$};
    \draw[anyon auxiliary,topological arrow]
      (1.5,-1.05)--(4.5,-2.25);
    \node[above=6pt] at (3,-1.65) {$|x_-r\rangle_X$};
    \node[below=6pt] at (3,-1.65) {$s^{-1}$};
    \node[anyon junction] at (1.5,-1.05) {};
    \node[anyon junction] at (4.5,-2.25) {};
    \node[left=6pt] at (1.5,-1.05) {$\cI_f$};
    \node[right=6pt] at (4.5,-2.25) {$\cJ_g^\dagger$};
  \end{tikzpicture}%
}

\tikzset{
  tlm leaf/.style={font=\footnotesize,below=5pt,inner sep=1pt},
  tlm site/.style={font=\footnotesize,above=5pt,inner sep=1pt},
  tlm object/.style={font=\scriptsize,inner sep=1pt},
  tlm auxiliary/.style={anyon auxiliary,topological arrow=0.92}
}

\newcommand{\TwoLegLocalSingleModule}[3]{%
  \draw[anyon module,reversed topological arrow] (#1,0)--(-0.5,0);
  \draw[anyon module,reversed topological arrow] (3.9,0)--(#1,0);
  \node[tlm site] at ({(#1-0.5)/2},0) {$#2$};
  \node[tlm site] at ({(#1+3.9)/2},0) {$#3$};
  \node[anyon junction] at (#1,0) {};
}

\newcommand{\TwoLegLocalSplitModule}[3]{%
  \foreach \xa/\xb in {-0.5/0.65,0.65/2.75,2.75/3.9} {
    \draw[anyon module,reversed topological arrow] (\xb,0)--(\xa,0);
  }
  \node[tlm site] at (0.075,0) {$#1$};
  \node[tlm site] at (1.7,0) {$#2$};
  \node[tlm site] at (3.325,0) {$#3$};
  \node[anyon junction] at (0.65,0) {};
  \node[anyon junction] at (2.75,0) {};
}

\newcommand{\TwoLegLocalRho}[2]{%
  \draw[anyon leg,reversed topological arrow] (#1,0)--(#1,-1.9);
  \node[anyon junction] at (#1,0) {};
  \node[font=\footnotesize,left=4pt] at (#1,-0.32) {$\rho$};
  \node[tlm leaf] at (#1,-1.9) {$#2$};
}

\newcommand{\TwoLegLocalXLabel}[1]{%
  \node[tlm leaf] at (#1,-1.9) {$|x_-r\rangle_X$};
  \node[font=\scriptsize,right=4pt] at (#1,-1.5) {$s^{-1}$};
}

\newcommand{\TwoLegLocalState}[1]{%
  \begin{tikzpicture}[anyon diagram,x=1cm,y=0.8cm,
    baseline=-1.02cm,font=\footnotesize]
    \path[use as bounding box] (-0.65,-2.95) rectangle (4.05,0.45);
    \ifcase#1
      \TwoLegLocalSingleModule{1.7}{g_{i-1}}{g_i}
      \TwoLegLocalRho{1.7}{(\bde_{q_-},x_-)}
      \node[tlm object] at (1.7,-2.72) {$g_{i-1}\triangleleft\rho$};
    \or
      \TwoLegLocalSingleModule{0.65}{g_{i-1}}{g_i}
      \TwoLegLocalRho{0.65}{(\bde_{q_-s},x_-)}
      \draw[tlm auxiliary] (0.65,-0.65)--(2.75,-1.15)--(2.75,-1.9);
      \node[anyon junction] at (0.65,-0.65) {};
      \TwoLegLocalXLabel{2.75}
      \node[tlm object] at (1.7,-2.72)
        {$g_{i-1}\triangleleft(\rho\otimes X)$};
    \or
      \TwoLegLocalSplitModule{g_{i-1}}{g_i s}{g_i}
      \TwoLegLocalRho{0.65}{(\bde_{q_-s},x_-)}
      \draw[tlm auxiliary] (2.75,0)--(2.75,-1.9);
      \TwoLegLocalXLabel{2.75}
      \node[tlm object] at (1.7,-2.72)
        {$(g_{i-1}\triangleleft\rho)\triangleleft X$};
    \or
      \TwoLegLocalSplitModule{g_i s}{g_i}{g_{i+1}}
      \draw[tlm auxiliary] (0.65,0)--(0.65,-1.9);
      \TwoLegLocalXLabel{0.65}
      \TwoLegLocalRho{2.75}{(\bde_{q_+},x_+)}
      \node[tlm object] at (1.7,-2.72)
        {$(g_i s\triangleleft X)\triangleleft\rho$};
    \or
      \TwoLegLocalSingleModule{2.75}{g_i s}{g_{i+1}}
      \TwoLegLocalRho{2.75}{(\bde_{q_+},x_+)}
      \draw[tlm auxiliary] (2.75,-0.65)--(0.65,-1.15)--(0.65,-1.9);
      \node[anyon junction] at (2.75,-0.65) {};
      \TwoLegLocalXLabel{0.65}
      \node[tlm object] at (1.7,-2.72)
        {$g_i s\triangleleft(X\otimes\rho)$};
    \or
      \TwoLegLocalSingleModule{1.7}{g_i s}{g_{i+1}}
      \TwoLegLocalRho{1.7}{(\bde_{s^{-1}q_+},x_+)}
      \node[tlm object] at (1.7,-2.72) {$g_i s\triangleleft\rho$};
    \fi
  \end{tikzpicture}%
}

\newcommand{\TwoLegLocalMove}[1]{%
  \text{(#1)}:\quad
  \ifcase#1\or
    \TwoLegLocalState{0}\;\xrightarrow{\ \cI_f\ }\;\TwoLegLocalState{1}%
  \or
    \TwoLegLocalState{1}\;\xrightarrow{\ (\alpha^{\cM})^{-1}\ }\;\TwoLegLocalState{2}%
  \or
    \TwoLegLocalState{3}\;\xrightarrow{\ \alpha^{\cM}\ }\;\TwoLegLocalState{4}%
  \or
    \TwoLegLocalState{4}\;\xrightarrow{\ \cJ_g^\dagger\ }\;\TwoLegLocalState{5}%
  \fi
}

\newcommand{\NonabelianAnyonChain}[1][]{%
  \begin{tikzpicture}[anyon diagram,#1]
    \node[font=\footnotesize,anchor=south west,inner sep=0pt] at (-0.82,0.85)
      {$\cC=\Vec_{G_L}^{\omega}\boxtimes\Vec_{G_R}^{-\omega}$};
    \node[text=black,font=\footnotesize,anchor=south east,inner sep=0pt]
      at (6.82,0.85) {$\cM=\Vec_G$};
    \foreach \xa/\xb in {-0.5/0,0/2,2/4,4/6,6/6.5} {
      \draw[anyon module,reversed topological arrow] (\xb,0)--(\xa,0);
    }
    \node[anyon continuation] at (-0.82,0) {$\cdots$};
    \node[anyon continuation] at (6.82,0) {$\cdots$};
    \foreach \x in {0,2,4,6} {
      \draw[anyon leg,reversed topological arrow] (\x,0) -- (\x,-0.95);
      \node[anyon rho label] at (\x,-0.95) {$\rho$};
      \node[anyon junction] (nonabelian-junction-\x) at (\x,0) {};
    }
    \node[anyon module label] at (1,0) {$g_{i-1}$};
    \node[anyon module label] at (3,0) {$g_i\in G$};
    \node[anyon module label] at (5,0) {$g_{i+1}$};
    \node[anyon junction label] at (nonabelian-junction-2)
      {$x_{i,i-1}$};
    \node[anyon junction label] at (nonabelian-junction-4)
      {$x_{i+1,i}\in G$};
    \node[font=\footnotesize,anchor=north west,inner sep=0pt] at (-0.82,-1.65)
      {$\cC^*_{\cM}=\cZ(\Vec_G^\omega)$};
  \end{tikzpicture}%
}

\newcommand{\AnyonTwoLeg}[2][]{%
  \begin{tikzpicture}[anyon diagram,baseline=-0.99cm,#1]
    \foreach \x in {0,1.8} {
      \draw[anyon leg,reversed topological arrows={0.17/0.83}]
        (\x,0) -- (\x,-1.8);
      \node[text=black,above=4pt,inner sep=1pt] at (\x,0) {$\rho$};
      \node[anyon rho label] at (\x,-1.8) {$\rho$};
    }
    \draw[anyon auxiliary,
      postaction={decorate},
      decoration={markings,mark=at position 0.55 with {
        \arrow{Stealth[length=4.2pt,width=3.5pt]}
      }}] (0,-0.65) -- (1.8,-1.15);
    \node[text=black,above=4pt,inner sep=1pt] at (0.9,-0.9) {$#2$};
    \node[anyon junction] at (0,-0.65) {};
    \node[anyon junction] at (1.8,-1.15) {};
  \end{tikzpicture}%
}

\newcommand{\AnyonLeftSymmetryMove}[2][]{%
  \begin{tikzpicture}[anyon diagram,x=0.84cm,baseline=-0.32cm,#1]
    \foreach \xa/\xb in {0/1.2,1.2/2.9,2.9/4.6,4.6/5.8} {
      \draw[anyon module,reversed topological arrow] (\xb,0)--(\xa,0);
    }
    \node[anyon continuation] at (-0.2,0) {$\cdots$};
    \node[anyon continuation] at (6,0) {$\cdots$};
    \draw[draw=anyonJunction,line width=1.05pt,topological arrow=0.62]
      (4.6,0) .. controls (4.6,1.25) and (1.2,1.25) .. (1.2,0);
    \node[text=black,above=2pt] at (2.9,0.94) {$#2$};
    \node[anyon junction] at (1.2,0) {};
    \node[anyon junction] at (4.6,0) {};
    \draw[anyon leg,reversed topological arrow] (2.9,0) -- (2.9,-1.05);
    \node[anyon junction] at (2.9,0) {};
    \node[anyon rho label] at (2.9,-1.05) {$\rho$};
    \node[anyon junction label] at (2.9,-0.15) {$x_{i+1,i}$};
    \node[anyon module label] at (0.45,0) {$#2g_i$};
    \node[anyon module label] at (2.05,0) {$g_i$};
    \node[anyon module label] at (3.75,0) {$g_{i+1}$};
    \node[anyon module label] at (5.35,0) {$#2g_{i+1}$};
  \end{tikzpicture}%
}

\newcommand{\AnyonRightSymmetryMove}[2][]{%
  \begin{tikzpicture}[anyon diagram,x=0.92cm,baseline=-0.8cm,#1]
    \foreach \xa/\xb in {0/1.5,1.5/3.2,3.2/4.9,4.9/6.4} {
      \draw[anyon module,reversed topological arrow] (\xb,0)--(\xa,0);
    }
    \node[anyon continuation] at (-0.2,0) {$\cdots$};
    \node[anyon continuation] at (6.6,0) {$\cdots$};
    \draw[anyon leg,reversed topological arrows={0.3/0.86}]
      (3.2,0) -- (3.2,-2.05);
    \node[anyon junction] at (3.2,0) {};
    \node[anyon junction label] at (3.2,-0.1) {$x_{i+1,i}$};
    \node[anyon rho label] at (3.2,-2.05) {$\rho$};
    \draw[draw=anyonJunction,line width=1.05pt,
      preaction={draw=white,line width=4.5pt},
      postaction={decorate},
      decoration={markings,mark=at position 0.57 with {
        \arrow{Stealth[length=4.2pt,width=3.5pt]}
      }}]
      (1.5,0) .. controls (1.5,-1.7) and (4.9,-1.7) .. (4.9,0);
    \node[anyon junction] at (1.5,0) {};
    \node[anyon junction] at (4.9,0) {};
    \node[text=black,left=3pt] at (1.95,-1.05) {$#2$};
    \node[text=black,left=3pt] at (3.2,-0.75) {$q$};
    \node[text=black,left=3pt] at (3.2,-1.7) {$q^h$};
    \node[text=black,anchor=north west,font=\footnotesize]
      at (3.65,-1.45) {$c_{#2,\rho}$};
    \node[anyon module label] at (0.6,0) {$g_i#2^{-1}$};
    \node[anyon module label] at (2.35,0) {$g_i$};
    \node[anyon module label] at (4.05,0) {$g_{i+1}$};
    \node[anyon module label] at (5.8,0) {$g_{i+1}#2^{-1}$};
  \end{tikzpicture}%
}

\newcommand{\AnyonTwoLegSlidePanel}[1]{%
  \draw[anyon module] (0.15,0)--(0,0);
  \draw[anyon module] (3.4,0)--(3.25,0);
  \foreach \xa/\xb in {0.15/0.9,0.9/2.5,2.5/3.25} {
    \draw[anyon module,reversed topological arrow] (\xb,0)--(\xa,0);
  }
  \node[anyon continuation] at (-0.18,0) {$\cdots$};
  \node[anyon continuation] at (3.58,0) {$\cdots$};
  \foreach \x in {0.9,2.5} {
    \draw[anyon leg,reversed topological arrows={0.1/0.9}]
      (\x,0)--(\x,-2.7);
    \node[anyon rho label] at (\x,-2.7) {$\rho$};
  }
  \draw[anyon auxiliary,
    postaction={decorate},
    decoration={markings,mark=at position 0.28 with {
      \arrow{Stealth[length=4.2pt,width=3.5pt]}
    }}] (0.9,-0.95)--(2.5,-1.65);
  \draw[draw=anyonJunction,line width=1.05pt,
    preaction={draw=white,line width=4.5pt}]
    (0.15,0) .. controls (0.15,#1) and (0.4,#1) .. (0.65,#1)
    --(2.75,#1)
    .. controls (3,#1) and (3.25,#1) .. (3.25,0);
  \draw[draw=anyonJunction,line width=1.05pt,
    -{Stealth[length=4.2pt,width=3.5pt]}]
    (1.95,#1)--(2.2,#1);
  \foreach \x in {0.15,0.9,2.5,3.25} {
    \node[anyon junction] at (\x,0) {};
  }
  \node[anyon junction] at (0.9,-0.95) {};
  \node[anyon junction] at (2.5,-1.65) {};
  \node[text=black,left=3pt,inner sep=0pt]
    at (0.78,-0.95) {$f$};
  \node[text=black,right=3pt,inner sep=0pt]
    at (2.62,-1.65) {$g^\dagger$};
  \node[text=black,above=3pt,inner sep=0pt]
    at (1.7,-1.3) {$X$};
  \node[text=black,below=3pt,inner sep=0pt]
    at (1.7,#1) {$h$};
  \node[anyon module label,font=\scriptsize]
    at (0.48,0) {$g_{i-1}$};
  \node[anyon module label,font=\scriptsize]
    at (1.7,0) {$g_i$};
  \node[anyon module label,font=\scriptsize]
    at (2.98,0) {$g_{i+1}$};
}

\newcommand{\AnyonTwoLegSymmetrySlide}[1][]{%
  \begin{tikzpicture}[anyon diagram,font=\small,#1]
    \AnyonTwoLegSlidePanel{-0.45}
    \begin{scope}[shift={(4.2,0)}]
      \AnyonTwoLegSlidePanel{-2.15}
    \end{scope}
    \node[text=black,inner sep=0pt] at (3.8,-1.3) {$=$};
  \end{tikzpicture}%
}

\definecolor{wzwFlowBlue}{HTML}{355C7D}
\newcommand{\WZWFlowDiagram}[1][\linewidth]{%
\resizebox{#1}{!}{%
\begin{tikzpicture}[
  x=1em,y=1em,
  wzw theory/.style={
    draw=black!65,fill=black!2,line width=0.055em,
    rounded corners=0.2em,align=center,font=\small,
    text width=9em,minimum height=5em,inner sep=0.4em
  },
  wzw fixed point/.style={wzw theory,draw=wzwFlowBlue,fill=wzwFlowBlue!6},
  wzw arrow/.style={-{Stealth[length=0.68em,width=0.45em]},line width=0.075em},
  wzw open arrow/.style={wzw arrow,dashed,draw=black!55},
  wzw label/.style={font=\footnotesize,align=center,fill=white,inner sep=0.3em}
]
\node[wzw theory] (lat) at (0,0)
  {Lattice Hamiltonian\\[0.3em]$H_{\text{lattice}}(\beta,k)$};
\node[wzw theory] (cont) at (0,-10)
  {Continuum WZ\\sigma model\\[0.3em]$H(\beta,k)$,\quad $\beta\neq\beta_*$};
\node[wzw fixed point] (wzw) at (16,-10)
  {Critical WZW model\\[0.3em]$G_k$};

\draw[wzw open arrow] (lat.east) -- (wzw.north)
  node[midway,above=0.4em,sloped,wzw label]
  {lattice IR};
\draw[wzw arrow] (lat.south) -- (cont.north)
  node[midway,right=0.5em,wzw label,align=left]
  {BO projection\\+ smooth-field limit};
\draw[wzw arrow] (cont.east) -- (wzw.west)
  node[midway,below=0.4em,wzw label] {Continuum\\RG\\$\beta\longrightarrow\beta_*$};
\end{tikzpicture}%
}%
}

\newcommand{\TransgressionPrismDiagram}{%
\resizebox{\linewidth}{!}{%
\begin{tikzpicture}[x=1cm,y=1cm,font=\small,
  line cap=round,line join=round,
  prism argument/.style={draw=anyonJunction,
    -{Stealth[length=4pt]},line width=1.05pt},
  prism label/.style={fill=white,inner sep=1pt}]
  \coordinate (a0) at (0,0.5);
  \coordinate (a1) at (2.7,0.9);
  \coordinate (a2) at (1.2,0);
  \coordinate (b0) at (0,2.9);
  \coordinate (b1) at (2.7,3.3);
  \coordinate (b2) at (1.2,2.4);
  \fill[anyonModule!6] (a0)--(a2)--(b2)--(b0)--cycle;
  \fill[anyonModule!11] (a2)--(a1)--(b1)--(b2)--cycle;
  \fill[anyonModule!6] (b0)--(b1)--(b2)--cycle;
  \draw[anyon module,dashed] (a0)--(a1) (a0)--(b1);
  \draw[anyon module] (a0)--(a2)--(a1)
    (b0)--(b1)--(b2)--cycle
    (a0)--(b2)--(a1);
  \foreach \a/\b in {a0/b0,a1/b1,a2/b2} {
    \draw[prism argument] (\a)--(\b);
  }
  \foreach \v in {a0,a1,a2,b0,b1,b2} {
    \fill[anyonInk] (\v) circle (1.7pt);
  }
  \node[left=4pt] at (a0) {$1$};
  \node[right=4pt] at (a1) {$x$};
  \node[below=4pt] at (a2) {$xy$};
  \node[left=4pt] at (b0) {$q$};
  \node[above right=3pt] at (b1) {$qx$};
  \node[below right=3pt,prism label] at (b2) {$qxy$};
  \node at (1.35,-0.85) {$P_q(x,y)$};
  \node[font=\large] at (3.65,1.55) {$=$};

  \begin{scope}[xshift=4.5cm,yshift=0.35cm]
    \coordinate (v0) at (0,0);
    \coordinate (v1) at (2.3,0.15);
    \coordinate (v2) at (2.8,1.65);
    \coordinate (v3) at (0.7,2.6);
    \fill[anyonModule!6] (v0)--(v1)--(v3)--cycle;
    \fill[anyonModule!11] (v1)--(v2)--(v3)--cycle;
    \draw[anyon module,dashed] (v0)--(v2);
    \draw[anyon module] (v0)--(v3) (v1)--(v3);
    \draw[prism argument] (v0)--node[below=4pt] {$q$}(v1);
    \draw[prism argument] (v1)--node[right=4pt] {$x$}(v2);
    \draw[prism argument] (v2)--node[above=4pt] {$y$}(v3);
    \foreach \v in {v0,v1,v2,v3} {
      \fill[anyonInk] (\v) circle (1.7pt);
    }
    \node[below left=4pt] at (v0) {$1$};
    \node[below right=4pt] at (v1) {$q$};
    \node[right=4pt] at (v2) {$qx$};
    \node[above=4pt] at (v3) {$qxy$};
    \node at (1.4,-1.2) {$\mathcal T_1$};
    \node at (1.4,-1.8) {$\omega(q,x,y)$};
  \end{scope}
  \node[font=\large] at (8,1.55) {$-$};

  \begin{scope}[xshift=8.85cm,yshift=0.35cm]
    \coordinate (v0) at (0,0);
    \coordinate (v1) at (2.3,0.15);
    \coordinate (v2) at (2.8,1.65);
    \coordinate (v3) at (0.7,2.6);
    \fill[anyonModule!6] (v0)--(v1)--(v3)--cycle;
    \fill[anyonModule!11] (v1)--(v2)--(v3)--cycle;
    \draw[anyon module,dashed] (v0)--(v2);
    \draw[anyon module] (v0)--(v3) (v1)--(v3);
    \draw[prism argument] (v0)--node[below=4pt] {$x$}(v1);
    \draw[prism argument] (v1)--node[right=4pt] {${}^x q$}(v2);
    \draw[prism argument] (v2)--node[above=4pt] {$y$}(v3);
    \foreach \v in {v0,v1,v2,v3} {
      \fill[anyonInk] (\v) circle (1.7pt);
    }
    \node[below left=4pt] at (v0) {$1$};
    \node[below right=4pt] at (v1) {$x$};
    \node[right=4pt] at (v2) {$qx$};
    \node[above=4pt] at (v3) {$qxy$};
    \node at (1.4,-1.2) {$\mathcal T_2$};
    \node at (1.4,-1.8) {$\omega(x,{}^x q,y)$};
  \end{scope}
  \node[font=\large] at (12.35,1.55) {$+$};

  \begin{scope}[xshift=13.2cm,yshift=0.35cm]
    \coordinate (v0) at (0,0);
    \coordinate (v1) at (2.3,0.15);
    \coordinate (v2) at (2.8,1.65);
    \coordinate (v3) at (0.7,2.6);
    \fill[anyonModule!6] (v0)--(v1)--(v3)--cycle;
    \fill[anyonModule!11] (v1)--(v2)--(v3)--cycle;
    \draw[anyon module,dashed] (v0)--(v2);
    \draw[anyon module] (v0)--(v3) (v1)--(v3);
    \draw[prism argument] (v0)--node[below=4pt] {$x$}(v1);
    \draw[prism argument] (v1)--node[right=4pt] {$y$}(v2);
    \draw[prism argument] (v2)--node[above=4pt] {${}^{xy}q$}(v3);
    \foreach \v in {v0,v1,v2,v3} {
      \fill[anyonInk] (\v) circle (1.7pt);
    }
    \node[below left=4pt] at (v0) {$1$};
    \node[below right=4pt] at (v1) {$x$};
    \node[right=4pt] at (v2) {$xy$};
    \node[above=4pt] at (v3) {$qxy$};
    \node at (1.4,-1.2) {$\mathcal T_3$};
    \node at (1.4,-1.8) {$\omega(x,y,{}^{xy}q)$};
  \end{scope}
\end{tikzpicture}%
}%
}

\newcommand{\TransgressionPrismSweepDiagram}{%
\resizebox{0.4\linewidth}{!}{%
\begin{tikzpicture}[x=1cm,y=1cm,font=\small,line cap=round,line join=round]
  \fill[black!3] (0.3313,2.9061)--(6.6403,2.7329)--(7.0000,8.4447)--(0.0000,8.5572)--cycle;
  \fill[anyonJunction!18] (0.3313,2.9061)--(3.1669,0.0000)--(3.1384,6.6287)--(0.0000,8.5572)--cycle;
  \fill[anyonJunction!21] (0.3313,2.9061)--(0.6299,2.6001)--(0.8255,3.2074)--(0.2935,3.5511)--cycle;
  \fill[anyonJunction!21] (0.6299,2.6001)--(0.9423,2.2799)--(1.2688,2.9067)--(0.8255,3.2074)--cycle;
  \fill[anyonJunction!21] (0.9423,2.2799)--(1.2694,1.9447)--(1.6369,2.6319)--(1.2688,2.9067)--cycle;
  \fill[anyonJunction!21] (1.2694,1.9447)--(1.6123,1.5933)--(1.9485,2.3614)--(1.6369,2.6319)--cycle;
  \fill[anyonJunction!21] (1.6123,1.5933)--(1.9721,1.2245)--(2.2271,2.0701)--(1.9485,2.3614)--cycle;
  \fill[anyonJunction!21] (1.9721,1.2245)--(2.3502,0.8370)--(2.4999,1.7283)--(2.2271,2.0701)--cycle;
  \fill[anyonJunction!21] (2.3502,0.8370)--(2.7479,0.4294)--(2.7992,1.3004)--(2.4999,1.7283)--cycle;
  \fill[anyonJunction!21] (2.7479,0.4294)--(3.1669,0.0000)--(3.1637,0.7410)--(2.7992,1.3004)--cycle;
  \fill[anyonJunction!21] (0.2935,3.5511)--(0.8255,3.2074)--(0.9602,3.8485)--(0.2547,4.2124)--cycle;
  \fill[anyonJunction!21] (0.8255,3.2074)--(1.2688,2.9067)--(1.5014,3.5651)--(0.9602,3.8485)--cycle;
  \fill[anyonJunction!21] (1.2688,2.9067)--(1.6369,2.6319)--(1.8994,3.3342)--(1.5014,3.5651)--cycle;
  \fill[anyonJunction!21] (1.6369,2.6319)--(1.9485,2.3614)--(2.1874,3.1208)--(1.8994,3.3342)--cycle;
  \fill[anyonJunction!21] (1.9485,2.3614)--(2.2271,2.0701)--(2.4065,2.8840)--(2.1874,3.1208)--cycle;
  \fill[anyonJunction!21] (2.2271,2.0701)--(2.4999,1.7283)--(2.6042,2.5767)--(2.4065,2.8840)--cycle;
  \fill[anyonJunction!21] (2.4999,1.7283)--(2.7992,1.3004)--(2.8339,2.1420)--(2.6042,2.5767)--cycle;
  \fill[anyonJunction!21] (2.7992,1.3004)--(3.1637,0.7410)--(3.1605,1.5047)--(2.8339,2.1420)--cycle;
  \fill[anyonJunction!21] (0.2547,4.2124)--(0.9602,3.8485)--(1.0308,4.5227)--(0.2150,4.8904)--cycle;
  \fill[anyonJunction!21] (0.9602,3.8485)--(1.5014,3.5651)--(1.6351,4.2546)--(1.0308,4.5227)--cycle;
  \fill[anyonJunction!21] (1.5014,3.5651)--(1.8994,3.3342)--(2.0531,4.0528)--(1.6351,4.2546)--cycle;
  \fill[anyonJunction!21] (1.8994,3.3342)--(2.1874,3.1208)--(2.3271,3.8751)--(2.0531,4.0528)--cycle;
  \fill[anyonJunction!21] (2.1874,3.1208)--(2.4065,2.8840)--(2.5106,3.6727)--(2.3271,3.8751)--cycle;
  \fill[anyonJunction!21] (2.4065,2.8840)--(2.6042,2.5767)--(2.6632,3.3903)--(2.5106,3.6727)--cycle;
  \fill[anyonJunction!21] (2.6042,2.5767)--(2.8339,2.1420)--(2.8519,2.9605)--(2.6632,3.3903)--cycle;
  \fill[anyonJunction!21] (2.8339,2.1420)--(3.1605,1.5047)--(3.1571,2.2922)--(2.8519,2.9605)--cycle;
  \fill[anyonJunction!21] (0.2150,4.8904)--(1.0308,4.5227)--(1.0342,5.2295)--(0.1742,5.5860)--cycle;
  \fill[anyonJunction!21] (1.0308,4.5227)--(1.6351,4.2546)--(1.6652,4.9750)--(1.0342,5.2295)--cycle;
  \fill[anyonJunction!21] (1.6351,4.2546)--(2.0531,4.0528)--(2.0934,4.7887)--(1.6652,4.9750)--cycle;
  \fill[anyonJunction!21] (2.0531,4.0528)--(2.3271,3.8751)--(2.3651,4.6276)--(2.0934,4.7887)--cycle;
  \fill[anyonJunction!21] (2.3271,3.8751)--(2.5106,3.6727)--(2.5378,4.4423)--(2.3651,4.6276)--cycle;
  \fill[anyonJunction!21] (2.5106,3.6727)--(2.6632,3.3903)--(2.6764,4.1765)--(2.5378,4.4423)--cycle;
  \fill[anyonJunction!21] (2.6632,3.3903)--(2.8519,2.9605)--(2.8527,3.7618)--(2.6764,4.1765)--cycle;
  \fill[anyonJunction!21] (2.8519,2.9605)--(3.1571,2.2922)--(3.1536,3.1046)--(2.8527,3.7618)--cycle;
  \fill[anyonJunction!21] (0.1742,5.5860)--(1.0342,5.2295)--(0.9675,5.9679)--(0.1323,6.2997)--cycle;
  \fill[anyonJunction!21] (1.0342,5.2295)--(1.6652,4.9750)--(1.5870,5.7255)--(0.9675,5.9679)--cycle;
  \fill[anyonJunction!21] (1.6652,4.9750)--(2.0934,4.7887)--(2.0159,5.5430)--(1.5870,5.7255)--cycle;
  \fill[anyonJunction!21] (2.0934,4.7887)--(2.3651,4.6276)--(2.2972,5.3821)--(2.0159,5.5430)--cycle;
  \fill[anyonJunction!21] (2.3651,4.6276)--(2.5378,4.4423)--(2.4855,5.1987)--(2.2972,5.3821)--cycle;
  \fill[anyonJunction!21] (2.5378,4.4423)--(2.6764,4.1765)--(2.6420,4.9425)--(2.4855,5.1987)--cycle;
  \fill[anyonJunction!21] (2.6764,4.1765)--(2.8527,3.7618)--(2.8358,4.5519)--(2.6420,4.9425)--cycle;
  \fill[anyonJunction!21] (2.8527,3.7618)--(3.1536,3.1046)--(3.1499,3.9432)--(2.8358,4.5519)--cycle;
  \fill[anyonJunction!21] (0.1323,6.2997)--(0.9675,5.9679)--(0.8282,6.7365)--(0.0894,7.0322)--cycle;
  \fill[anyonJunction!21] (0.9675,5.9679)--(1.5870,5.7255)--(1.3965,6.5055)--(0.8282,6.7365)--cycle;
  \fill[anyonJunction!21] (1.5870,5.7255)--(2.0159,5.5430)--(1.8152,6.3169)--(1.3965,6.5055)--cycle;
  \fill[anyonJunction!21] (2.0159,5.5430)--(2.2972,5.3821)--(2.1184,6.1423)--(1.8152,6.3169)--cycle;
  \fill[anyonJunction!21] (2.2972,5.3821)--(2.4855,5.1987)--(2.3492,5.9481)--(2.1184,6.1423)--cycle;
  \fill[anyonJunction!21] (2.4855,5.1987)--(2.6420,4.9425)--(2.5574,5.6955)--(2.3492,5.9481)--cycle;
  \fill[anyonJunction!21] (2.6420,4.9425)--(2.8358,4.5519)--(2.7999,5.3371)--(2.5574,5.6955)--cycle;
  \fill[anyonJunction!21] (2.8358,4.5519)--(3.1499,3.9432)--(3.1462,4.8091)--(2.7999,5.3371)--cycle;
  \fill[anyonJunction!21] (0.0894,7.0322)--(0.8282,6.7365)--(0.6142,7.5339)--(0.0453,7.7845)--cycle;
  \fill[anyonJunction!21] (0.8282,6.7365)--(1.3965,6.5055)--(1.0899,7.3138)--(0.6142,7.5339)--cycle;
  \fill[anyonJunction!21] (1.3965,6.5055)--(1.8152,6.3169)--(1.4860,7.1119)--(1.0899,7.3138)--cycle;
  \fill[anyonJunction!21] (1.8152,6.3169)--(2.1184,6.1423)--(1.8220,6.9125)--(1.4860,7.1119)--cycle;
  \fill[anyonJunction!21] (2.1184,6.1423)--(2.3492,5.9481)--(2.1229,6.6970)--(1.8220,6.9125)--cycle;
  \fill[anyonJunction!21] (2.3492,5.9481)--(2.5574,5.6955)--(2.4184,6.4432)--(2.1229,6.6970)--cycle;
  \fill[anyonJunction!21] (2.5574,5.6955)--(2.7999,5.3371)--(2.7438,6.1239)--(2.4184,6.4432)--cycle;
  \fill[anyonJunction!21] (2.7999,5.3371)--(3.1462,4.8091)--(3.1424,5.7038)--(2.7438,6.1239)--cycle;
  \fill[anyonJunction!21] (0.0453,7.7845)--(0.6142,7.5339)--(0.3236,8.3583)--(0.0000,8.5572)--cycle;
  \fill[anyonJunction!21] (0.6142,7.5339)--(1.0899,7.3138)--(0.6639,8.1492)--(0.3236,8.3583)--cycle;
  \fill[anyonJunction!21] (1.0899,7.3138)--(1.4860,7.1119)--(1.0220,7.9292)--(0.6639,8.1492)--cycle;
  \fill[anyonJunction!21] (1.4860,7.1119)--(1.8220,6.9125)--(1.3995,7.6972)--(1.0220,7.9292)--cycle;
  \fill[anyonJunction!21] (1.8220,6.9125)--(2.1229,6.6970)--(1.7980,7.4523)--(1.3995,7.6972)--cycle;
  \fill[anyonJunction!21] (2.1229,6.6970)--(2.4184,6.4432)--(2.2193,7.1935)--(1.7980,7.4523)--cycle;
  \fill[anyonJunction!21] (2.4184,6.4432)--(2.7438,6.1239)--(2.6653,6.9194)--(2.2193,7.1935)--cycle;
  \fill[anyonJunction!21] (2.7438,6.1239)--(3.1424,5.7038)--(3.1384,6.6287)--(2.6653,6.9194)--cycle;
  \fill[anyonJunction!25] (3.1669,0.0000)--(6.6403,2.7329)--(7.0000,8.4447)--(3.1384,6.6287)--cycle;
  \fill[anyonJunction!28] (3.1669,0.0000)--(3.6746,0.3995)--(3.6505,1.2744)--(3.1637,0.7410)--cycle;
  \fill[anyonJunction!28] (3.6746,0.3995)--(4.1584,0.7801)--(4.0444,1.6824)--(3.6505,1.2744)--cycle;
  \fill[anyonJunction!28] (4.1584,0.7801)--(4.6197,1.1431)--(4.3937,2.0071)--(4.0444,1.6824)--cycle;
  \fill[anyonJunction!28] (4.6197,1.1431)--(5.0603,1.4898)--(4.7383,2.2820)--(4.3937,2.0071)--cycle;
  \fill[anyonJunction!28] (5.0603,1.4898)--(5.4815,1.8211)--(5.1124,2.5348)--(4.7383,2.2820)--cycle;
  \fill[anyonJunction!28] (5.4815,1.8211)--(5.8844,2.1382)--(5.5456,2.7895)--(5.1124,2.5348)--cycle;
  \fill[anyonJunction!28] (5.8844,2.1382)--(6.2703,2.4418)--(6.0624,3.0670)--(5.5456,2.7895)--cycle;
  \fill[anyonJunction!28] (6.2703,2.4418)--(6.6403,2.7329)--(6.6813,3.3841)--(6.0624,3.0670)--cycle;
  \fill[anyonJunction!28] (3.1637,0.7410)--(3.6505,1.2744)--(3.6342,2.1190)--(3.1605,1.5047)--cycle;
  \fill[anyonJunction!28] (3.6505,1.2744)--(4.0444,1.6824)--(3.9651,2.5386)--(3.6342,2.1190)--cycle;
  \fill[anyonJunction!28] (4.0444,1.6824)--(4.3937,2.0071)--(4.2348,2.8340)--(3.9651,2.5386)--cycle;
  \fill[anyonJunction!28] (4.3937,2.0071)--(4.7383,2.2820)--(4.5100,3.0583)--(4.2348,2.8340)--cycle;
  \fill[anyonJunction!28] (4.7383,2.2820)--(5.1124,2.5348)--(4.8494,3.2560)--(4.5100,3.0583)--cycle;
  \fill[anyonJunction!28] (5.1124,2.5348)--(5.5456,2.7895)--(5.3048,3.4659)--(4.8494,3.2560)--cycle;
  \fill[anyonJunction!28] (5.5456,2.7895)--(6.0624,3.0670)--(5.9193,3.7219)--(5.3048,3.4659)--cycle;
  \fill[anyonJunction!28] (6.0624,3.0670)--(6.6813,3.3841)--(6.7234,4.0518)--(5.9193,3.7219)--cycle;
  \fill[anyonJunction!28] (3.1605,1.5047)--(3.6342,2.1190)--(3.6257,2.9395)--(3.1571,2.2922)--cycle;
  \fill[anyonJunction!28] (3.6342,2.1190)--(3.9651,2.5386)--(3.9202,3.3572)--(3.6257,2.9395)--cycle;
  \fill[anyonJunction!28] (3.9651,2.5386)--(4.2348,2.8340)--(4.1428,3.6306)--(3.9202,3.3572)--cycle;
  \fill[anyonJunction!28] (4.2348,2.8340)--(4.5100,3.0583)--(4.3766,3.8231)--(4.1428,3.6306)--cycle;
  \fill[anyonJunction!28] (4.5100,3.0583)--(4.8494,3.2560)--(4.6957,3.9866)--(4.3766,3.8231)--cycle;
  \fill[anyonJunction!28] (4.8494,3.2560)--(5.3048,3.4659)--(5.1666,4.1675)--(4.6957,3.9866)--cycle;
  \fill[anyonJunction!28] (5.3048,3.4659)--(5.9193,3.7219)--(5.8444,4.4064)--(5.1666,4.1675)--cycle;
  \fill[anyonJunction!28] (5.9193,3.7219)--(6.7234,4.0518)--(6.7665,4.7367)--(5.8444,4.4064)--cycle;
  \fill[anyonJunction!28] (3.1571,2.2922)--(3.6257,2.9395)--(3.6253,3.7420)--(3.1536,3.1046)--cycle;
  \fill[anyonJunction!28] (3.6257,2.9395)--(3.9202,3.3572)--(3.9102,4.1459)--(3.6253,3.7420)--cycle;
  \fill[anyonJunction!28] (3.9202,3.3572)--(4.1428,3.6306)--(4.1187,4.4040)--(3.9102,4.1459)--cycle;
  \fill[anyonJunction!28] (4.1428,3.6306)--(4.3766,3.8231)--(4.3404,4.5805)--(4.1187,4.4040)--cycle;
  \fill[anyonJunction!28] (4.3766,3.8231)--(4.6957,3.9866)--(4.6554,4.7283)--(4.3404,4.5805)--cycle;
  \fill[anyonJunction!28] (4.6957,3.9866)--(5.1666,4.1675)--(5.1355,4.8944)--(4.6554,4.7283)--cycle;
  \fill[anyonJunction!28] (5.1666,4.1675)--(5.8444,4.4064)--(5.8408,5.1203)--(5.1355,4.8944)--cycle;
  \fill[anyonJunction!28] (5.8444,4.4064)--(6.7665,4.7367)--(6.8107,5.4396)--(5.8408,5.1203)--cycle;
  \fill[anyonJunction!28] (3.1536,3.1046)--(3.6253,3.7420)--(3.6333,4.5328)--(3.1499,3.9432)--cycle;
  \fill[anyonJunction!28] (3.6253,3.7420)--(3.9102,4.1459)--(3.9363,4.9124)--(3.6333,4.5328)--cycle;
  \fill[anyonJunction!28] (3.9102,4.1459)--(4.1187,4.4040)--(4.1650,5.1605)--(3.9363,4.9124)--cycle;
  \fill[anyonJunction!28] (4.1187,4.4040)--(4.3404,4.5805)--(4.4052,5.3349)--(4.1650,5.1605)--cycle;
  \fill[anyonJunction!28] (4.3404,4.5805)--(4.6554,4.7283)--(4.7329,5.4831)--(4.4052,5.3349)--cycle;
  \fill[anyonJunction!28] (4.6554,4.7283)--(5.1355,4.8944)--(5.2163,5.6467)--(4.7329,5.4831)--cycle;
  \fill[anyonJunction!28] (5.1355,4.8944)--(5.8408,5.1203)--(5.9116,5.8628)--(5.2163,5.6467)--cycle;
  \fill[anyonJunction!28] (5.8408,5.1203)--(6.8107,5.4396)--(6.8562,6.1610)--(5.9116,5.8628)--cycle;
  \fill[anyonJunction!28] (3.1499,3.9432)--(3.6333,4.5328)--(3.6501,5.3180)--(3.1462,4.8091)--cycle;
  \fill[anyonJunction!28] (3.6333,4.5328)--(3.9363,4.9124)--(4.0007,5.6642)--(3.6501,5.3180)--cycle;
  \fill[anyonJunction!28] (3.9363,4.9124)--(4.1650,5.1605)--(4.2855,5.9070)--(4.0007,5.6642)--cycle;
  \fill[anyonJunction!28] (4.1650,5.1605)--(4.4052,5.3349)--(4.5758,6.0911)--(4.2855,5.9070)--cycle;
  \fill[anyonJunction!28] (4.4052,5.3349)--(4.7329,5.4831)--(4.9337,6.2530)--(4.5758,6.0911)--cycle;
  \fill[anyonJunction!28] (4.7329,5.4831)--(5.2163,5.6467)--(5.4133,6.4246)--(4.9337,6.2530)--cycle;
  \fill[anyonJunction!28] (5.2163,5.6467)--(5.9116,5.8628)--(6.0595,6.6334)--(5.4133,6.4246)--cycle;
  \fill[anyonJunction!28] (5.9116,5.8628)--(6.8562,6.1610)--(6.9028,6.9019)--(6.0595,6.6334)--cycle;
  \fill[anyonJunction!28] (3.1462,4.8091)--(3.6501,5.3180)--(3.6766,6.1044)--(3.1424,5.7038)--cycle;
  \fill[anyonJunction!28] (3.6501,5.3180)--(4.0007,5.6642)--(4.1064,6.4090)--(3.6766,6.1044)--cycle;
  \fill[anyonJunction!28] (4.0007,5.6642)--(4.2855,5.9070)--(4.4860,6.6504)--(4.1064,6.4090)--cycle;
  \fill[anyonJunction!28] (4.2855,5.9070)--(4.5758,6.0911)--(4.8594,6.8538)--(4.4860,6.6504)--cycle;
  \fill[anyonJunction!28] (4.5758,6.0911)--(4.9337,6.2530)--(5.2638,7.0404)--(4.8594,6.8538)--cycle;
  \fill[anyonJunction!28] (4.9337,6.2530)--(5.4133,6.4246)--(5.7311,7.2278)--(5.2638,7.0404)--cycle;
  \fill[anyonJunction!28] (5.4133,6.4246)--(6.0595,6.6334)--(6.2871,7.4312)--(5.7311,7.2278)--cycle;
  \fill[anyonJunction!28] (6.0595,6.6334)--(6.9028,6.9019)--(6.9508,7.6628)--(6.2871,7.4312)--cycle;
  \fill[anyonJunction!28] (3.1424,5.7038)--(3.6766,6.1044)--(3.7136,6.8992)--(3.1384,6.6287)--cycle;
  \fill[anyonJunction!28] (3.6766,6.1044)--(4.1064,6.4090)--(4.2581,7.1553)--(3.7136,6.8992)--cycle;
  \fill[anyonJunction!28] (4.1064,6.4090)--(4.4860,6.6504)--(4.7744,7.3981)--(4.2581,7.1553)--cycle;
  \fill[anyonJunction!28] (4.4860,6.6504)--(4.8594,6.8538)--(5.2646,7.6286)--(4.7744,7.3981)--cycle;
  \fill[anyonJunction!28] (4.8594,6.8538)--(5.2638,7.0404)--(5.7305,7.8477)--(5.2646,7.6286)--cycle;
  \fill[anyonJunction!28] (5.2638,7.0404)--(5.7311,7.2278)--(6.1741,8.0563)--(5.7305,7.8477)--cycle;
  \fill[anyonJunction!28] (5.7311,7.2278)--(6.2871,7.4312)--(6.5967,8.2551)--(6.1741,8.0563)--cycle;
  \fill[anyonJunction!28] (6.2871,7.4312)--(6.9508,7.6628)--(7.0000,8.4447)--(6.5967,8.2551)--cycle;
  \draw[anyonRho!35,dashed,line width=0.5pt] (6.6997,3.6758)--(6.3349,3.6557)--(5.9972,3.6373)--(5.6844,3.6205)--(5.3944,3.6054)--(5.1249,3.5921)--(4.8739,3.5807)--(4.6391,3.5712)--(4.4183,3.5638)--(4.2091,3.5584)--(4.0095,3.5552)--(3.8171,3.5541)--(3.6296,3.5554)--(3.4448,3.5590)--(3.2604,3.5650)--(3.0741,3.5734)--(2.8837,3.5844)--(2.6871,3.5979)--(2.4818,3.6139)--(2.2658,3.6327)--(2.0369,3.6540)--(1.7930,3.6781)--(1.5320,3.7050)--(1.2518,3.7345)--(0.9504,3.7669)--(0.6260,3.8021)--(0.2765,3.8401);
  \draw[anyonRho!58,line width=0.6pt] (0.2765,3.8401)--(0.4800,3.7232)--(0.6710,3.6134)--(0.8497,3.5102)--(1.0163,3.4130)--(1.1713,3.3213)--(1.3151,3.2344)--(1.4483,3.1516)--(1.5716,3.0722)--(1.6856,2.9954)--(1.7912,2.9203)--(1.8891,2.8461)--(1.9803,2.7718)--(2.0657,2.6964)--(2.1462,2.6190)--(2.2230,2.5384)--(2.2971,2.4537)--(2.3695,2.3636)--(2.4415,2.2668)--(2.5143,2.1621)--(2.5891,2.0482)--(2.6673,1.9234)--(2.7502,1.7861)--(2.8395,1.6347)--(2.9367,1.4670)--(3.0437,1.2810)--(3.1623,1.0742);
  \draw[anyonRho!58,line width=0.6pt] (3.1623,1.0742)--(3.3270,1.2723)--(3.4756,1.4507)--(3.6105,1.6116)--(3.7339,1.7570)--(3.8479,1.8888)--(3.9543,2.0085)--(4.0550,2.1177)--(4.1516,2.2179)--(4.2457,2.3102)--(4.3388,2.3960)--(4.4323,2.4764)--(4.5277,2.5525)--(4.6262,2.6253)--(4.7292,2.6958)--(4.8379,2.7650)--(4.9534,2.8338)--(5.0768,2.9031)--(5.2094,2.9738)--(5.3520,3.0466)--(5.5055,3.1223)--(5.6710,3.2017)--(5.8490,3.2854)--(6.0404,3.3741)--(6.2456,3.4684)--(6.4653,3.5688)--(6.6997,3.6758);
  \draw[anyonRho!35,dashed,line width=0.5pt] (6.7543,4.5432)--(6.3229,4.5111)--(5.9328,4.4812)--(5.5809,4.4537)--(5.2640,4.4286)--(4.9789,4.4060)--(4.7222,4.3862)--(4.4906,4.3691)--(4.2806,4.3550)--(4.0889,4.3439)--(3.9118,4.3359)--(3.7460,4.3311)--(3.5877,4.3297)--(3.4336,4.3317)--(3.2800,4.3372)--(3.1233,4.3463)--(2.9601,4.3591)--(2.7868,4.3756)--(2.6000,4.3959)--(2.3963,4.4200)--(2.1723,4.4480)--(1.9248,4.4800)--(1.6505,4.5158)--(1.3465,4.5556)--(1.0096,4.5994)--(0.6371,4.6472)--(0.2262,4.6988);
  \draw[anyonRho!58,line width=0.6pt] (0.2262,4.6988)--(0.4918,4.5739)--(0.7378,4.4595)--(0.9641,4.3552)--(1.1712,4.2603)--(1.3594,4.1739)--(1.5295,4.0952)--(1.6822,4.0231)--(1.8185,3.9566)--(1.9397,3.8944)--(2.0468,3.8354)--(2.1413,3.7783)--(2.2246,3.7217)--(2.2981,3.6642)--(2.3635,3.6045)--(2.4223,3.5411)--(2.4762,3.4723)--(2.5269,3.3966)--(2.5761,3.3123)--(2.6257,3.2175)--(2.6776,3.1103)--(2.7337,2.9885)--(2.7962,2.8499)--(2.8672,2.6918)--(2.9494,2.5113)--(3.0453,2.3051)--(3.1580,2.0693);
  \draw[anyonRho!58,line width=0.6pt] (3.1580,2.0693)--(3.3273,2.2971)--(3.4728,2.4966)--(3.5985,2.6715)--(3.7078,2.8249)--(3.8039,2.9595)--(3.8895,3.0777)--(3.9675,3.1817)--(4.0401,3.2735)--(4.1099,3.3549)--(4.1791,3.4277)--(4.2498,3.4934)--(4.3242,3.5537)--(4.4043,3.6099)--(4.4921,3.6634)--(4.5894,3.7156)--(4.6983,3.7678)--(4.8204,3.8213)--(4.9575,3.8773)--(5.1112,3.9369)--(5.2828,4.0014)--(5.4739,4.0717)--(5.6855,4.1489)--(5.9187,4.2338)--(6.1742,4.3274)--(6.4526,4.4303)--(6.7543,4.5432);
  \draw[anyonRho!32,dashed,line width=0.55pt] (1.8936,2.8632)--(1.9188,2.9225)--(1.9434,2.9826)--(1.9672,3.0435)--(1.9903,3.1051)--(2.0128,3.1676)--(2.0345,3.2308)--(2.0554,3.2948)--(2.0757,3.3596)--(2.0952,3.4252)--(2.1139,3.4915)--(2.1319,3.5587)--(2.1491,3.6266)--(2.1656,3.6953)--(2.1812,3.7648)--(2.1961,3.8350)--(2.2102,3.9060)--(2.2235,3.9778)--(2.2360,4.0503)--(2.2477,4.1236)--(2.2585,4.1977)--(2.2686,4.2725)--(2.2778,4.3480)--(2.2861,4.4243)--(2.2936,4.5014)--(2.3003,4.5792)--(2.3061,4.6577)--(2.3111,4.7369)--(2.3152,4.8169)--(2.3184,4.8976)--(2.3208,4.9790)--(2.3222,5.0611)--(2.3228,5.1439)--(2.3225,5.2274)--(2.3213,5.3116);
  \draw[anyonRho!32,dashed,line width=0.55pt] (3.4658,2.8201)--(3.4638,2.8752)--(3.4618,2.9313)--(3.4599,2.9883)--(3.4580,3.0462)--(3.4562,3.1051)--(3.4544,3.1649)--(3.4527,3.2256)--(3.4511,3.2872)--(3.4495,3.3498)--(3.4480,3.4134)--(3.4465,3.4778)--(3.4451,3.5432)--(3.4437,3.6095)--(3.4424,3.6767)--(3.4412,3.7448)--(3.4400,3.8139)--(3.4389,3.8838)--(3.4379,3.9547)--(3.4369,4.0265)--(3.4360,4.0992)--(3.4352,4.1728)--(3.4344,4.2473)--(3.4337,4.3226)--(3.4330,4.3989)--(3.4324,4.4760)--(3.4319,4.5540)--(3.4315,4.6329)--(3.4311,4.7126)--(3.4308,4.7932)--(3.4306,4.8746)--(3.4304,4.9569)--(3.4303,5.0400)--(3.4303,5.1239)--(3.4303,5.2087);
  \draw[anyonRho!32,dashed,line width=0.55pt] (5.0480,2.7766)--(5.0195,2.8378)--(4.9917,2.8997)--(4.9648,2.9623)--(4.9386,3.0256)--(4.9133,3.0897)--(4.8887,3.1546)--(4.8650,3.2201)--(4.8421,3.2864)--(4.8201,3.3535)--(4.7989,3.4212)--(4.7786,3.4897)--(4.7592,3.5590)--(4.7406,3.6289)--(4.7229,3.6996)--(4.7061,3.7710)--(4.6902,3.8432)--(4.6752,3.9160)--(4.6611,3.9896)--(4.6479,4.0638)--(4.6357,4.1388)--(4.6243,4.2145)--(4.6140,4.2909)--(4.6045,4.3680)--(4.5960,4.4458)--(4.5885,4.5243)--(4.5819,4.6035)--(4.5763,4.6833)--(4.5717,4.7638)--(4.5681,4.8451)--(4.5654,4.9269)--(4.5638,5.0095)--(4.5631,5.0927)--(4.5635,5.1766)--(4.5648,5.2611);
  \draw[anyonRho,line width=0.9pt] (0.9423,2.2799)--(0.9854,2.3520)--(1.0273,2.4246)--(1.0679,2.4976)--(1.1073,2.5710)--(1.1454,2.6448)--(1.1823,2.7191)--(1.2179,2.7938)--(1.2521,2.8690)--(1.2851,2.9446)--(1.3168,3.0206)--(1.3472,3.0971)--(1.3762,3.1740)--(1.4040,3.2513)--(1.4303,3.3291)--(1.4554,3.4073)--(1.4791,3.4860)--(1.5014,3.5651)--(1.5224,3.6446)--(1.5419,3.7245)--(1.5601,3.8049)--(1.5769,3.8857)--(1.5923,3.9669)--(1.6063,4.0486)--(1.6189,4.1307)--(1.6301,4.2132)--(1.6398,4.2961)--(1.6481,4.3795)--(1.6549,4.4633)--(1.6603,4.5475)--(1.6642,4.6322)--(1.6667,4.7172)--(1.6677,4.8027)--(1.6672,4.8886)--(1.6652,4.9750);
  \draw[anyonRho,line width=0.9pt] (1.6123,1.5933)--(1.6569,1.6842)--(1.7001,1.7750)--(1.7420,1.8656)--(1.7826,1.9560)--(1.8218,2.0463)--(1.8597,2.1365)--(1.8962,2.2265)--(1.9314,2.3165)--(1.9653,2.4062)--(1.9978,2.4959)--(2.0289,2.5855)--(2.0587,2.6749)--(2.0872,2.7643)--(2.1143,2.8536)--(2.1400,2.9427)--(2.1644,3.0318)--(2.1874,3.1208)--(2.2090,3.2098)--(2.2293,3.2987)--(2.2481,3.3875)--(2.2656,3.4762)--(2.2817,3.5649)--(2.2965,3.6536)--(2.3098,3.7422)--(2.3217,3.8308)--(2.3322,3.9194)--(2.3413,4.0079)--(2.3490,4.0965)--(2.3553,4.1850)--(2.3601,4.2735)--(2.3635,4.3620)--(2.3655,4.4505)--(2.3660,4.5391)--(2.3651,4.6276);
  \draw[anyonRho,line width=0.9pt] (2.3502,0.8370)--(2.3702,0.9444)--(2.3896,1.0510)--(2.4083,1.1570)--(2.4264,1.2623)--(2.4438,1.3670)--(2.4606,1.4710)--(2.4768,1.5744)--(2.4924,1.6771)--(2.5073,1.7793)--(2.5216,1.8809)--(2.5353,1.9819)--(2.5483,2.0824)--(2.5607,2.1823)--(2.5725,2.2817)--(2.5837,2.3805)--(2.5942,2.4789)--(2.6042,2.5767)--(2.6135,2.6741)--(2.6221,2.7710)--(2.6302,2.8675)--(2.6376,2.9635)--(2.6444,3.0590)--(2.6506,3.1542)--(2.6561,3.2489)--(2.6610,3.3433)--(2.6653,3.4372)--(2.6689,3.5308)--(2.6719,3.6240)--(2.6743,3.7169)--(2.6760,3.8094)--(2.6771,3.9017)--(2.6775,3.9936)--(2.6773,4.0851)--(2.6764,4.1765);
  \draw[anyonRho,line width=0.9pt] (4.1584,0.7801)--(4.1431,0.8889)--(4.1284,0.9970)--(4.1141,1.1044)--(4.1004,1.2110)--(4.0871,1.3170)--(4.0743,1.4222)--(4.0620,1.5268)--(4.0501,1.6307)--(4.0388,1.7340)--(4.0279,1.8366)--(4.0175,1.9386)--(4.0076,2.0401)--(3.9981,2.1409)--(3.9892,2.2412)--(3.9807,2.3409)--(3.9726,2.4400)--(3.9651,2.5386)--(3.9580,2.6367)--(3.9514,2.7343)--(3.9453,2.8314)--(3.9397,2.9280)--(3.9345,3.0242)--(3.9298,3.1199)--(3.9256,3.2151)--(3.9219,3.3099)--(3.9187,3.4043)--(3.9159,3.4983)--(3.9136,3.5919)--(3.9118,3.6851)--(3.9105,3.7779)--(3.9097,3.8704)--(3.9094,3.9626)--(3.9096,4.0544)--(3.9102,4.1459);
  \draw[anyonRho,line width=0.9pt] (5.0603,1.4898)--(5.0176,1.5839)--(4.9761,1.6778)--(4.9360,1.7714)--(4.8971,1.8648)--(4.8595,1.9579)--(4.8233,2.0508)--(4.7883,2.1434)--(4.7546,2.2358)--(4.7222,2.3280)--(4.6912,2.4200)--(4.6614,2.5118)--(4.6329,2.6033)--(4.6057,2.6947)--(4.5798,2.7858)--(4.5552,2.8768)--(4.5320,2.9677)--(4.5100,3.0583)--(4.4893,3.1488)--(4.4700,3.2391)--(4.4520,3.3293)--(4.4353,3.4194)--(4.4199,3.5093)--(4.4059,3.5991)--(4.3932,3.6888)--(4.3818,3.7783)--(4.3718,3.8678)--(4.3631,3.9571)--(4.3557,4.0464)--(4.3498,4.1356)--(4.3452,4.2247)--(4.3419,4.3137)--(4.3400,4.4027)--(4.3396,4.4916)--(4.3404,4.5805);
  \draw[anyonRho,line width=0.9pt] (5.8844,2.1382)--(5.8396,2.2135)--(5.7961,2.2892)--(5.7539,2.3652)--(5.7131,2.4416)--(5.6735,2.5183)--(5.6353,2.5954)--(5.5984,2.6728)--(5.5629,2.7505)--(5.5287,2.8286)--(5.4959,2.9071)--(5.4644,2.9859)--(5.4343,3.0650)--(5.4056,3.1445)--(5.3783,3.2243)--(5.3524,3.3045)--(5.3279,3.3850)--(5.3048,3.4659)--(5.2831,3.5471)--(5.2629,3.6287)--(5.2441,3.7106)--(5.2267,3.7929)--(5.2108,3.8755)--(5.1963,3.9585)--(5.1833,4.0418)--(5.1718,4.1255)--(5.1617,4.2095)--(5.1532,4.2939)--(5.1461,4.3786)--(5.1405,4.4637)--(5.1365,4.5492)--(5.1339,4.6349)--(5.1329,4.7211)--(5.1335,4.8076)--(5.1355,4.8944);
  \fill[anyonJunction!60] (0.1742,5.5860)--(6.8107,5.4396)--(3.1536,3.1046)--cycle;
  \fill[anyonRho!24] (0.1742,5.5860)--(0.4637,5.4639)--(0.7311,5.3531)--(0.9763,5.2530)--(1.1996,5.1628)--(1.4013,5.0817)--(1.5822,5.0087)--(1.7433,4.9429)--(1.8856,4.8829)--(2.0104,4.8278)--(2.1192,4.7760)--(2.2135,4.7264)--(2.2949,4.6774)--(2.3651,4.6276)--(2.4259,4.5757)--(2.4791,4.5199)--(2.5266,4.4586)--(2.5703,4.3903)--(2.6123,4.3131)--(2.6545,4.2251)--(2.6991,4.1243)--(2.7483,4.0084)--(2.8045,3.8750)--(2.8702,3.7213)--(2.9481,3.5442)--(3.0414,3.3400)--(3.1536,3.1046)--(3.1536,3.1046)--(3.3267,3.3325)--(3.4731,3.5305)--(3.5973,3.7027)--(3.7031,3.8522)--(3.7942,3.9822)--(3.8738,4.0951)--(3.9449,4.1933)--(4.0102,4.2789)--(4.0725,4.3537)--(4.1342,4.4197)--(4.1979,4.4784)--(4.2659,4.5315)--(4.3404,4.5805)--(4.4239,4.6267)--(4.5184,4.6717)--(4.6261,4.7169)--(4.7491,4.7634)--(4.8892,4.8127)--(5.0483,4.8659)--(5.2281,4.9243)--(5.4301,4.9889)--(5.6556,5.0609)--(5.9057,5.1411)--(6.1813,5.2305)--(6.4829,5.3298)--(6.8107,5.4396)--(6.8107,5.4396)--(6.3516,5.4053)--(5.9391,5.3733)--(5.5698,5.3437)--(5.2401,5.3167)--(4.9463,5.2923)--(4.6847,5.2708)--(4.4516,5.2521)--(4.2430,5.2366)--(4.0551,5.2242)--(3.8839,5.2151)--(3.7254,5.2094)--(3.5755,5.2072)--(3.4303,5.2087)--(3.2857,5.2138)--(3.1376,5.2227)--(2.9819,5.2355)--(2.8149,5.2522)--(2.6324,5.2729)--(2.4306,5.2977)--(2.2058,5.3265)--(1.9542,5.3594)--(1.6723,5.3965)--(1.3566,5.4377)--(1.0037,5.4830)--(0.6106,5.5324)--(0.1742,5.5860)--cycle;
  \draw[anyonJunction,line width=0.9pt] (0.1742,5.5860)--(6.8107,5.4396)--(3.1536,3.1046)--(0.1742,5.5860);
  \draw[anyonRho,line width=1.05pt] (0.1742,5.5860)--(0.4637,5.4639)--(0.7311,5.3531)--(0.9763,5.2530)--(1.1996,5.1628)--(1.4013,5.0817)--(1.5822,5.0087)--(1.7433,4.9429)--(1.8856,4.8829)--(2.0104,4.8278)--(2.1192,4.7760)--(2.2135,4.7264)--(2.2949,4.6774)--(2.3651,4.6276)--(2.4259,4.5757)--(2.4791,4.5199)--(2.5266,4.4586)--(2.5703,4.3903)--(2.6123,4.3131)--(2.6545,4.2251)--(2.6991,4.1243)--(2.7483,4.0084)--(2.8045,3.8750)--(2.8702,3.7213)--(2.9481,3.5442)--(3.0414,3.3400)--(3.1536,3.1046)--(3.1536,3.1046)--(3.3267,3.3325)--(3.4731,3.5305)--(3.5973,3.7027)--(3.7031,3.8522)--(3.7942,3.9822)--(3.8738,4.0951)--(3.9449,4.1933)--(4.0102,4.2789)--(4.0725,4.3537)--(4.1342,4.4197)--(4.1979,4.4784)--(4.2659,4.5315)--(4.3404,4.5805)--(4.4239,4.6267)--(4.5184,4.6717)--(4.6261,4.7169)--(4.7491,4.7634)--(4.8892,4.8127)--(5.0483,4.8659)--(5.2281,4.9243)--(5.4301,4.9889)--(5.6556,5.0609)--(5.9057,5.1411)--(6.1813,5.2305)--(6.4829,5.3298)--(6.8107,5.4396)--(6.8107,5.4396)--(6.3516,5.4053)--(5.9391,5.3733)--(5.5698,5.3437)--(5.2401,5.3167)--(4.9463,5.2923)--(4.6847,5.2708)--(4.4516,5.2521)--(4.2430,5.2366)--(4.0551,5.2242)--(3.8839,5.2151)--(3.7254,5.2094)--(3.5755,5.2072)--(3.4303,5.2087)--(3.2857,5.2138)--(3.1376,5.2227)--(2.9819,5.2355)--(2.8149,5.2522)--(2.6324,5.2729)--(2.4306,5.2977)--(2.2058,5.3265)--(1.9542,5.3594)--(1.6723,5.3965)--(1.3566,5.4377)--(1.0037,5.4830)--(0.6106,5.5324)--(0.1742,5.5860)--(0.1742,5.5860);
  \draw[anyonRho!35,dashed,line width=0.5pt] (6.8691,6.3665)--(6.4233,6.3384)--(6.0201,6.3122)--(5.6564,6.2880)--(5.3288,6.2660)--(5.0339,6.2462)--(4.7685,6.2288)--(4.5290,6.2139)--(4.3118,6.2015)--(4.1135,6.1917)--(3.9303,6.1847)--(3.7587,6.1805)--(3.5950,6.1793)--(3.4355,6.1810)--(3.2766,6.1859)--(3.1145,6.1939)--(2.9457,6.2051)--(2.7664,6.2195)--(2.5732,6.2373)--(2.3626,6.2585)--(2.1310,6.2831)--(1.8751,6.3111)--(1.5916,6.3425)--(1.2774,6.3774)--(0.9294,6.4158)--(0.5447,6.4576)--(0.1204,6.5029);
  \draw[anyonRho!58,line width=0.6pt] (0.1204,6.5029)--(0.3939,6.3934)--(0.6473,6.2931)--(0.8807,6.2016)--(1.0943,6.1183)--(1.2886,6.0425)--(1.4641,5.9733)--(1.6219,5.9100)--(1.7628,5.8515)--(1.8880,5.7968)--(1.9987,5.7449)--(2.0964,5.6946)--(2.1826,5.6448)--(2.2586,5.5942)--(2.3262,5.5416)--(2.3870,5.4857)--(2.4427,5.4250)--(2.4951,5.3583)--(2.5459,5.2839)--(2.5972,5.2002)--(2.6508,5.1055)--(2.7088,4.9978)--(2.7735,4.8752)--(2.8471,4.7353)--(2.9322,4.5753)--(3.0318,4.3923)--(3.1489,4.1828);
  \draw[anyonRho!58,line width=0.6pt] (3.1489,4.1828)--(3.3252,4.3852)--(3.4766,4.5623)--(3.6072,4.7173)--(3.7207,4.8531)--(3.8204,4.9721)--(3.9091,5.0766)--(3.9899,5.1685)--(4.0651,5.2496)--(4.1374,5.3215)--(4.2090,5.3857)--(4.2822,5.4437)--(4.3591,5.4968)--(4.4420,5.5463)--(4.5328,5.5934)--(4.6336,5.6394)--(4.7462,5.6854)--(4.8725,5.7325)--(5.0143,5.7817)--(5.1732,5.8342)--(5.3507,5.8909)--(5.5481,5.9527)--(5.7667,6.0205)--(6.0074,6.0951)--(6.2710,6.1772)--(6.5582,6.2675)--(6.8691,6.3665);
  \draw[anyonRho!35,dashed,line width=0.5pt] (6.9295,7.3255)--(6.5399,7.3101)--(6.1790,7.2960)--(5.8447,7.2831)--(5.5347,7.2716)--(5.2467,7.2614)--(4.9782,7.2526)--(4.7271,7.2454)--(4.4909,7.2397)--(4.2672,7.2356)--(4.0536,7.2331)--(3.8477,7.2323)--(3.6471,7.2333)--(3.4494,7.2360)--(3.2521,7.2406)--(3.0529,7.2471)--(2.8493,7.2555)--(2.6389,7.2658)--(2.4195,7.2781)--(2.1886,7.2924)--(1.9440,7.3088)--(1.6834,7.3272)--(1.4045,7.3478)--(1.1054,7.3704)--(0.7837,7.3951)--(0.4375,7.4220)--(0.0648,7.4510);
  \draw[anyonRho!58,line width=0.6pt] (0.0648,7.4510)--(0.2803,7.3617)--(0.4827,7.2777)--(0.6723,7.1986)--(0.8493,7.1241)--(1.0141,7.0537)--(1.1671,6.9869)--(1.3089,6.9232)--(1.4403,6.8621)--(1.5618,6.8029)--(1.6744,6.7450)--(1.7789,6.6877)--(1.8762,6.6303)--(1.9674,6.5721)--(2.0534,6.5122)--(2.1355,6.4498)--(2.2146,6.3842)--(2.2920,6.3142)--(2.3690,6.2391)--(2.4469,6.1577)--(2.5270,6.0690)--(2.6108,5.9718)--(2.6999,5.8647)--(2.7958,5.7463)--(2.9005,5.6150)--(3.0159,5.4691)--(3.1441,5.3065);
  \draw[anyonRho!58,line width=0.6pt] (3.1441,5.3065)--(3.3228,5.4623)--(3.4836,5.6023)--(3.6294,5.7283)--(3.7625,5.8420)--(3.8852,5.9448)--(3.9998,6.0381)--(4.1080,6.1232)--(4.2117,6.2011)--(4.3126,6.2728)--(4.4124,6.3394)--(4.5126,6.4018)--(4.6147,6.4607)--(4.7202,6.5171)--(4.8304,6.5716)--(4.9466,6.6251)--(5.0700,6.6783)--(5.2020,6.7318)--(5.3435,6.7863)--(5.4957,6.8424)--(5.6595,6.9007)--(5.8359,6.9618)--(6.0256,7.0261)--(6.2292,7.0943)--(6.4475,7.1666)--(6.6808,7.2436)--(6.9295,7.3255);
  \draw[anyonRho!32,dashed,line width=0.55pt] (2.3213,5.3116)--(2.3192,5.3965)--(2.3162,5.4820)--(2.3123,5.5682)--(2.3075,5.6551)--(2.3018,5.7427)--(2.2952,5.8309)--(2.2876,5.9197)--(2.2792,6.0092)--(2.2698,6.0993)--(2.2595,6.1900)--(2.2482,6.2813)--(2.2360,6.3733)--(2.2229,6.4658)--(2.2089,6.5589)--(2.1939,6.6526)--(2.1780,6.7469)--(2.1612,6.8417)--(2.1434,6.9370)--(2.1247,7.0330)--(2.1050,7.1294)--(2.0844,7.2263)--(2.0629,7.3238)--(2.0404,7.4218)--(2.0170,7.5202)--(1.9926,7.6192)--(1.9674,7.7186)--(1.9411,7.8184)--(1.9140,7.9187)--(1.8859,8.0195)--(1.8569,8.1206)--(1.8270,8.2222)--(1.7961,8.3242)--(1.7643,8.4266)--(1.7316,8.5294);
  \draw[anyonRho!32,dashed,line width=0.55pt] (3.4303,5.2087)--(3.4304,5.2942)--(3.4306,5.3806)--(3.4309,5.4677)--(3.4312,5.5557)--(3.4316,5.6444)--(3.4321,5.7338)--(3.4327,5.8241)--(3.4333,5.9150)--(3.4340,6.0067)--(3.4348,6.0991)--(3.4356,6.1922)--(3.4365,6.2861)--(3.4375,6.3806)--(3.4386,6.4758)--(3.4397,6.5716)--(3.4410,6.6681)--(3.4422,6.7653)--(3.4436,6.8630)--(3.4450,6.9614)--(3.4466,7.0604)--(3.4481,7.1600)--(3.4498,7.2601)--(3.4515,7.3608)--(3.4534,7.4621)--(3.4552,7.5638)--(3.4572,7.6661)--(3.4592,7.7689)--(3.4613,7.8722)--(3.4635,7.9760)--(3.4657,8.0802)--(3.4681,8.1849)--(3.4704,8.2900)--(3.4729,8.3955)--(3.4754,8.5013);
  \draw[anyonRho!32,dashed,line width=0.55pt] (4.5648,5.2611)--(4.5672,5.3462)--(4.5706,5.4320)--(4.5749,5.5185)--(4.5804,5.6055)--(4.5868,5.6932)--(4.5943,5.7815)--(4.6028,5.8704)--(4.6124,5.9599)--(4.6230,6.0500)--(4.6346,6.1407)--(4.6473,6.2320)--(4.6610,6.3239)--(4.6758,6.4163)--(4.6917,6.5092)--(4.7086,6.6028)--(4.7265,6.6968)--(4.7456,6.7914)--(4.7656,6.8866)--(4.7868,6.9822)--(4.8090,7.0784)--(4.8323,7.1750)--(4.8566,7.2722)--(4.8820,7.3698)--(4.9085,7.4679)--(4.9360,7.5665)--(4.9646,7.6656)--(4.9943,7.7650)--(5.0250,7.8650)--(5.0568,7.9653)--(5.0896,8.0661)--(5.1235,8.1673)--(5.1584,8.2688)--(5.1945,8.3708)--(5.2315,8.4731);
  \draw[anyonRho,line width=0.9pt,-{Stealth[length=3.8pt,width=2.8pt]}] (1.6652,4.9750)--(1.6617,5.0617)--(1.6567,5.1489)--(1.6502,5.2365)--(1.6421,5.3245)--(1.6326,5.4129)--(1.6215,5.5017)--(1.6089,5.5909)--(1.5947,5.6806)--(1.5790,5.7706)--(1.5617,5.8611)--(1.5428,5.9519)--(1.5224,6.0432)--(1.5004,6.1349)--(1.4768,6.2269)--(1.4517,6.3194)--(1.4249,6.4122)--(1.3965,6.5055)--(1.3666,6.5991)--(1.3350,6.6932)--(1.3018,6.7876)--(1.2669,6.8824)--(1.2305,6.9776)--(1.1924,7.0732)--(1.1526,7.1691)--(1.1112,7.2655)--(1.0682,7.3622)--(1.0235,7.4593)--(0.9771,7.5568)--(0.9291,7.6546)--(0.8794,7.7528)--(0.8281,7.8514)--(0.7750,7.9503)--(0.7203,8.0496)--(0.6639,8.1492);
  \draw[anyonRho,line width=0.9pt,-{Stealth[length=3.8pt,width=2.8pt]}] (2.3651,4.6276)--(2.3627,4.7162)--(2.3588,4.8048)--(2.3534,4.8935)--(2.3466,4.9822)--(2.3383,5.0710)--(2.3284,5.1598)--(2.3171,5.2486)--(2.3042,5.3376)--(2.2898,5.4266)--(2.2739,5.5157)--(2.2564,5.6049)--(2.2374,5.6942)--(2.2168,5.7836)--(2.1946,5.8731)--(2.1708,5.9627)--(2.1454,6.0524)--(2.1184,6.1423)--(2.0898,6.2323)--(2.0595,6.3225)--(2.0276,6.4128)--(1.9940,6.5032)--(1.9588,6.5939)--(1.9218,6.6847)--(1.8832,6.7756)--(1.8428,6.8668)--(1.8007,6.9582)--(1.7569,7.0498)--(1.7112,7.1415)--(1.6638,7.2336)--(1.6146,7.3258)--(1.5636,7.4183)--(1.5108,7.5110)--(1.4561,7.6040)--(1.3995,7.6972);
  \draw[anyonRho,line width=0.9pt,-{Stealth[length=3.8pt,width=2.8pt]}] (2.6764,4.1765)--(2.6749,4.2675)--(2.6727,4.3582)--(2.6698,4.4487)--(2.6663,4.5390)--(2.6621,4.6290)--(2.6572,4.7188)--(2.6517,4.8084)--(2.6454,4.8978)--(2.6385,4.9870)--(2.6309,5.0761)--(2.6225,5.1650)--(2.6135,5.2537)--(2.6037,5.3423)--(2.5933,5.4308)--(2.5821,5.5191)--(2.5701,5.6074)--(2.5574,5.6955)--(2.5440,5.7836)--(2.5298,5.8716)--(2.5148,5.9596)--(2.4991,6.0476)--(2.4826,6.1355)--(2.4652,6.2234)--(2.4471,6.3113)--(2.4282,6.3992)--(2.4084,6.4872)--(2.3878,6.5751)--(2.3664,6.6632)--(2.3441,6.7513)--(2.3209,6.8395)--(2.2968,6.9278)--(2.2719,7.0162)--(2.2460,7.1048)--(2.2193,7.1935);
  \draw[anyonRho,line width=0.9pt,-{Stealth[length=3.8pt,width=2.8pt]}] (3.9102,4.1459)--(3.9114,4.2371)--(3.9130,4.3280)--(3.9152,4.4186)--(3.9179,4.5089)--(3.9211,4.5990)--(3.9248,4.6888)--(3.9290,4.7784)--(3.9337,4.8678)--(3.9390,4.9569)--(3.9448,5.0459)--(3.9511,5.1347)--(3.9580,5.2233)--(3.9654,5.3117)--(3.9734,5.4000)--(3.9819,5.4882)--(3.9910,5.5762)--(4.0007,5.6642)--(4.0109,5.7520)--(4.0217,5.8398)--(4.0331,5.9274)--(4.0450,6.0151)--(4.0576,6.1026)--(4.0708,6.1902)--(4.0846,6.2777)--(4.0990,6.3653)--(4.1140,6.4528)--(4.1297,6.5404)--(4.1460,6.6280)--(4.1630,6.7156)--(4.1807,6.8034)--(4.1990,6.8912)--(4.2180,6.9791)--(4.2377,7.0671)--(4.2581,7.1553);
  \draw[anyonRho,line width=0.9pt,-{Stealth[length=3.8pt,width=2.8pt]}] (4.3404,4.5805)--(4.3427,4.6693)--(4.3464,4.7581)--(4.3515,4.8468)--(4.3581,4.9356)--(4.3660,5.0243)--(4.3754,5.1131)--(4.3862,5.2018)--(4.3985,5.2905)--(4.4122,5.3793)--(4.4274,5.4681)--(4.4441,5.5569)--(4.4623,5.6458)--(4.4819,5.7347)--(4.5031,5.8237)--(4.5258,5.9127)--(4.5501,6.0019)--(4.5758,6.0911)--(4.6032,6.1804)--(4.6321,6.2698)--(4.6626,6.3592)--(4.6947,6.4489)--(4.7284,6.5386)--(4.7638,6.6285)--(4.8008,6.7185)--(4.8394,6.8087)--(4.8797,6.8990)--(4.9217,6.9895)--(4.9655,7.0802)--(5.0109,7.1711)--(5.0581,7.2621)--(5.1070,7.3534)--(5.1577,7.4449)--(5.2102,7.5366)--(5.2646,7.6286);
  \draw[anyonRho,line width=0.9pt,-{Stealth[length=3.8pt,width=2.8pt]}] (5.1355,4.8944)--(5.1391,4.9816)--(5.1443,5.0691)--(5.1510,5.1570)--(5.1593,5.2453)--(5.1692,5.3339)--(5.1806,5.4228)--(5.1937,5.5121)--(5.2083,5.6018)--(5.2246,5.6918)--(5.2424,5.7822)--(5.2619,5.8729)--(5.2830,5.9639)--(5.3058,6.0554)--(5.3302,6.1471)--(5.3562,6.2393)--(5.3839,6.3317)--(5.4133,6.4246)--(5.4443,6.5177)--(5.4771,6.6113)--(5.5115,6.7052)--(5.5475,6.7994)--(5.5853,6.8940)--(5.6248,6.9889)--(5.6661,7.0842)--(5.7090,7.1798)--(5.7537,7.2758)--(5.8001,7.3722)--(5.8482,7.4689)--(5.8981,7.5659)--(5.9498,7.6633)--(6.0032,7.7610)--(6.0584,7.8591)--(6.1153,7.9575)--(6.1741,8.0563);
  \fill[anyonRho,fill opacity=0.045] (0.0000,8.5572)--(7.0000,8.4447)--(3.1384,6.6287)--cycle;
  \draw[black!38,dashed,line width=0.6pt] (0.3313,2.9061)--(6.6403,2.7329);
  \draw[black,line width=0.9pt] (0.3313,2.9061)--(3.1669,0.0000);
  \draw[black,line width=0.9pt] (3.1669,0.0000)--(6.6403,2.7329);
  \draw[black,line width=0.9pt] (0.0000,8.5572)--(7.0000,8.4447);
  \draw[black,line width=0.9pt] (7.0000,8.4447)--(3.1384,6.6287);
  \draw[black,line width=0.9pt] (3.1384,6.6287)--(0.0000,8.5572);
  \draw[black,line width=0.9pt] (0.3313,2.9061)--(0.0000,8.5572);
  \draw[black,line width=0.9pt] (6.6403,2.7329)--(7.0000,8.4447);
  \draw[black,line width=0.9pt] (3.1669,0.0000)--(3.1384,6.6287);
  \fill[black] (0.3313,2.9061) circle (1.6pt);
  \node[left=3pt,fill=white,inner sep=1pt] at (0.3313,2.9061) {$1$};
  \fill[black] (6.6403,2.7329) circle (1.6pt);
  \node[right=3pt,fill=white,inner sep=1pt] at (6.6403,2.7329) {$x$};
  \fill[black] (3.1669,0.0000) circle (1.6pt);
  \node[below=3pt,fill=white,inner sep=1pt] at (3.1669,0.0000) {$xy$};
  \fill[black] (0.0000,8.5572) circle (1.6pt);
  \node[above left=3pt,fill=white,inner sep=1pt] at (0.0000,8.5572) {$q$};
  \fill[black] (7.0000,8.4447) circle (1.6pt);
  \node[above right=3pt,fill=white,inner sep=1pt] at (7.0000,8.4447) {$qx$};
  \fill[black] (3.1384,6.6287) circle (1.6pt);
  \node[below right=3pt,fill=white,inner sep=1pt] at (3.1384,6.6287) {$qxy$};
  \node[anchor=west,text=black] at (8.2000,7.9072) {$P_q(x,y)$};
  \draw[black,line width=0.65pt] (8.0000,7.9072)--(7.6000,7.9072)--(6.9295,7.3255);
  \fill[black] (6.9295,7.3255) circle (1.35pt);
  \node[anchor=west,text=anyonRho] at (8.2000,6.3072) {$\Phi_q\Delta[1,x,xy]$};
  \draw[anyonRho,line width=0.65pt] (8.0000,6.3072)--(7.6000,6.3072)--(3.3782,4.8127);
  \fill[anyonRho] (3.3782,4.8127) circle (1.35pt);
  \node[anchor=west,text=anyonJunction] at (8.2000,4.1072) {$W_q[x,xy]$};
  \draw[anyonJunction,line width=0.65pt] (8.0000,4.1072)--(7.6000,4.1072)--(4.7452,4.4828);
  \fill[anyonJunction] (4.7452,4.4828) circle (1.35pt);
\end{tikzpicture}%
}%
}

\newcommand{\RadialGaugeSUtwo}{%
\resizebox{\linewidth}{!}{%
\begin{tikzpicture}[x=1cm,y=0.9cm,font=\small,
  line cap=round,line join=round,
  >={Stealth[length=4pt]}]
  \begin{scope}
    \coordinate (l0) at (0,3.8);
    \coordinate (l1) at (0,0);
    \coordinate (l2) at (4.6,-0.2);
    \coordinate (l3) at (5.1,2.7);
    \node at (2.45,4.6) {(a)};
    \fill[anyonRho!6] (l0)--(l1)--(l3)--cycle;
    \fill[anyonRho!20] (l1)--(l2)--(l3)--cycle;
    \draw[draw=anyonMuted!75,line width=0.7pt] (l0)--(l2);
    \draw[draw=anyonMuted!75,line width=0.7pt] (l1)--(l0)--(l3);
    \draw[draw=anyonRho,line width=1pt] (l1)--(l3);
    \draw[draw=anyonModule,line width=1pt,->] (l1)--(l2);
    \draw[draw=anyonJunction,line width=1pt,->]
      (l2)--(l3) node[midway,right=5pt] {$s$};
    \node[text=anyonMuted] at (2.6,2.7) {$\Delta[1,g_{i-1},g_i s]$};
    \node[text=anyonRho] at (2.45,0.35) {$\Delta[g_{i-1},g_i,g_i s]$};
    \node[above left=4pt] at (l0) {$1$};
    \node[below left=5pt] at (l1) {$g_{i-1}$};
    \node[below right=5pt] at (l2) {$g_i$};
    \node[above right=4pt] at (l3) {$g_i s$};
    \foreach \p in {l0,l1,l2,l3} {\fill[anyonInk] (\p) circle (1.6pt);}
  \end{scope}
  \begin{scope}[xshift=8cm]
    \coordinate (r0) at (0,3.8);
    \coordinate (r1) at (0,0);
    \coordinate (r2) at (4.6,-0.2);
    \coordinate (r3) at (5.1,2.7);
    \node at (2.45,4.6) {(b)};
    \fill[anyonJunction!6] (r0)--(r1)--(r3)--cycle;
    \fill[anyonJunction!20] (r1)--(r2)--(r3)--cycle;
    \draw[draw=anyonMuted!75,line width=0.7pt] (r0)--(r2);
    \draw[draw=anyonMuted!75,line width=0.7pt] (r1)--(r0)--(r3);
    \draw[draw=anyonJunction,line width=1pt] (r1)--(r3);
    \draw[draw=anyonModule,line width=1pt,->] (r2)--(r3);
    \draw[draw=anyonJunction,line width=1pt,->]
      (r2)--(r1) node[midway,below=5pt] {$s$};
    \node[text=anyonMuted] at (2.4,2.9) {$\Delta[1,g_i s,g_{i+1}]$};
    \node[text=anyonJunction] at (2.45,0.35) {$\Delta[g_i s,g_i,g_{i+1}]$};
    \node[above left=4pt] at (r0) {$1$};
    \node[below left=5pt] at (r1) {$g_i s$};
    \node[below right=5pt] at (r2) {$g_i$};
    \node[above right=4pt] at (r3) {$g_{i+1}$};
    \foreach \p in {r0,r1,r2,r3} {\fill[anyonInk] (\p) circle (1.6pt);}
  \end{scope}
\end{tikzpicture}
}
}

\newcommand{\bC}{\mathbb C}
\newcommand{\bZ}{\mathbb Z}
\newcommand{\bR}{\mathbb R}

\newcommand{\cA}{\mathcal A}
\newcommand{\cB}{\mathcal B}
\newcommand{\cC}{\mathcal C}
\newcommand{\cD}{\mathcal D}
\newcommand{\cG}{\mathcal G}
\newcommand{\cH}{\mathcal H}
\newcommand{\cI}{\mathcal I}
\newcommand{\cJ}{\mathcal J}

\newcommand{\cM}{\mathcal M}
\newcommand{\cT}{\mathcal T}
\newcommand{\cU}{\mathcal U}

\newcommand{\cZ}{\mathcal Z}

\newcommand{\Hom}{\operatorname{Hom}}
\newcommand{\End}{\operatorname{End}}
\newcommand{\Fun}{\operatorname{Fun}}
\newcommand{\Irr}{\operatorname{Irr}}

\newcommand{\Ad}{\operatorname{Ad}}

\newcommand{\Rep}{\operatorname{Rep}}

\renewcommand{\Vec}{\operatorname{Vec}}
\newcommand{\sVec}{\operatorname{sVec}}

\newcommand{\WZW}{\mathrm{WZW}}

\newcommand{\bde}{\boldsymbol{\delta}}

\newcommand{\dd}{\mathrm d}
\newcommand{\tr}{\operatorname{tr}}
\newcommand{\one}{\mathbf 1}
\newcommand{\unit}{\one}
\newcommand{\ket}[1]{\lvert #1\rangle}
\newcommand{\bra}[1]{\langle #1\rvert}

\newcommand{\be}{\begin{equation}}
\newcommand{\ee}{\end{equation}}
\newcommand{\bea}{\begin{equation}\begin{aligned}}
\newcommand{\eea}{\end{aligned}\end{equation}}
\newcommand{\ba}{\begin{aligned}}
\newcommand{\ea}{\end{aligned}}

\begin{document}

\title{Non-abelian lattice chiral symmetry from continuous anyon chains}

\author{Christian Copetti}
\thanks{On leave of absence from the University of Oxford.}
\affiliation{Mathematical Institute, University of Oxford,
Andrew Wiles Building, Radcliffe Observatory Quarter,
Woodstock Road, Oxford OX2 6GG, United Kingdom}

\begin{abstract}
We construct (1+1)d quantum rotor lattice models with an exact non-abelian chiral symmetry and a tensor-product Hilbert space.
Our construction applies the anyon-chain formalism to continuous symmetry categories with nontrivial
't Hooft anomalies. Our models generalize the modified Villain model and descend directly from the categorical input of the anyon chain. We provide
evidence that our Hamiltonian flows to the critical Wess--Zumino--Witten model in the continuum
limit.
\end{abstract}

\maketitle


\section{Introduction}
\label{sec:introduction}
Lattice regularization provides a natural setting for non-perturbative studies of
quantum systems \cite{Kogut:1974ag}. This approach is particularly useful when the discretization further
preserves the microscopic symmetry of the underlying continuum theory.
The relation between the lattice and continuum symmetry can often be nontrivial \cite{Cheng:2022sgb} and involve a map (a tensor functor) between the lattice and continuum symmetry
structures~\cite{Seiberg:2025bqy,Antinucci:2025fjp}.
However, it has long been appreciated that severe obstructions can forbid the exact lattice realization of symmetries, even in this more general sense~\cite{Nielsen:1980rz,Nielsen:1981xu, Nielsen:1981hk}. These results are particularly restrictive for
continuous symmetries with a perturbative 't Hooft anomaly, which include
chiral symmetries. 
These ideas have a long history rooted in the context of Euclidean lattice theories \cite{Ginsparg:1981bj,Luscher:1998pqa}. Here we consider a Hamiltonian description with continuous time instead.
In one spatial dimension, interacting generalizations of
the Nielsen--Ninomiya theorem prohibit a lattice anomalous compact symmetry under
the following locality and finiteness
requirements~\cite{Kapustin:2024rrm,Liu:2026atf}:
\begin{description}
  \item[A1] A tensor-product Hilbert space
  $\cH=\bigotimes_i\cH_i$ on a one-dimensional lattice, with
  finite-dimensional local factors $\cH_i$.
  \item[A2] Quantized, Hermitian, quasi-local charges
  $Q=\sum_i q_i$, whose exponentials implement the action
  of the compact symmetry group.
\end{description}
The obstruction reflects the infinite order of a perturbative anomaly. In
a finite-dimensional spin chain, locality places torsion restrictions on the lattice
anomaly class~\cite{Liu:2026atf}. Similar no-go results exist for the parent SPT phases \cite{kapustin2020local}.
Relaxing these assumptions leads to the known lattice realizations of chiral
symmetries: domain-wall fermions use an additional spatial dimension and realize
the chiral theory on its boundary~\cite{Kaplan:1992bt}, and a similar mild locality violation leads to overlap fermions \cite{Neuberger:1997fp,Hernandez:1998et}. It is also possible 
to exactly retain a noncompact axial symmetry, whose charge is not
quantized~\cite{Chatterjee:2024sz,Gioia:2025bhl,Lew-Smith:2026jqb}.\footnote{With a vector
symmetry, one can instead allow the lattice vector and axial charges to
generate the Onsager algebra rather than commute~\cite{Vernier:2018han,Chatterjee:2024sz,Kourkoulou:2026lzx}.}
It remains unclear whether such non-compact realizations match the continuum 't Hooft anomaly by themselves.
In this work we instead relax A1 by allowing for an infinite-dimensional local Hilbert space $\cH_i$, which invalidates the aforementioned torsion constraint on the possible anomalies.
This idea has been explored successfully in lattice realizations of \emph{abelian} chiral symmetries in
$(1+1)$ dimensions~\cite{Sulejmanpasic:2019ytl,Fazza:2022fss,Cheng:2022sgb,Berkowitz:2023pnz,Thorngren:2026ydw,Seifnashri:2026ema,
Dharanikota:2026vik,Lu:2026itw}
and $(3+1)$ dimensions~\cite{Thorngren:2026ydw,Lu:2026jnq}, as well as commuting projector models for (2+1)d integer quantum Hall SPTs \cite{DeMarco:2021erp}.

The bosonic version of these theories describes a compact continuum boson (or axion) $\phi$. The naive site variable $\phi_i$ must be supplemented by an integer-valued Villain field $n_{i+1,i}$ on links~\cite{Villain:1974ir,Gorantla:2021svj}. These additional degrees of
freedom are crucial to realize the anomalous symmetry
$U(1)_m\times U(1)_w$ exactly, but are integrated out in the continuum limit.

The chiral symmetries relevant to Standard
Model fermions are however \emph{non-abelian}. Extending the Villain procedure to
this setting requires identifying the correct additional degrees of freedom and their
Hamiltonian, and is presently an unsolved problem for interacting systems.\footnote{
While this work was in preparation \cite{Shao:2026rht} appeared, describing a quantum Hamiltonian lattice realization of free fermions in which the full non-abelian symmetry is preserved.
Furthermore, Euclidean proposals to implement non-abelian Villain models have
appeared previously in \cite{Chen:2024ddr,Zhang:2024sgm,Onoda:2026ijj}.}
A (1+1)-dimensional realization of this idea would provide a natural lattice discretization of e.g. the Wess--Zumino--Witten model.\footnote{There are several known lattice systems which flow to the critical WZW model in the continuum limit, such as the spin-1/2 Heisenberg chain \cite{Affleck:1987ch}. In these models, however, the anomalous $G_L \times G_R$ is not exact at the lattice level and the anomaly-free vector symmetry is the only surviving continuous symmetry.}

Our proposed quantum lattice Hamiltonian acts on a Hilbert space of rotors $g_i \in G$ on lattice sites and Villain rotors $x_{i+1, i} \in G$ on links. We use the shorthand $q_{i+1, i} = g_i^{-1} g_{i+1}$ to denote the lattice derivative. The quantum Hamiltonian is:
\be \label{eq: result}
H = \frac{1}{2\beta}\sum_{i \, \in \, \text{sites}}  \tr P_i^2  - \frac{\beta}{2} \sum_{\ell \, \in \, \text{links}} \Delta_{x_\ell, q_{\ell}} \, .
\ee
The operator $P_i^a = J_i^a + k\cB_i^a$ is the twisted lattice momentum and
$\Delta_{x_\ell, q_{\ell}} = \tr D^2_{\ell}$ a covariant Laplacian for the link rotor at fixed $q_\ell$. The first term is the kinetic term for the site rotors $g_i$, while the second provides a Villain potential. This expression is valid in a local chart containing the identity, and transition functions are needed to define the Hamiltonian globally.
The Hamiltonian \eqref{eq: result} is not exactly solvable, but we will provide evidence that, in the continuum limit, it flows to the critical Wess--Zumino--Witten model at level $k$: WZW$_k$. See \cite{Falceto:1992bf,Alekseev:1992jra,Alekseev:1996jz} for previous approaches to the lattice discretization of the WZW model.

To derive \eqref{eq: result}, we approach this problem via the formalism of anyon chains \cite{Feiguin:2006ydp,Aasen:2016dop,Buican:2017rxc,Aasen:2020jwb}. This setting makes any (categorical) symmetry $\cC$ manifest on the lattice and has led to multiple applications ranging from phases of matter \cite{Bhardwaj:2024kvy} to scattering amplitudes \cite{Copetti:2024rqj,Copetti:2024dcz}.
The underlying Hilbert space, however, need not tensor factorize $\cH \neq \bigotimes \cH_i$. Factorization can be recovered from a suitable choice of categorical data or from
an enlargement by (possibly infinite) ancillary degrees of
freedom~\cite{Bunner:2026yhn,Wen:2026ncw}.

The second ingredient is a categorical description of continuous symmetries and their
anomalies, which has recently been developed in e.g.~\cite{Jia:2025vrj,Jia:2026vcr,Stockall:2025ngz}, following earlier work which used the SymTFT construction \cite{Gaiotto:2020iye, Freed:2022qnc, Apruzzi:2021nmk} to describe them \cite{Antinucci:2024zjp,Brennan:2024fgj,Bonetti:2024cjk}. Our main results are:

\vspace{2mm }\noindent \textbf{Villain Hamiltonians from first principles.} Our methods identify the Villain degrees of freedom to add to a lattice model, and how to build Hamiltonians. 

\vspace{2mm}\noindent \textbf{Non-abelian lattice chiral Hamiltonians.} This gives a new class of Villain lattice models with exact non-abelian chiral symmetry and a tensor-factorized Hilbert space.   

\vspace{2mm} Our approach should extend to higher-dimensional systems, and the resulting models might be amenable to future quantum simulations. We detail several open directions in the Discussion.

\section{Anyon chain lattice models}
\label{sec:anyon-chain-lattice-models}
We review the anyon-chain formalism following~\cite{Aasen:2020jwb}, using standard
fusion category techniques~\cite{Etingof:2002vpd}. An anyon chain is specified
by a triple $(\cC,\cM,\rho)$. Here $\cC$ describes the lattice symmetry and
$\cM$ is a left $\cC$-module category, with action
$\triangleright:\cC\boxtimes\cM\to\cM$. The chain object $\rho$ belongs
to the dual category
\be
  \cC^*_{\cM}=\Fun_{\cC}(\cM,\cM),
\ee
whose objects are $\cC$-module endofunctors. In our conventions, the dual category $\cC^*_\cM$ acts on $\cM$ from the right, with action denoted by
$\triangleleft$. 
The module category $\cM$ can be presented as the category of
right modules over a separable algebra object $\cA\in\cC$, and the dual
category consists of $\cA$-bimodules~\cite{Ostrik:2001xnt}.
Choosing $\cA$ amounts to implementing a generalized gauging, with $\cM$ describing the gauging interface \cite{Bhardwaj:2017xup}.
Equivalently, one can choose an algebra in $\cC^*_{\cM}$ and reverse the
construction.

\begin{figure}[tbp]
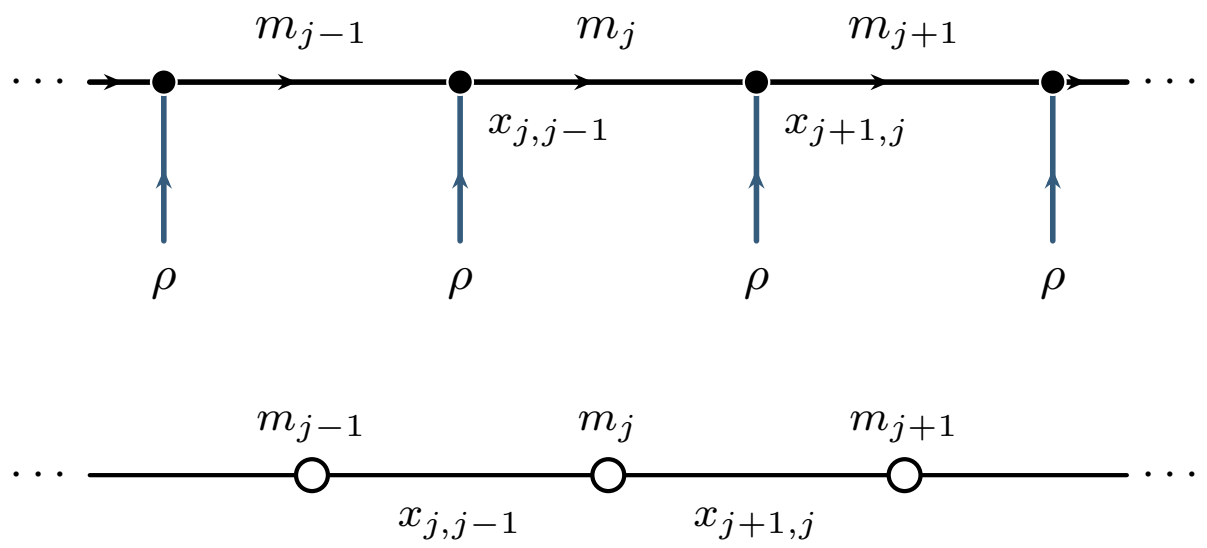

  \centering
  \AnyonChain
  \caption{A basis state of the anyon chain. Horizontal segments carry
  module labels $m_i \in \cM$, downward legs are labeled $\rho \in \cC^*_\cM$, and the vertices are
  junction vectors $x_{i+1,i}\in V[m_i,m_{i+1}]$.
  Below, the corresponding physical lattice, where $m_i$ live on lattice sites and $x_\ell$ on the corresponding links.}
  \label{fig:anyon-chain}
\end{figure}

States in the anyon chain Hilbert space are obtained by specifying module labels $m_i$ and junction vectors $x_{i+1, i}$ as shown in Figure \ref{fig:anyon-chain}. Such an assignment must be consistent and leads to the junction Hilbert spaces
\be
  V[m_i,m_{i+1}]
  =\Hom_{\cM}(m_i\triangleleft\rho,m_{i+1}) \, ,
  \label{eq:chain-junction-space}
\ee
which can be zero.
A basis state
$\ket{\{m_i\},\{x_{i+1,i}\}}$ is represented by the diagram in
Fig.~\ref{fig:anyon-chain}. On the physical lattice, the labels $m_i$ are physical site variables,
while $x_{i+1,i}$ labels
the physical link data. The latter can naturally be interpreted as \emph{Villain fields} which enrich the naive lattice model to allow an exact implementation of the symmetry.
The allowed triples are correlated, so
\eqref{eq:chain-junction-space} does not factorize over physical sites and
links.

\begin{figure}[tbp]
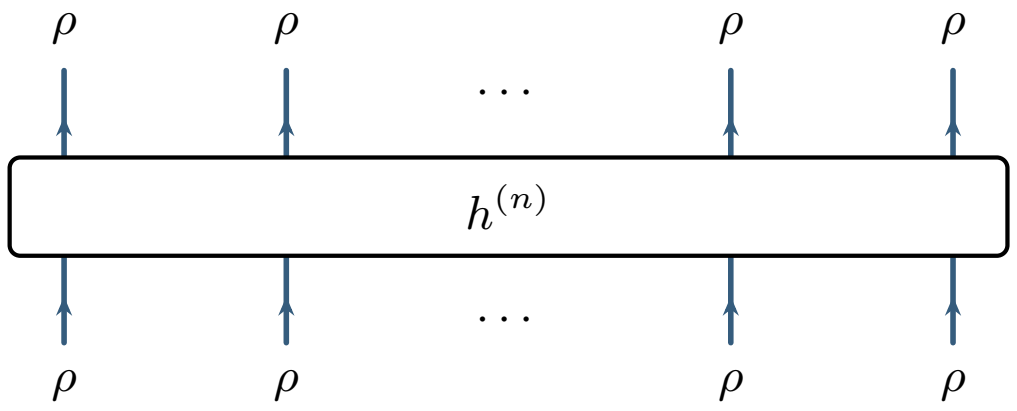

  \centering
  \AnyonLegTerm
  \caption{An $n$-leg term $h^{(n)}\in\End(\rho^{\otimes n})$.
  The box represents an element in $\End(\rho^{\otimes n})$. Choosing a basis of this space gives a basis of symmetric terms.}
  \label{fig:n-leg-term}
\end{figure}

The symmetry category $\cC$ acts on the chain from \emph{above}. Its action
is determined by the module-category data.
Local symmetric interactions are also specified categorically. An
\emph{$n$-leg} interaction is an endomorphism
\be
K^{(n)} \in \End\left( \rho^{\otimes n} \right) \, ,
\ee
represented in Fig.~\ref{fig:n-leg-term}. A Hermitian operator is
described by $h^{(n)} = K^{(n)}+K^{(n)\dagger}$.
We denote its lattice insertions by $h_i^{(n)}$ and their sum by
\be
H^{(n)}=\sum_i h_i^{(n)} \, .
\ee
In this work, we restrict our attention to one-leg terms
$K^{(1)}$ and two-leg terms $K^{(2)}$:
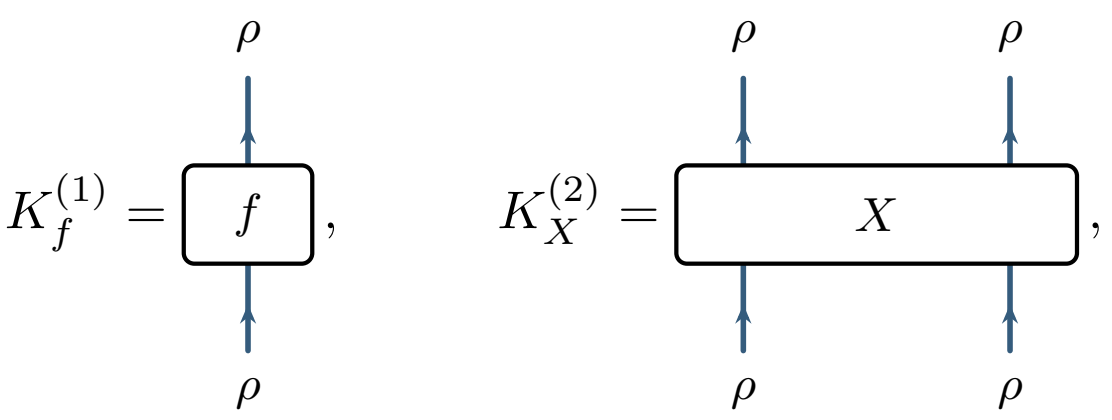
\bea
K^{(1)}_f&=\AnyonBoxTerm{1}{f} \, , \qquad
& K^{(2)}_X&=\AnyonBoxTerm{2}{X} \, ,
\eea
the former do not induce any interaction between adjacent Villain data and they can be interpreted as a generalization of the Villain potential.
Our strategy for constructing chiral lattice models from this setup has four steps:
\begin{enumerate}
    \item Determine $\cM$ by flat-gauging the largest anomaly-free (vector) subgroup in $\cC$. This gives the desired light variables on sites. 
    \item Choose $\rho$ to be the regular object in $\cC^*_\cM$. This leads to a tensor-product Hilbert space, with Villain data encoded in the junction spaces $V[m_i, m_{i+1}] \simeq V$ for $\rho$.
    \item Construct the general one- and two-leg terms. They implement finite parallel transport on the Villain fields and site rotors, respectively. Taking an infinitesimal limit leads to the lattice Villain potential and kinetic term.
    \item To describe the continuum limit, consider the regime in which site fields $m_i$ are light variables and diagonalize the one-leg (Villain) potential explicitly.
\end{enumerate}
We begin by illustrating these steps in the simpler case of an abelian symmetry in Section \ref{sec:villain-model}, where we recover the known (modified) Villain Hamiltonian \cite{Gorantla:2021svj}. We move on to the non-abelian construction in Section \ref{sec:non-abelian-chiral-symmetry}.

\section{Villain model from anyon chain}
\label{sec:villain-model}
As a first example, we describe the modified Villain model in the anyon-chain
formalism. This is an alternative derivation of a well-known lattice model, but will serve as an introduction to our methodology for the technically more complex non-abelian case.

The (modified) Villain model is described by compact rotors $\phi_i\in\bR/2\pi\bZ$ on sites and $n_\ell\in\bZ$
on oriented links $\ell=(i+1,i)$. We consider the domain $\phi_i \in [0, \, 2 \pi)$ and use the notation
$d\phi_{i+1,i}=\phi_{i+1}-\phi_i$ for the lattice exterior derivative.
We use the ordinary remainder modulo $2\pi$,
\be
  [x]_{2\pi}=x-2\pi\left\lfloor\frac{x}{2\pi}\right\rfloor
  \in[0,2\pi),
  \label{eq:modular-remainder}
\ee
with zero represented by zero. The Hamiltonian is
\be
H_{\text{Villain}} = \frac{1}{2\beta}\sum_{i \, \in \, \text{sites}}\pi_i^2
  +\frac{\beta}{2}\sum_{\ell \, \in \, \text{links}}
  \left(\frac{[d\phi_\ell]_{2\pi}}{2\pi}-n_\ell\right)^2.
  \label{eq:villain-hamiltonian}
\ee
This is the tensor-product presentation obtained by a locality-preserving
Villain disentangler, with a different branch convention from
 \cite{Thorngren:2026ydw}.
Locally we have the canonical commutation relations
\be
[\pi_i, \phi_j] = i \delta_{ij} \, , \quad [n_\ell, \widetilde{\phi}_{\ell'}] = i \delta_{\ell, \ell'} \, ,
\ee
where $\widetilde\phi_\ell$ is the rotor conjugate to $n_\ell$. Globally,
only exponentials of the rotors are well-defined variables. The momentum and
winding symmetry $U(1)_m\times U(1)_w$ has charges
\be
  Q_m=\sum_i\pi_i, \qquad
  Q_w=\sum_\ell\left(n_\ell-
    \frac{[d\phi_\ell]_{2\pi}}{2\pi}\right).
  \label{eq:villain-charges}
\ee
On a periodic chain, $\sum_\ell d\phi_\ell=0$, so
\be
  Q_w=\sum_\ell\left(n_\ell+
  \left\lfloor\frac{d\phi_\ell}{2\pi}\right\rfloor\right)
  \in\bZ.
\ee
The integer-valued winding charge therefore generates a compact symmetry. In the anyon-chain description, we start from the symmetry category
\be
\cC = \left[ \Vec_{U(1)_m} \boxtimes \Vec_{U(1)_w} \right]^{\omega} \, .
\ee
The notation $\Vec_G^\omega$ denotes the continuous analogue of the twisted group category~\cite{Jia:2025vrj,Jia:2026vcr}.
The anomaly class $[\omega]\in H^4(BG,\bZ) \simeq H^3_{\text{Borel}}(G,U(1))$ corresponds to the anomaly polynomial
\be
  I_4=\frac{d A_m}{2\pi}\wedge\frac{d A_w}{2\pi} \, .
\ee
To reproduce \eqref{eq:villain-hamiltonian} using the anyon chain we must first determine $\cM$. The guiding principle is that we would like the site variables $m_i$ to describe the continuum light fields $\phi_i$: in this way, the split between light and heavy degrees of freedom is baked into the construction.
The corresponding module category is $\cM = \Vec_{U(1)}$ and describes a topological interface obtained by flat-gauging a maximal anomaly-free subgroup, $U(1)_w \subset \cC$ \cite{Argurio:2024ewp}.\footnote{Alternatively, to obtain the T-dual description we gauge $U(1)_m$.} Its simple objects are labeled by:
\be
\phi \, \in \, [0, 2\pi) \, .
\ee
The dual category is $\Vec_{\bR}$\footnote{Similar tools have been used in Euclidean studies of the Villain models see e.g. \cite{Argurio:2026txf}.}. This can be seen by applying the methods of \cite{Tachikawa:2017gyf}, which extend to the flat gauging of continuous abelian groups. We denote its simple elements by $t \in \bR$. Its right action on $\cM$ is given by:
\be
\phi \triangleleft t = [\phi + t]_{2\pi} \, .
\ee
To have a tensor-factorized Hilbert space {we require the junction spaces to have the same dimension, i.e. $\rho \otimes t \simeq \rho$, then} the natural choice for the chain object $\rho$ is the regular object:
\be \label{eq: rhovill}
\rho = \text{Reg}(\Vec_\bR) = \int^\oplus_{\bR} t \, . 
\ee
From this we can determine the Villain data. Decompose $t \in \bR$ as $t = u + 2 \pi n$, with $u \in [0, 2 \pi)$, then:
\be
\Hom(\phi_i \triangleleft t_{i+1, i} , \phi_{i+1}) = \delta(u_{i+1, i}, [d \phi_{i+1, i}]_{2\pi} ) \, \bC \, ,
\ee
where we use $\delta(x, y)$ and $\delta_{ij}$ to denote Dirac and Kronecker deltas, respectively.
Combining this with \eqref{eq: rhovill} leads to
\be
\Hom(\phi_i \triangleleft \rho, \phi_{i+1}) = \bigoplus_n  \bC \, .
\ee
Thus, the Villain data are described by an integer $n_{i+1,i} \, \in \, \bZ$ for each lattice link, and the total Hilbert space is tensor factorized.

\begin{figure}[tbp]
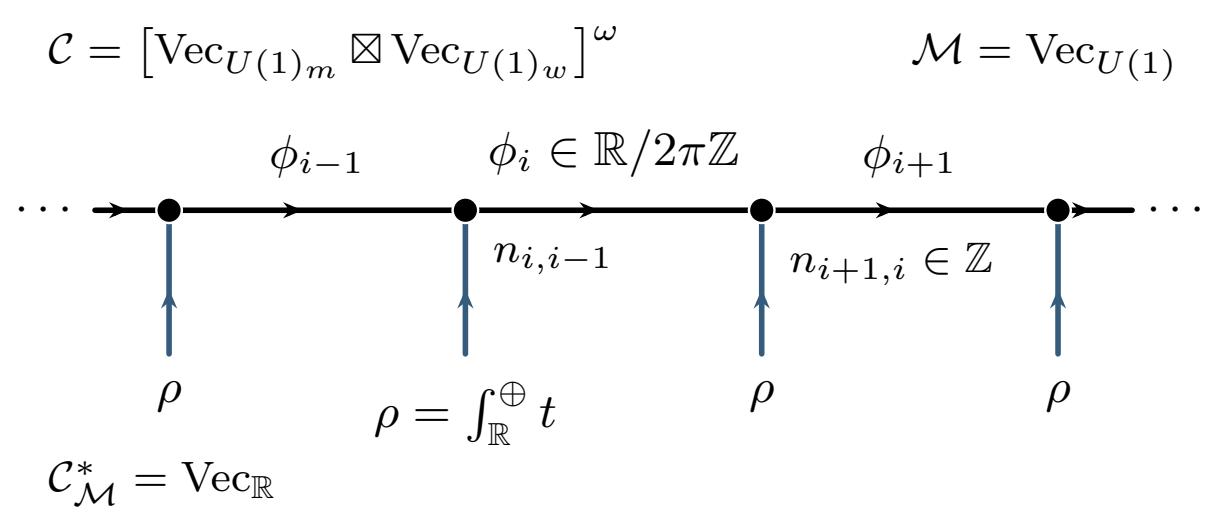

  \centering
  \VillainAnyonChain
  \caption{The Villain anyon chain. Module segments carry site rotors
  $\phi_i\in\bR/2\pi\bZ$, and junction vectors carry independent integers
  $n_{i+1,i}\in\bZ$.}
  \label{fig:villain-anyon-chain}
\end{figure}

\vspace{2mm} \noindent \textbf{Hamiltonians.} Let us now describe the one- and two-leg Hamiltonians in this model. 

\vspace{2mm} \noindent \emph{One-leg terms.} The vector space $\End(\rho)$ consists of functions $f$ on $\bR$:
\be
t \to f(t) \, ,
\ee
and Hermitian operators are identified with real functions. Computing the matrix elements in a state $\ket{\phi_i, \phi_{i+1}, n_{i+1,i}}$ leads to the identification:
\be
t = [d\phi]_{2\pi} + 2\pi n \, ,
\ee
thus
\be
h_{i+1,i}^{(1)} = f([d\phi_{i+1,i}]_{2\pi} + 2\pi n_{i+1,i}) \, ,
\ee
is the Villain potential. To match the original lattice model we choose $f(t) = (t/2\pi)^2$, and map $n \to -n$.

\vspace{2mm} \noindent \emph{Two-leg terms.} We choose the following (crossed) basis of $\End(\rho \otimes \rho)$:
\be
K^{(2)}_{t} = \AnyonTwoLeg{t} \, .
\label{eq:villain-two-leg-diagram}
\ee
In the junction basis with trivial right $\Vec_{\bR}$-module associators,
evaluating its matrix elements between states
$\ket{\psi} = \ket{\phi_{i-1}, \phi_i, \phi_{i+1}, n_{i,i-1}, n_{i+1,i}}$
and
$\ket{\psi'} = \ket{\phi_{i-1}, \phi_i', \phi_{i+1}, n_{i,i-1}', n_{i+1,i}'}$
leads to:
\bea
\bra{\psi'} K^{(2)}_{t,i} \ket{\psi}
  ={}&\delta\left(\phi_i', [\phi_i+t]_{2\pi}\right)\\
  &\times\delta_{n_{i,i-1}',\,n_{i,i-1}+c_{i,i-1}(t)}\\
  &\times\delta_{n_{i+1,i}',\,n_{i+1,i}+c_{i+1,i}(t)}
\eea
where
\bea
c_{i,i-1}(t)
  &=\left\lfloor\frac{[d\phi_{i,i-1}]_{2\pi}+t}{2\pi}\right\rfloor,\\
c_{i+1,i}(t)
  &=\left\lfloor\frac{[d\phi_{i+1,i}]_{2\pi}-t}{2\pi}\right\rfloor.
\eea
These identifications follow from expanding the $\rho$ operators in their real labels and decomposing $t = [d \phi]_{2\pi} + 2 \pi n$. 
To obtain a kinetic term, we Taylor expand around $t=0$. We can do this within a branch and drop the $c_{i+1, i}(t)$ terms. 
The oriented diagram then acts as $e^{it\pi_i}$, so the Hermitian two-leg term is
\be
h^{(2)}_{t,i}
  =e^{it\pi_i}+e^{-it\pi_i} 
  =2\unit-t^2\pi_i^2+O(t^4).
\ee
This expansion is understood on smooth wavefunctions supported away from
the branch cuts. The positive kinetic term is therefore
\bea
h_i^{(2)} &=\pi_i^2,\\
H^{(2)}&=\sum_i h_i^{(2)}=\sum_i\pi_i^2 \, .
\eea
Assembling the one- and two-leg terms gives:
\be
H = \frac{1}{2 \beta} H^{(2)} + \frac{\beta}{2} H^{(1)} = H_{\text{Villain}} \, ,
\ee
reproducing the original Villain Hamiltonian~\eqref{eq:villain-hamiltonian}.

\vspace{2mm} \noindent \textbf{Several compact bosons.} 
The construction extends to $D$ compact bosons $\phi^\alpha$, with symmetry
\be
\cC_D = \left( \left[ \Vec_{U(1)_m} \boxtimes \Vec_{U(1)_w} \right]^\omega \right)^D \, .
\label{eq:multi-boson-symmetry}
\ee
The continuum data are a target metric $G_{\alpha\beta}$ and a constant
antisymmetric $B$-field. To obtain them, start from the dual category
$\cC^*_{\cM}=\Vec_{\bR^D}$ and gauge an algebra $\cA_\Lambda$ corresponding to a full-rank lattice $2\pi\Lambda$, where
\be
\Lambda=\bigoplus_{\alpha=1}^D\bZ e_\alpha\subset\bR^D.
\ee
The gauging algebra is supported on $2\pi\Lambda$, and module labels lie
on $T_\Lambda=\bR^D/(2\pi\Lambda)$. The regular chain object is
$\rho=\int_{\bR^D}^{\oplus}t$, giving independent Villain integers
$n_{i+1,i}^\alpha$ on each link. Fixing a Euclidean quadratic form on
$\bR^D$ and writing a target point as $\phi^\alpha e_\alpha$, with
$\phi^\alpha\sim\phi^\alpha+2\pi$, defines a metric
\be
G_{\alpha\beta}=(e_\alpha,e_\beta).
\ee
The $B$-field is instead encoded by a choice of discrete torsion, which modifies the multiplication in the algebra $\cA_\Lambda$:
\be
[\tau]\in H^2(\bZ^D,U(1))\simeq U(1)^{D(D-1)/2},
\ee
with representative
\be
\tau(n, m) = \exp\left( \frac{i}{2} B_{\alpha \beta} n^\alpha m^\beta \right) \, ,
\ee
with $B_{\alpha\beta}=-B_{\beta\alpha}$.
With the same conventions of~\eqref{eq:villain-hamiltonian}, define
\be
t_{i+1,i}^\alpha
  =[d\phi_{i+1,i}^\alpha]_{2\pi} -2 \pi n_{i+1,i}^\alpha.
\ee
We choose the one- and two-leg terms:
\be \label{eq: D-boslegs}
h^{(1)}_t = \frac{1}{4 \pi^2}\left( t , \, t \right) \, , \qquad K^{(2)}_{[t]} = \int_{(s,s)= t^2} ds \, K^{(2)}_s\,  .
\ee
Evaluating the one- and two-leg terms as in the single boson analysis and taking the appropriate limit gives the $D$-boson Hamiltonian:
\bea
H_{D\text{-boson}}={}&\frac{1}{2\beta}\sum_{i \, \in \, \text{sites}}
  G^{\alpha\beta}\Pi_{i,\alpha}\Pi_{i,\beta}\\
&+\frac{\beta}{8 \pi^2}\sum_{\ell \, \in \, \text{links}}
  G_{\alpha\beta}\, t_{\ell}^\alpha \, t_{\ell}^\beta,
\eea
where $G^{\alpha\beta}$ is the inverse metric and 
\be
\Pi_{i,\alpha}=\pi_{i,\alpha}
  -\frac{B_{\alpha\beta}}{8\pi^2}
   \left(t_{i,i-1}^\beta+t_{i+1,i}^\beta\right) \, ,
\ee
is the twisted momentum.
Thus the torsion modifies
the two-leg kinetic term, while the one-leg term remains the quadratic
Villain potential.\footnote{While this work was in preparation, the same Hamiltonian has been analyzed in depth in \cite{Choi:2026ruw}.}
The anyon chain does not force on us the presence of the metric in one- and two-leg operators: we have made the choice \eqref{eq: D-boslegs} to recover the Villain discretization of the continuum relativistic Hamiltonian.

\section{Non-abelian chiral symmetry from \texorpdfstring{anyon chains}{anyon chains}}
\label{sec:non-abelian-chiral-symmetry}

We now come to the main construction of this work: a Villain lattice Hamiltonian with exact non-abelian chiral symmetry. While in spirit the steps we take to derive equation \eqref{eq: result} are entirely parallel to the abelian case we have just studied, they are technically complex. Thus, we will provide only the main formulas in the text, and refer the reader to the Appendices for details.

Let $G$ be a compact simple and simply connected Lie group, e.g. $G=SU(2)$. We consider an anyon chain based on:
\be \label{eq: lattice-symmetry}
\cC = \Vec_{G_L}^\omega \boxtimes \Vec_{G_R}^{-\omega} \, , \quad G_L = G_R = G \, ,
\ee
with opposite perturbative 't Hooft anomalies for the left and right symmetries $[\omega] = k c_2$ (for $G=SU(N)$) in $H^4(BG,\bZ)$,
the integral cohomology of the classifying space~\cite{Jia:2025vrj,Jia:2026vcr}.
We denote the elements of $\cC$ by $(h_L, h_R)$ and use the shorthand notation
\be
g^h \equiv h g h^{-1} \, , \qquad  {}^h g \equiv h^{-1} g h \, ,
\ee
to denote conjugation throughout.
The microscopic symmetry is $G_L \times G_R$, which is different from the faithfully acting symmetry in the WZW model
\be
G_{\WZW}=(G_L \times G_R) / Z(G) \, .
\ee
We will argue that $G_{\WZW}$ is the faithfully acting symmetry on the light rotors $g_i$, thus implementing the lattice/continuum map:
\be
\cC  \overset{f}{\longrightarrow} \Vec_{G_{\WZW}}^\Omega \, , \quad f^*(\Omega) = (\omega, -\omega) \, .
\ee

\begin{figure}[tbp]
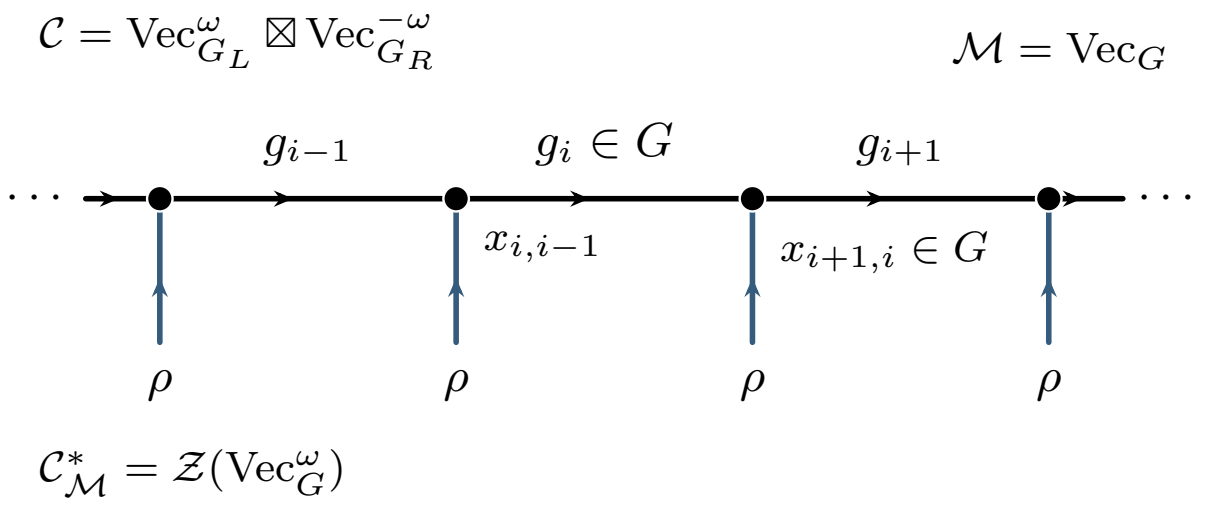

  \centering
  \NonabelianAnyonChain
  \caption{The non-abelian anyon chain. Horizontal segments carry site rotors
  $g_i\in G$, junctions carry independent Villain rotors
  $x_{i+1,i}\in G$, whose grade $q_{i+1,i}=g_i^{-1}g_{i+1}$ is fixed by the adjacent sites. The vertical legs carry the regular chain object $\rho = \text{Reg}(\cC^*_\cM)$.
 }
  \label{fig:non-abelian-anyon-chain}
\end{figure}

\vspace{2mm} \noindent \textbf{Anyon chain data.} Let us now describe the anyon chain data $(\cC, \cM, \rho)$ for the non-abelian model. The module category $\cM$ can be obtained by flat-gauging a maximal anomaly-free (vector) symmetry:\footnote{Other choices of modules lead to different valid chiral lattice models, but they will not have a clean split between Villain data and continuum light fields.}
\be
\Vec_{G_V} \sim (h, h) \in \cC \, .
\ee
The corresponding module category is $\cM = \Vec_{G}$, with simple objects described by equivalence classes
\be
[g] = \left\{ (g_L, g_R) \in G_L \times G_R \, ,  (g_L, g_R) \sim ( g_L h , g_R h) \right\} \, ,
\ee
the representative can be chosen as $(g ,1)$. The symmetry acts on $\cM$ by:
\be
(h_L, h_R) \cdot [g] = [h_L g h_R^{-1}]
\ee
which is the well-known $G_{\WZW}$ action on the Wess--Zumino--Witten fields. We will henceforth denote the site rotor $[g]_i$ by $g_i$. 
The dual category is also simple, and is described by the Drinfeld center of a single chiral symmetry:\footnote{Write $\cD=\Vec_G^\omega$. The two chiral factors act on $\cM=\cD$ by left and right multiplication, so $\cC\simeq\cD\boxtimes\cD^{\rm rev}$. Hence $\cC^*_\cM=\operatorname{Fun}_{\cD\boxtimes\cD^{\rm rev}}(\cD,\cD)$ is the category of bimodule endofunctors, which is known to be equivalent to the Drinfeld center $\cZ(\cD)$~\cite{Ostrik:2001xnt,Etingof:2002vpd}.}
\be \label{eq: catequiv}
\cC^*_\cM \simeq \cZ(\Vec_G^\omega) \, .
\ee
For local computations, we will use a description of \eqref{eq: catequiv} as representations of a distributional Drinfeld double $D(G)^\omega$
\be
\cZ(\Vec_G^\omega) = \Rep \left( D(G)^\omega \right) \, .
\ee
A discussion of its mathematical structure and global definition can be found in Appendix \ref{app:drinfeld-double}. Using the distributional basis
\be
(\bde_q,x),\qquad q,x\in G ,
\ee
the product on $D(G)^\omega$ takes the form:
\be
(\bde_q,x)\star(\bde_p,z)
 =\delta(p, {}^x q)\theta_q(x,z)^{-1}(\bde_q,xz) \, ,
\ee
where $\delta(g,h)$ is a Dirac delta with respect to the Haar measure on $G$ and we have defined the transgression:
\be
\theta_q(x,y)
  =\frac{\omega(q,x,y)\,
    \omega\!\left(x,y,{}^{xy} q\right)}
  {\omega\!\left(x,{}^x q,y\right)} \, .
\ee
The anyon chain object leading to a tensor-factorized Hilbert space is again the regular object
\be
\rho=\operatorname{Reg}\left(\cZ(\Vec_G^\omega)\right).
\ee
In the twisted-double description, this is the left regular representation
of $D(G)^\omega$, which is isomorphic to the quantum double itself:
\be
\rho \simeq  D(G)^\omega \, .
\ee
The object $(\bde_q, x)$ has grade $q$. It acts on $\cM$ via the forgetful action which retains only the grade, $F(\bde_q, x) = q$, thus sending
$g\triangleleft q=gq$, while $x$ labels the junction multiplicity.
Thus, the junction grading is fixed to be $q_{i+1,i}=g_i^{-1}g_{i+1}$, leaving $x_{i+1, i}$ unconstrained. The Villain Hilbert space is
\be
\Hom_{\cM}(g_i\triangleleft\rho,g_{i+1}) \simeq L^2(G)_x \, .
\ee
The Hilbert space factorizes into site
rotors $g_i\in G$ and independent Villain rotors $x_{i+1,i}\in G$
\be
\cH = \bigotimes_{i \, \in \, \text{sites}} L^2(G)_{g_i} \bigotimes_{\ell \, \in \, \text{links}} L^2(G)_{x_\ell} \, ,
\ee
as depicted in
Fig.~\ref{fig:non-abelian-anyon-chain}. 
In the position eigenbasis states are given by:
\be
|\Psi\rangle =  \ket{ \{ g_i \}_i , \{ x_{\ell} \}_\ell } \, .
\ee
The setup is parallel to the abelian case using the dual position basis where $n_\ell$ is conjugate to a compact rotor $\widetilde{\phi}_\ell$.

\vspace{2mm} \noindent \textbf{Left and right symmetry action.} Having defined the anyon chain Hilbert space, we derive the left and right symmetry action on its Hilbert space.

\vspace{2mm} \noindent \emph{Left symmetry action.} Our choice of equivalence between $\cC^*_\cM$ and $\cZ(\Vec_G^\omega)$ leads to a transparent left action. The grade $q_{i+1,i}=g_i^{-1}g_{i+1}$ remains invariant, and the local action on junctions is encoded in the diagram:
\be
\cU_L(h)_{i+1,i}=\AnyonLeftSymmetryMove{h} \, .
\ee
Its evaluation uses the forgetful functor $F:\cZ(\Vec_G^\omega)\to\Vec_G^\omega$, which retains the grade $q$ and forgets the regular multiplicity $x$:
\be
(\bde_q,x)\longmapsto q \, .
\ee
Set $m=hg_i$ and $q=q_{i+1,i}$ for compactness. On this graded component, the bimodule associator is the ordinary associator $\alpha$ of $\Vec_G^\omega$:
\be
\begin{tikzcd}[row sep=2.5em,column sep=2em,font=\small]
(h^{-1}\triangleright m)\triangleleft\rho
  \arrow[r,"F"] \arrow[d,"\varphi_L"']
& (h^{-1}\otimes m)\otimes q \arrow[d,"\alpha"]
\\
h^{-1}\triangleright(m\triangleleft\rho) \arrow[r,"F"']
& h^{-1}\otimes(m\otimes q)
\end{tikzcd}
\ee
Thus the phase is
\be
\varphi_L(h;g_i,g_{i+1})
  =\omega\!\left(h^{-1},hg_i,q_{i+1,i}\right),
\ee
independently of $x_{i+1,i}$. The local action is
\bea
&\cU_L(h)_{i+1,i}\ket{g_i,g_{i+1},x_{i+1,i}}\\
&\quad=\omega\!\left(h^{-1},hg_i,q_{i+1,i}\right)
  \ket{hg_i,hg_{i+1},x_{i+1,i}},
\eea
and the global action is
\bea
&U_L(h)\ket{\{g_i\},\{x_{i+1,i}\}}\\
&\quad=\prod_i\omega\!\left(h^{-1},hg_i,q_{i+1,i}\right)
  \ket{\{hg_i\},\{x_{i+1,i}\}} \, .
\label{eq:left-symmetry-action}
\eea
By using the pentagon identity it is possible to check that (on the periodic chain)
$U_L(h)U_L(h')=U_L(hh')$. Notice that this is the standard form of an anomalous symmetry generator \cite{Chen:2011pg}, but for continuous $G$.
On an interval, the endpoint factors instead determine the anomaly class $[\omega]$ by the Else--Nayak construction~\cite{Else:2014vma}.

\vspace{2mm} \noindent \emph{Right symmetry action.} Right multiplication $g_i\mapsto g_i h^{-1}$ conjugates the junctions' grade $q_{\ell} \mapsto q_\ell^h$, thus acting nontrivially on the Villain data.
Set $m=g_i h^{-1}$ and write $F(\rho) \in \Vec_G^\omega$ for the graded object obtained by applying the forgetful functor $F : \cZ(\cD) \to \cD$ to $\rho$. The local move contains an additional half-braiding:
\be
\begin{tikzcd}[row sep=2.8em,column sep=2em,font=\small]
(m\otimes h)\otimes F(\rho)
  \arrow[r,"\alpha"] \arrow[d,dashed,"\cU_R(h)"']
& m\otimes(h\otimes F(\rho))
  \arrow[d,"\unit_m\otimes c_{h,\rho}"]
\\
(m\otimes F(\rho))\otimes h
& m\otimes(F(\rho)\otimes h) \arrow[l,"\alpha^{-1}"]
\end{tikzcd}
\ee
Here $c_{h,\rho}:h\otimes F(\rho)\to F(\rho)\otimes h$ sends the component of grade $q$ to that of grade $q^h$. The diagram above can be understood by moving the symmetry line below the chain, where it crosses the $\rho$ line:
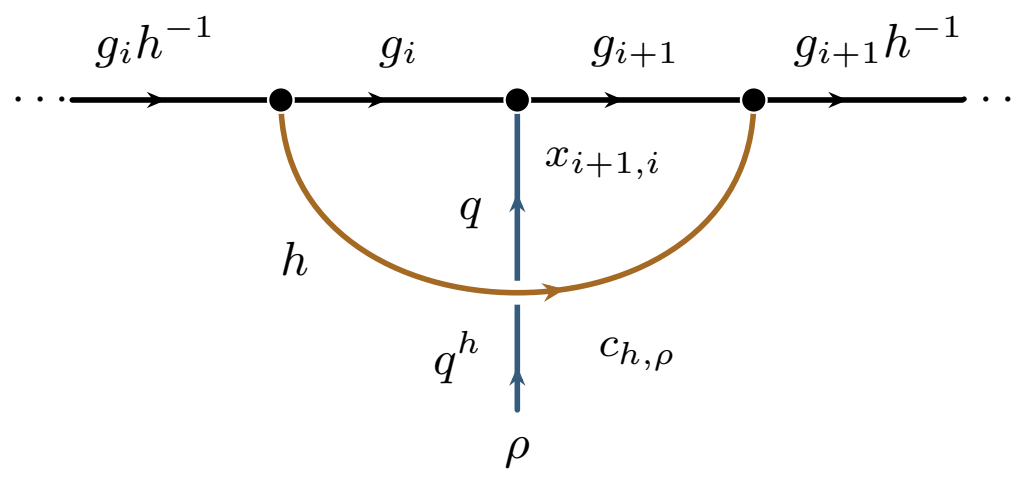
\be
\AnyonRightSymmetryMove{h}
\ee
In the double notation, the half-braiding is generated by left multiplication by the element
\be
\widetilde{h} = \int_G d\mu(p)   (\bde_p, h) \, ,
\ee
which gives
\be
\widetilde{h}\star(\bde_q,x)
  =\theta_{q^h}(h,x)^{-1}\,(\bde_{q^h},hx) \, .
\ee
Thus the grade is conjugated and the Villain label $x$ is translated on the left. The two associators contribute
\be
\varphi_R(h;g_i,g_{i+1})
  =\frac{\omega(g_i h^{-1},h,q_{i+1,i})}
  {\omega(g_i h^{-1},q^h_{i+1,i},h)} \, ,
\ee
while the local symmetry action also needs the correct half-braiding. The final action is
\bea
&\cU_R(h)_{i+1,i}\ket{g_i,g_{i+1},x_{i+1,i}}\\
&\quad=\varphi_R(h;g_i,g_{i+1})\,
  \theta_{q_{i+1,i}^h}(h,x_{i+1,i})^{-1}\\
&\qquad\times\ket{g_i h^{-1},g_{i+1}h^{-1},h x_{i+1,i}} \, ,
\eea
while globally
\bea \label{eq:right-symmetry-action}
&U_R(h)\ket{\{g_i\},\{x_{i+1,i}\}}\\
&\quad=\prod_i\left[
  \varphi_R(h;g_i,g_{i+1})\,
  \theta_{q^h_{i+1,i}}(h,x_{i+1,i})^{-1}\right]\\
&\qquad\times\ket{\{g_i h^{-1}\},\{h x_{i+1,i}\}} \, .
\eea
Again, it can be checked (see Appendix \ref{app: EN}) that the symmetry operators compose correctly, and reproduce the Else--Nayak anomaly.

\section{The non-abelian Villain model}
Having set up the anyon-chain description of the non-abelian model, we explore the allowed interaction terms, described by one- and two-leg Hamiltonians. We will focus on presenting an accessible local form of the lattice interactions by working in a fixed chart containing the identity. Our choice of chart is convenient to provide a simple and explicit form of the kinetic terms and the Villain potential.
The full derivations can be found in Appendix \ref{app:onetwoleg}.

\vspace{2mm} \noindent \textbf{One-leg terms: the Villain potential.} We start by determining the form of one-leg terms. Since $\rho$ is the regular object in the center, and $\cZ(\Vec_G^\omega) = \Rep(D(G)^\omega) \equiv \Rep(A)$, it follows that
\be
\End(\rho) = \End_A ({}_A A) = A^{\text{op}} \, ,
\ee
where op denotes the opposite algebra.
The action of $\End(\rho)\simeq A^{\rm op}$ on elements of the double is by right multiplication, $R_b(a)=a\star b$. This commutes with the half-braiding, since left and right multiplication of the Drinfeld double commute.\footnote{This follows from the identity
\be
\theta_{q^h}(h,x)\,\theta_{q^h}(hx,z)
=\theta_q(x,z)\,\theta_{q^h}(h,xz),
\ee
by direct computation.
}
The right action also commutes with the phases $\varphi_L$ and $\varphi_R$ appearing in the definition of the symmetry operators, since they are independent of $x$. Thus, the one-leg terms are $\cC$-symmetric.
To build smooth (non-distributional) Hamiltonians we employ the integrated elements:
\be
{\widetilde{z}} = \int_G d \mu(p) (\bde_p, z) \, ,
\ee
and define the right actions:
\be
\label{one-leg-right-multiplication}
\ba
R_{\widetilde z}(\bde_q,x)&=\theta_q(x,z)^{-1}(\bde_q,xz)\,,\\
R_{\widetilde z}^\dagger(\bde_q,x)&=\theta_q(xz^{-1},z)\,(\bde_q,xz^{-1})\,,
\ea
\ee
which on Villain wavefunctions $\psi_q(x) \in L^2(G)$ read:
\be
\left(R^\dagger_{\widetilde{z}}\psi_q\right)(x)
  =\theta_q(x,z)\,\psi_q(xz) \, .
\ee
A Hermitian one-leg term is $h_z^{(1)}= R_{\widetilde{z}}+R_{\widetilde{z}}^\dagger$. 
The operator $R_{\widetilde{z}}$ describes parallel transport on the Villain fiber. A canonical interaction term can be extracted from an isotropic limit in which $z$ approaches the identity, and provides a natural definition for a Laplacian.
To obtain this, we average $h_{z}^{(1)}$ over the conjugacy class $[z]$, using the normalized Haar measure:
\be
h_{[z]}^{(1)}=\int_G d\mu(u)\,h_{z^u}^{(1)} \, .
\ee
We then take $[z]\to[e]$, with $z = e^{i \epsilon_a T^a}$ and expand around the identity. 
To simplify the formula we use the basis
\be \label{eq: tildebasis}
\widetilde{|\Psi\rangle} = \prod_{\ell \, \in \, \text{links}} \nu_{q_\ell}(x_{\ell}) |\Psi\rangle \, , 
\ee
where $\theta_q(x,y) = \nu_q(x) \nu_{{}^x q}(y)/\nu_q(xy)$ is a local trivialization of the transgression $\theta_q$. Such trivialization exists always for fixed $q$, but cannot be chosen uniformly for all $q \in G$, as the transgression describes a nontrivial class in $H^2(G//G,U(1))$.
For example, when $G=SU(2)$, a trivialization can be taken in any chart which contains the identity and does not contain $q = -1$.
See Appendices \ref{app:drinfeld-double} and \ref{app: seconeleg} for more details. The same basis will be used to later to simplify the two-leg terms.

The linearization defines a covariant derivative
\be
R_{\widetilde{z}} = 1 - \epsilon^a \Ad_x(D)_a + O(\epsilon^2) \, , 
\ee
where $\Ad_x$ is the adjoint action and $D_a = \left( E_a - i \Gamma_{q, a} \right)$.
Explicitly, let $[T^a,T^b]=if^{abc}T^c$ and use the right-invariant frame
\be
E_a\psi(x)=\left.\frac{d}{dt}\psi(e^{itT^a}x)\right|_{t=0} \, ,
\qquad \Theta(E_a)=T^a \, ,
\ee
and also denote $q=e^{2\pi i\tau}$. We find, in the basis \eqref{eq: tildebasis} 
\be
\Gamma_q = - \frac{k}{2} \langle \tau , \Theta \rangle \, , \qquad \Gamma^a_q = - \frac{k}{2} \tau^a \, ,
\ee
where $\Theta = -i dx x^{-1}$ is the right-invariant Maurer--Cartan form and $\langle X, Y \rangle = \tr(XY)$. In the original basis the connection is not constant. 

For compact simple $G$ the class average gives\footnote{Parametrizing $z^u=e^{i\epsilon_a(u)T^a}$, the relevant Haar measure integrals are
\bea
\int_G d\mu(u)\,\epsilon_a(u)&=0,\\
\int_G d\mu(u)\,\epsilon_a(u)\epsilon_b(u)
  &=\frac{\epsilon^2}{\dim G}\,\delta_{ab}.
\eea}
\bea
h_{[z]}^{(1)}
  &=2\unit+\frac{\epsilon^2}{\dim G}\,\Delta_{x,q}
    +O(\epsilon^3),\\
\Delta_{x,q}&=\sum_a D_aD_a \, .
\eea
Unitarity then fixes the second-order term to be the square of the linearization.
The final one-leg Hamiltonian is therefore
\bea
&h_{\ell}^{(1)}&&=-\Delta_{x_{\ell},q_{\ell}} = \tr \left( -i E_{\ell} + \frac{k}{2} \tau_{\ell} \right)^2 \, , \\
&H^{(1)} &&=\sum_{\ell \, \in \, \text{links}} h_\ell^{(1)} \, , \qquad  \tau_\ell = \frac{1}{2\pi i}\log(q_\ell) \, .
\eea
This Villain potential is exactly solvable for fixed $\tau_{i+1,i}$, and describes a particle on $G$ in a constant magnetic field. We derive its spectrum and use it to study the continuum limit in Section~\ref{sec:continuum-limit-wzw}.

\paragraf{The $SU(2)$ case.} Let us work out the example of $G = SU(2)$ in more detail. We parametrize the $SU(2)$ link element as:
\be
q = e^{i \phi n \cdot \sigma} \, ,
\ee
with $n^a$ a unit vector and $\phi \in [0, \pi)$.\footnote{Remember that $\phi$ and $n^a$ are really functions of $g_i$ and $g_{i+1}$. The exact dependence is not illuminating and follows from performing the quaternionic multiplication.} We drop link labels for simplicity.
The logarithm gives $\tau=\frac{\phi}{2\pi}\,n_a\sigma^a$, i.e. $\tau^a=\frac{\phi}{\sqrt2\,\pi}\,n^a$ in the orthonormal basis $T^a=\sigma^a/\sqrt2$, and the potential in the invariant frame is:
\be
h^{(1)} = \tr\left( L + \frac{k \phi}{4 \pi} \, n \cdot\sigma  \right)^2  \, , \qquad L=L_aT^a \, ,
\ee
where $L_a$ are the $SU(2)$ generators.
It is also instructive to write the connection in the standard $S^3$ coordinates instead. We rotate into a basis in which $q = e^{i \phi \sigma_3}$, that is $n^a = (0,0,1)$ and parametrize
\be
x = e^{i \psi \sigma_3/2} e^{i \theta \sigma_2 /2} e^{i \varphi \sigma_3/2} \, ,
\ee
computing the Maurer--Cartan form gives
\be
\Gamma_q = -\frac{k \phi}{4 \pi}(d \psi + \cos(\theta) d\varphi) \, , \quad F_q = \frac{k \phi}{4 \pi} \sin(\theta) d\theta \wedge d\varphi \, .
\ee
This connection is in a gauge which is singular at the antipodal point $q= -1$.

\begin{figure}[t]
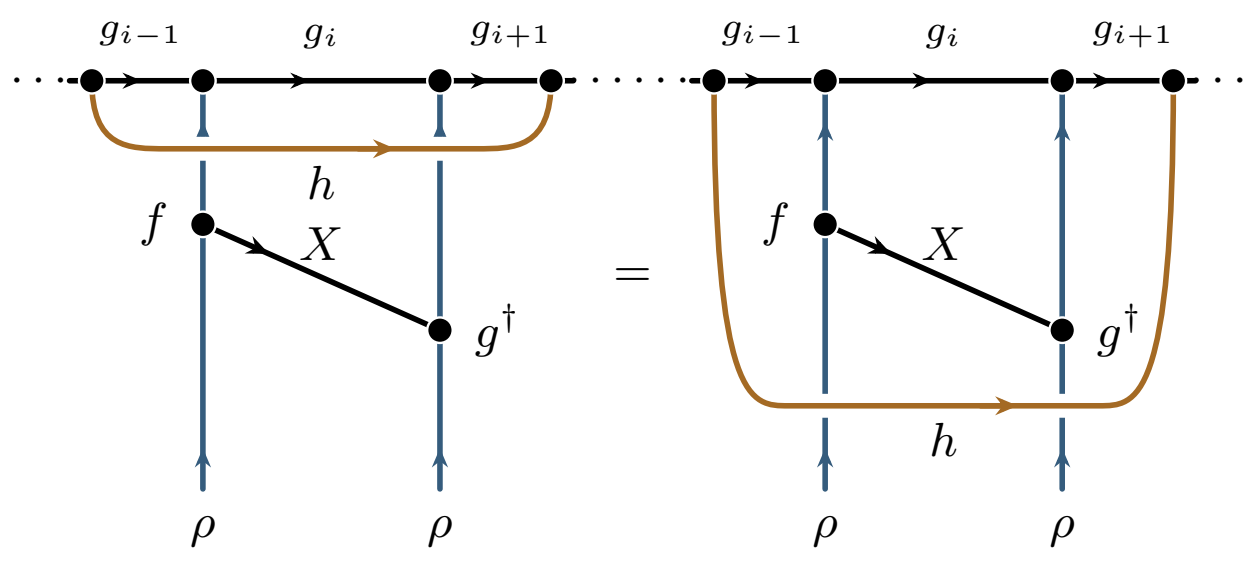

  \centering
  \AnyonTwoLegSymmetrySlide
  \caption{Sliding the right-symmetry line $h$ past the two-leg term
  $K_{X;(f,g)}$ to prove symmetry of the latter.}
  \label{fig: slide}
\end{figure}

\vspace{2mm} \noindent \textbf{Two-leg terms: the kinetic term.}
We now move to the description of the two-leg term. Given an object $X \in \cZ(\Vec_G^\omega)$ we choose a basis of two-leg operators
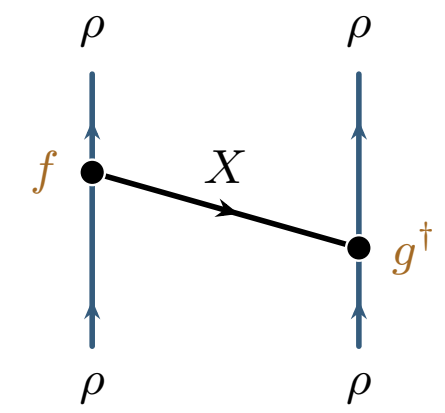
\be  \label{fig: 2-leg-wzw}
K_{X; (f,g)} \quad = \quad   \AnyonTwoLeg[execute at end picture={
    \node[text=anyonJunction,left=5pt,inner sep=1pt]
      at (0,-0.65) {$f$};
    \node[text=anyonJunction,right=5pt,inner sep=1pt]
      at (1.8,-1.15) {$g^\dagger$};
  }]{X} \, ,
\ee
where the vectors $f,g \in L^2(G)$ describe the junction maps
\be
\cI_f:\rho\longrightarrow\rho\otimes X,\qquad
\cJ_g:\rho\longrightarrow X\otimes\rho.
\ee
They can be expressed succinctly using the coproduct $\Delta$ on the Drinfeld double as
\be \label{eq: junction abstract}
\cI_f(a) =  \Delta(a) \cdot (1 \otimes \ket{f}_X) \, , \quad a \in D(G)^\omega \, ,
\ee
where $\Delta(a)$ acts on the left on the second factor.
These maps satisfy:
\be \label{eq: sliding move}
\cI_f \, c_{h, \rho} = c_{h, \rho \otimes X} \, \cI_f  \, ,
\ee
where $c_{h, X}$ is the half braiding in $\cZ(\Vec_G^\omega)$. An analogous identity holds for $\cJ_g$.
The two-leg operator can be expressed categorically as:
\be
K_{X; (f,g)} =   \left( \unit \otimes \cJ_g^\dagger \right)  \,  \alpha_{\rho, X, \rho} \, \left(\cI_f \otimes \unit\right) 
\ee
and symmetry follows from the sliding move:
\bea
 &K_{X; (f,g)} \,  c_{h, \rho \otimes \rho} = \\
  &\left( \unit \otimes \cJ_g^\dagger \right)  \,  \alpha_{\rho, X, \rho} \, c_{h,(\rho \otimes X) \otimes \rho}  \left(\cI_f \otimes \unit\right) = \\
  &\left( \unit \otimes \cJ_g^\dagger \right) c_{h, \rho \otimes (X \otimes \rho)}  \, \alpha_{\rho, X, \rho} \left(\cI_f \otimes \unit\right) = \\ 
  &c_{h, \rho \otimes \rho}  \, K_{X ; (f,g)} \,  ,
\eea
see Figure \ref{fig: slide} for a graphical representation. The proof of \eqref{eq: junction abstract} and \eqref{eq: sliding move} is provided in Appendix \ref{app:two-leg-evaluation}.
\begin{figure*}[t]
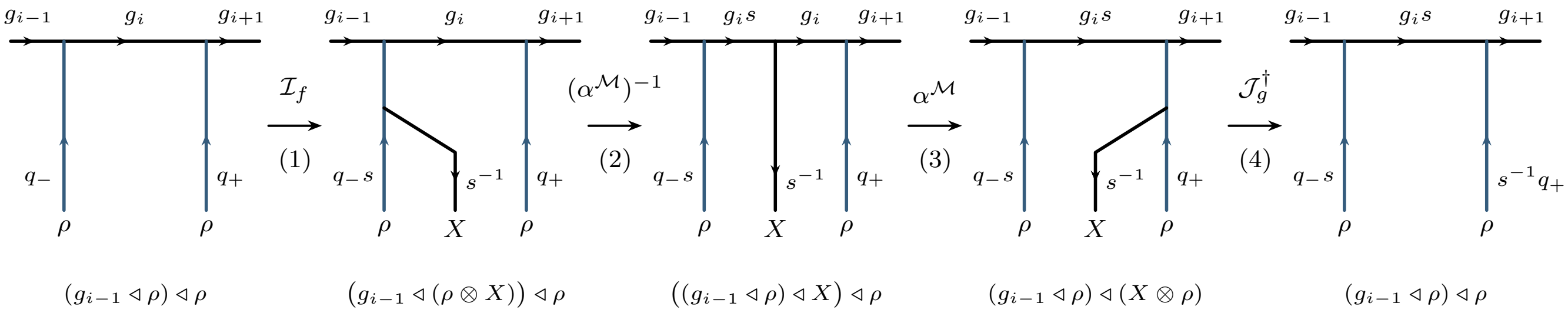

  \centering
  \TwoLegEvaluationOverview
  \caption{The four moves used to define the action of the two-leg operator $K_{X; (f,g)}$ on the anyon chain. (1) The line $X$ is emitted from $\rho$ via $\cI_f$. (2) The inverse module associator $(\alpha^\cM)^{-1}$ lets $X$ act on the anyon chain. (3) Using the module associator on the other $\rho$ leg removes $X$ from the chain. (4) $X$ is absorbed into $\rho$ via $\cJ_g^\dagger$.
  The labels beside the bulk lines indicate their image under the forgetful functor $F: \cZ(\Vec_G^\omega) \to \Vec_G^\omega$ when acting on the chain.}
  \label{fig: mod-ass}
\end{figure*}
The two-leg operator $K_{X; (f,g)} \in \End(\rho \otimes \rho)$ can be evaluated by the series of moves outlined in Figure \ref{fig: mod-ass}, with
\be
\alpha^{\cM}_{m, X, Y} = \alpha_{m, F(X), F(Y)} \, ,
\ee
the module associator, which is fixed in terms of the cocycle $\omega(g,h,l)$ by the forgetful functor $F: \cZ(\cD) \to \cD$. 
The abstract two-leg Hamiltonian is:
\be
h^{(2)}_{X; (f,g)} = K_{X; (f,g)} + K^\dagger_{X; (f,g)} \, ,
\ee
and we must evaluate its action on the anyon chain Hilbert space.
The operator acts on three sites and two links, that is on states of the form
\be
|\Psi\rangle = | g_{i-1}, g_i, g_{i+1}, x_{i,i-1}, x_{i+1,i} \rangle \, ,
\ee
to lighten the equations we denote $x_{i, i-1} = x_-$, $x_{i+1, i} = x_+$, with analogous notation for all link quantities. 
The standard choice for $X$ would be $X_{[z], \lambda}$, with $\lambda$ a projective irrep of the centralizer. This choice, however, does not lead to a smooth $[z] \to [e]$ limit, as the centralizer jumps discontinuously at the identity. We instead choose:
\be
X_{[z]} \equiv X_{[z], \text{Reg}} = A (\bde_z , 1) \, , \qquad A = D(G)^\omega \, ,
\ee
to carry the regular representation. We denote the corresponding two-leg term by $K_{[z]; (f,g)}$.
Evaluating the terms in Figure \ref{fig: mod-ass} explicitly gives the kernel (derived in Appendix \ref{app:onetwoleg}) 
\bea \label{eq: twoleglocal}
K_{[z]; (f,g)} \widetilde{|\Psi\rangle} &= \int_G d \mu(r) f(r) \overline{g(w r)} \, T_{s(r)} \, Z_{s(r)} \widetilde{\ket{\Psi}} \, , \\
Z_s &=  \omega(g_{i-1}, q_-, s) \omega(g_i s, s^{-1}, q_+) \, , \\
T_s &|..., g_i, ...\rangle =  |..., g_i s, ... \rangle \, ,\\
w = x_+^{-1} x_ - \, , &\quad s(r) = (z^{x_- r})^{-1} \, , \quad f,g \in L^2(G) \, .
\eea
Let us comment on its form: the two-leg operator is implementing a (right) translation on the site variable $g_i$, while leaving the link variables $x_{\ell}$ fixed. In order to be symmetric, the translation must be dressed by the $Z_s$ term, which gives rise to a covariant derivative in the infinitesimal limit. Its physical content encodes the WZW $B$-field on the lattice. This expression holds for $q_\pm$ in a chart containing the identity. Other charts are related by the transition functions of the multiplicative WZ gerbe \cite{Gawedzki:2002se,Gawedzki:2004tu,Gawedzki:2009jj}. We give a more complete treatment in Appendix \ref{app: chartwise}. 

At this point, we are free to choose $f, g$ to simplify the expression further. We choose isotropic junction data
\be
f(r) = g(r) = 1 \, ,
\ee
in the tilded basis.
Changing integration variables to $u = x_- r$ we have:
\be
K_{[z]; (1,1)} \widetilde{|\Psi\rangle} =  \int_G d \mu(u) 
T_{(z^u)^{-1}} \, Z_{(z^u)^{-1}} \,
 \widetilde{|\Psi\rangle} \, .
\ee                               
The lattice kinetic term follows from the small $[z]$ expansion of the above operator. For $z = e$, both $T_{(z^u)^{-1}}$ and $Z_{(z^u)^{-1}}$ collapse to the identity. As the integrand is unitary the expansion of the two-leg operator must again take the form:
\be
K_{X; (1,1)} + K^\dagger_{X; (1,1)} = 2 \unit - \frac{\epsilon^2}{\dim(G)} \tr P^2 + O(\epsilon^3) \, ,
\ee
with:
\be
 T_{z} Z_{z} \sim 1 + i \epsilon^a P_a + ... \, , \quad
P_a = J_a + k \cB_a \, ,
\ee
where $J_a$ is the generator of the right $G$ action on sites $g_i$, while:
\bea
k\cB_{a,i} &= -i \frac{\partial}{\partial \epsilon_a} \log Z_{s = e^{i \epsilon_b T^b}} \vert_{\epsilon_b = 0}  \\
&= \int_{\gamma[g_{i-1},g_i]}\iota_{\xi_-^a}B_k
   +\int_{\gamma[g_i,g_{i+1}]}\iota_{\xi_+^a}B_k \, ,
\label{eq:lattice-B}
\eea
is the lattice $B$-field. This is obtained from the integral of the WZW B-field on the geodesics $\gamma[g_i, g_{i+1}]$ connecting successive anyon chain sites. The precise form of the vector fields $\xi_\pm$ depend on the choice of cocycle representative $\omega$, and lead to equivalent Hamiltonians in the continuum limit. Details of the computation can be found in Appendix \ref{app:onetwoleg}. From the above, we conclude that the two-leg Hamiltonian is given by:
\bea
h^{(2)}_i &= \tr P^2_i = \sum_a \left( J_{a,i} + k \cB_{a,i} \right)^2 \, , \\
H^{(2)} &= \sum_{i \, \in \, \text{sites}} \tr P_i^2 \, .
\eea

\vspace{2mm} 
\noindent \textbf{The non-abelian chiral Hamiltonian.} We thus arrive at our final expression for the non-abelian chiral Hamiltonian:
\be \label{eq: chiral-ham}
H = \frac{1}{2 \beta }\sum_{i \, \in \, \text{sites}}  \tr P_i^2   - \frac{\beta}{2} \sum_{\ell \, \in \, \text{links}} \Delta_{x_\ell, q_\ell} \, .
\ee
Despite the complexity of the derivation the final result is remarkably simple: the expected kinetic term for the lattice $g_i$ fields is supplemented by a $q_\ell$-dependent Villain potential, which describes a rotor in a constant magnetic field. The anomaly has two roles in this Hamiltonian: first, it couples the bare lattice momentum for $g_i$ to the correct lattice B-field; second, it sources the background magnetic field for the auxiliary Villain rotors.

It is instructive to check consistency with the abelian Villain model. To do so, we restrict the rotor Hilbert spaces $g_i$ and $x_{\ell}$ to a single $U(1)$ Cartan generator $H$. We choose its embedding index to be one. Thus $g_i = e^{i \phi_i H}$, $x_{\ell} = e^{i \widetilde{ \phi_\ell} H}$, with both $\phi_i$ and $\widetilde{\phi_\ell}$ $2\pi$ periodic in this normalization.
To maintain this choice consistently, the translation operators are also restricted to the same Cartan. The two-leg term is very simple: the lattice B-field is trivial and the kinetic term reduces to
\be
\tr P_i^2 \mapsto  \frac12\,\pi_i^2 \, .
\ee
On the other hand, on links we have $q_\ell = e^{i d \phi_\ell H}$ exactly, which leads to $\tau_\ell = \frac{[d \phi_\ell]_{2\pi}}{2\pi} H$ upon taking logarithms. We diagonalize the momentum $E_\ell$ on the links, which gives rise to an integer Villain label $n_\ell$. Taking into account $\tr H^2 =2$, the one-leg potential becomes:
\be
h_\ell^{(1)} = \frac{1}{2} \left( n_\ell - \frac{k}{2\pi} [d \phi_\ell]_{2\pi} \right)^2 \, ,
\ee
in agreement with the Villain model \eqref{eq:villain-hamiltonian} for $k=1$.

\section{Evidence for a \texorpdfstring{WZW$_k$}{WZW-k} fixed point}
\label{sec:continuum-limit-wzw}
\begin{figure}[t]
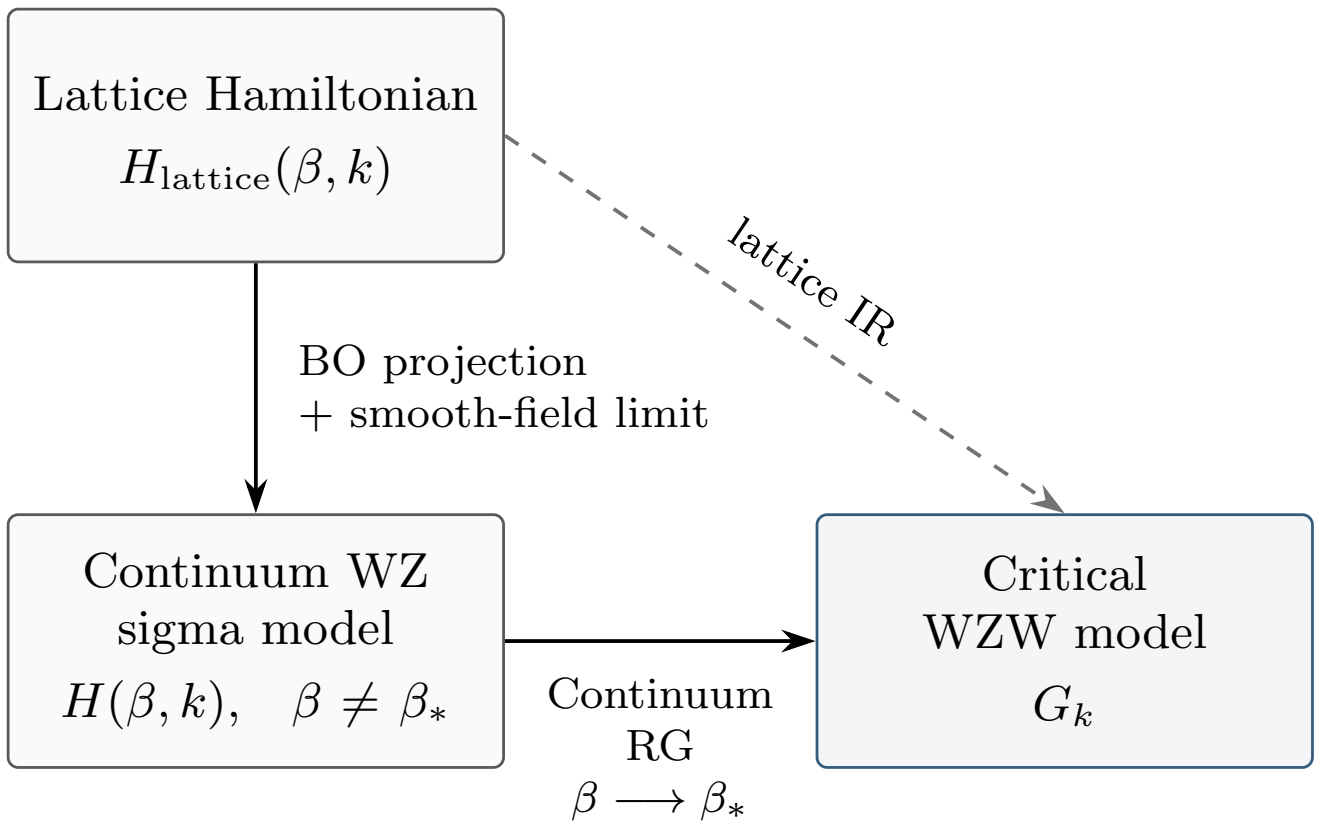

    \centering
    \WZWFlowDiagram
    \caption{The conjectured lattice-continuum flow for the non-abelian Villain model. The Born--Oppenheimer truncation leads to an anomalous principal chiral model, whose coupling $\beta$ flows in the continuum to the WZW$_k$ CFT.}
    \label{fig: lattice-to-WZW}
\end{figure}

The Hamiltonian \eqref{eq: chiral-ham} is not exactly solvable. To make progress we assume that, in the continuum limit, the $g_i$ fields remain light, and approach a smooth field $g(x)$ via $g_j \sim g(a j)$.
This suggests that we should perform a Born--Oppenheimer approximation, in which $H_{\text{heavy}} = \frac{\beta}{2} H^{(1)}$ and expand around smooth $g_i$, i.e. 
\be
q_{i+1,i} \sim \unit + i a \cJ + O(a^2) \, , \qquad \cJ =-i g^{-1}\partial_x g \, .
\ee
After performing the usual rescaling of the momenta $J_a \to a \pi_a$, with $\pi_a$ the continuum momentum, it also holds that $\cB_i \sim a g^*B$, with $g^* B$ denoting the contraction of the WZW B-field with $\cJ$, see \eqref{eq:app-lattice-B-continuum} in Appendix \ref{app:lattice-B}. 
The Hamiltonian \eqref{eq: chiral-ham} then splits into a heavy and light component:
\bea
&H_{\text{light}} &&= \frac{a^2}{2\beta} \sum_{i \, \in \, \text{sites}} \tr p^2 \, , \quad p = \pi + k g^* B \, , \\
&H_{\text{heavy}} &&= \frac{\beta}{2}\sum_{\ell \, \in \, \text{links}} \tr \left(-i E_\ell + a \frac{k}{4 \pi} \cJ |_{\ell} \right)^2 \, .
\eea
As $H_{\text{light}} \sim O(a^2)$ while $H_{\text{heavy}} \sim O(1)$, it is natural to take the Born--Oppenheimer ansatz:
\be
|\Psi \rangle =  |\Psi_{\text{light}}\rangle \otimes |\Psi_{\text{heavy}}\rangle \, ,
\ee
where the heavy wavefunction diagonalizes $H_{\text{heavy}}$ for \emph{fixed} light degrees of freedom $g_i$.
We are thus led to examine the spectrum of the Villain potential:
\be
h^{(1)} = \sum_{a} \left(L_a + \frac{k}{2} \tau_a \right)^2 \, , \quad  L_a = -i E_a \, ,
\ee
for fixed $\tau_a$.
Expanding the square gives a Zeeman-like Hamiltonian
\be
h^{(1)} = L^2 + k \tau^a L_a + \frac{k^2}{4} \tau^2 \, ,
\ee
which can be diagonalized in the Peter-Weyl basis $|R, \mu, \mu'\rangle \in \bigoplus_R V_{R} \otimes V_{R^*} \equiv L^2(G)$ for the Villain rotor, with spectrum:
\be
h^{(1)}|R, \mu, \mu'\rangle = \left( C_2(R) - \mu^2 + \left(\mu + \frac{k}{2}\tau\right)^2  \right) |R, \mu, \mu'\rangle \, .
\ee
The second weight $\mu'$ does not appear in the spectrum so each energy level is at least $\dim(R)$-fold degenerate.
To integrate out the Villain fields we minimize the effective potential:
\be \label{eq: vill-spec}
E_0(\tau) = \min_{R,\mu} \left( C_2(R) - \mu^2 + \left(\mu + \frac{k}{2}\tau\right)^2  \right) \, .
\ee
As $\tau \sim \frac{a}{2\pi} \cJ$ is small, the ground state is an s-wave in the basis \eqref{eq: tildebasis}. In the original basis it is instead given locally by:
\be
\Psi_{\text{heavy}}(\{ x_{\ell} \}) = \prod_\ell \nu_{q_\ell}(x_{\ell}) \, ,
\ee
leading to the effective light Hamiltonian
\be
H_{\text{eff}} =  \frac{a^2}{2\beta} \sum_{i \, \in \, \text{sites}} \tr p^2  + \frac{a^2 k^2 \beta}{32\pi^2} \sum_{\ell \, \in \, \text{links}} \tr \cJ|_\ell^2 \, .
\ee
After rescaling $H = H_{\text{eff}}/a$ we take the continuum limit:
\be \label{eq: cont-model}
H = \int dx \left[ \frac{1}{2\beta}  \tr p^2  - \frac{k^2 \beta}{32 \pi^2} \tr (g^{-1}\partial_x g )^2 \right] \, .
\ee
This is a version of the principal chiral model, with a WZW term. For arbitrary $\beta$, this is not the critical WZW Hamiltonian \cite{Witten:1983ar,Alekseev:2004np}. However the coefficient $\beta$ will generally undergo RG flow. 
In the continuum model \eqref{eq: cont-model} $\beta$ has a unique fixed point at \cite{Witten:1983ar}:
\be
\beta_* =  1  \, ,
\ee
corresponding to the WZW$_k$ CFT. Thus, we expect our lattice model to eventually flow to a critical Hamiltonian for every $\beta$, which should describe the WZW$_k$ CFT, see Figure \ref{fig: lattice-to-WZW}.
It would also be interesting to study the transformation of the lattice model under lattice renormalization (such as block spin renormalization) and study the behavior of the coupling $\beta$. It is not obvious that these two ways of computing the continuum limit commute. However, the symmetry and its anomaly strongly constrain the possibilities of the IR theory of the lattice model, in particular, since the lattice model has exact anomalous Lie group symmetry, the IR theory must be gapless. We hope to come back to this problem in the future.

\section{Discussion}
\label{sec:discussion}
In this work we have constructed a (1+1)d quantum lattice Hamiltonian with exact non-abelian lattice chiral symmetry. Our construction determines the additional Villain fields and leads to a proposal for its continuum fixed point. We conclude with a discussion of future and open directions. 

\vspace{2mm} \noindent \textbf{Proof of the continuum limit.} A future objective is an analytic proof or numerical test that our lattice model flows to the WZW$_k$ critical point, for example by performing numerical diagonalization of the low-lying spectrum and an estimate of the central charge.

\vspace{2mm} \noindent \textbf{Numerical truncations.} 
Our models require an infinite-dimensional on-site Hilbert space. This makes their numerical simulation challenging. A naive approach would be to truncate the rotor and Villain Hilbert space in the Peter--Weyl basis $L^2(G) = \bigoplus_R V_R \otimes V_{R^*}$ to a finite (but large) representation $R^{\text{max}}$. 
This breaks the anomalous symmetry explicitly, but hopefully such breaking could be small in e.g. the inverse Casimir $1/C_2(R^{\text{max}})$. 
The matter content of a chiral gauge theory must have zero anomalies in order to gauge the symmetry consistently. 
When this happens, it should be possible to devise an appropriate truncation scheme which preserves the anomaly-free chiral symmetry on a finite-dimensional tensor-factorized Hilbert space. {In a sense, this is a type of symmetric mass generation. See e.g. \cite{Wang:2013yta,Wang:2018ugf,Zeng:2022grc,Wang:2022ucy,Zakharov:2026nly} for studies of this mechanism in quantum lattice systems.}
A related truncation, which preserves a $\bZ_N$ chiral symmetry in abelian Villain models, has been studied recently in \cite{Lootens:2026kcg}.

\paragraf{Modular invariants and boundary conditions.} In this work we have not clarified whether the model flows to the \emph{diagonal} WZW CFT, or to a different modular invariant, nor how the spectrum of vertex operators can be extracted from the lattice model. This clearly involves a precise understanding of the excitations of Villain rotors.
Furthermore, it should be interesting to study the set of boundary conditions of the lattice model, which in the continuum limit encode the 't Hooft anomaly through their topology \cite{Choi:2025ebk,Copetti:2025sym,Drukker:2025dfm}.

\vspace{2mm} \noindent \textbf{Fermionic models.} The effort to define chiral lattice models was originally motivated by the prospect of regularizing Standard Model fermions. Our models so far have been purely bosonic in nature. In (1+1)d, a fermionic chiral lattice model can always be obtained by fermionization \cite{Karch:2019lnn,Seifnashri:2026ema} when the original model has an anomaly-free $\bZ_2$ symmetry. A more natural approach from the anyon chain perspective, which also can be generalized to higher-dimensional cases, is instead to replace the bosonic symmetry category $\Vec_G^\omega$ by its fermionic analogue $\sVec_G^\Omega$ \cite{Bhardwaj:2024ydc, Huang:2024ror, Wen:2024udn}. The anyon chain formulation should then incorporate microscopic fermions automatically.

\vspace{2mm} \noindent \textbf{Chiral gauge theories.} 
The final goal of any program studying lattice chiral symmetry is to provide a (3+1)d quantum Hamiltonian for a non-abelian chiral gauge theory, such as the Standard Model. A first step in this direction would be to study the lattice non-abelian gauge theory obtained by gauging an anomaly-free subgroup in our model. The resulting models should be comparable with the conjectured IR fixed points of gauged WZW models \cite{Dalley:1992yy,Kutasov:1993gq,Cherman:2019hbq,Komargodski:2020mxz,Delmastro:2021otj,Dempsey:2023fvm,Dempsey:2024alw}.

\vspace{2mm} \noindent \textbf{Higher-dimensional extensions.} An ambitious goal is to generalize our construction to (3+1)d systems. Anyon chains have been described for fusion 2-categories in (2+1) dimensions in e.g. \cite{Inamura:2023qzl, Inamura:2025cum}, while higher-dimensional generalizations have not yet explicitly been worked out.

\vspace{2mm} \noindent \textbf{Acknowledgments.} The author is grateful to Andrea Antinucci, Riccardo Argurio, Lukasz Fidkowski, Kansei Inamura, Sal Pace, Giovanni Rizi, and Shu-Heng Shao for discussions, and to Giovanni Galati for pointing out ref.~\cite{Chen:2024ddr} and discussion on non-abelian Villain fields. The author is especially thankful to Sakura Schafer-Nameki and Rui Wen for collaboration in the initial stages of the project and feedback on the draft.
The author is partially supported by the STFC grant ST/X000761/1.
The author thanks the Galileo Galilei Institute for hospitality during the program ``Defects and Extended Excitations in Quantum Field Theory, Quantum Matter and Statistical Models'', where the idea for this work was conceived. Commercially available Anthropic and OpenAI models have been used to simplify and check the expressions of lattice interactions and generate figures. The results were checked by the author. Typos in equations and text are his responsibility.

\bibliographystyle{ytphys}
 \let\bbb\bibitem\def\bibitem{\itemsep1.6pt\bbb}
\bibliography{refs}

\clearpage

\onecolumngrid

\begin{center} \LARGE \textbf{Appendix} \end{center}
\label{app:derivations}

\appendix
\addtocontents{toc}{\protect\AppendixContentsStart}
\vspace{2mm}\noindent\textbf{Contents}\par\vspace{1mm}
\begingroup
  \AppendixContentsfalse
  \let\FullContentsLine\contentsline
  \renewcommand{\contentsline}[4]{%
    \ifAppendixContents\FullContentsLine{#1}{#2}{#3}{#4}\fi}
  \makeatletter
  \let\appendix\appendix@toc
  \let\l@subsubsection\@gobbletwo
  \@starttoc{toc}
  \makeatother
\endgroup
\vspace{3mm}

\section{Categorical continuous groups}
\label{sec:sym-cont-cat}

\subsection{The category and its anomaly}
Let $G$ be a compact, connected Lie group. A perturbative bosonic anomaly for a
$G$ symmetry in (1+1)d is an integral class
\be
[\omega]\in H^4(BG,\bZ)
  \simeq H^3_{\rm Borel}(G,U(1)).
\label{eq:continuous-anomaly-class}
\ee
The second presentation uses measurable group cochains, or equivalently
locally smooth cochains with transition data~\cite{wagemann2015cocycle}.
For a compact, simple, simply connected group, the class is specified
by its integer level $k$. For $SU(2)$ our convention is $[\omega]=k c_2$.

We use the notation $\Vec_G^\omega$ for the corresponding continuous symmetry
category. Its simple objects (symmetry lines) are denoted by $\bde_g$, with fusion and associator:
\be
\bde_g\otimes\bde_h\simeq\bde_{gh},\qquad
\alpha_{g,h,l}:(\bde_g\otimes\bde_h)\otimes\bde_l
  \longrightarrow\bde_g\otimes(\bde_h\otimes\bde_l).
\ee
In a chosen local trivialization, $\alpha_{g,h,l}$ is described via scalar multiplication by
$\omega(g,h,l) \in U(1)$. The pentagon identity follows from the group cocycle identity:
\be
\omega(h,l,m)\,\omega(g,hl,m)\,\omega(g,h,l)
  =\omega(gh,l,m)\,\omega(g,h,lm),
\label{eq:continuous-cocycle-pentagon}
\ee
and we normalize:
\be
\omega(e,g,h) = \omega(g,e,h) = \omega(g,h,e) = 1 \, .
\ee
A change of fusion-junction phases by a two-cochain $\mu$ changes
$\omega$ by the coboundary
\be
\omega\longmapsto\omega\,\delta\mu,\qquad
(\delta\mu)(g,h,l)
  =\frac{\mu(h,l)\mu(g,hl)}{\mu(gh,l)\mu(g,h)}.
\ee
For a continuous $G$, general objects in $\Vec_G^\omega$ can be expressed as direct integrals:
\be
\Vec_G^\omega \ni X = \int_G^\oplus d\mu(g)  V_g(X) \, \bde_g \, ,
\ee
where $V_g(X)$ (the multiplicity space) is a (measurable) Hilbert space, meaning that $\Hom(X, \bde_g) = V_g(X)^\vee$.
The notation $\Vec_G^\omega$ should be understood in this setting; its
operator-algebraic formulation is denoted $\mathrm{Hilb}^k(G)$
in~\cite{Jia:2026vcr}. This construction can also be described as a multiplicative gerbe bundle, which directly connects with its appearance in WZW models. See e.g. \cite{Marin-Salvador:2025stc}, Example 3.12 and Remark 3.13 for a detailed presentation. The gerbe data enter in the fusion of general objects, which is twisted by the anomaly.

\subsection{Cocycles from the WZW term}
\label{app:cocycle}
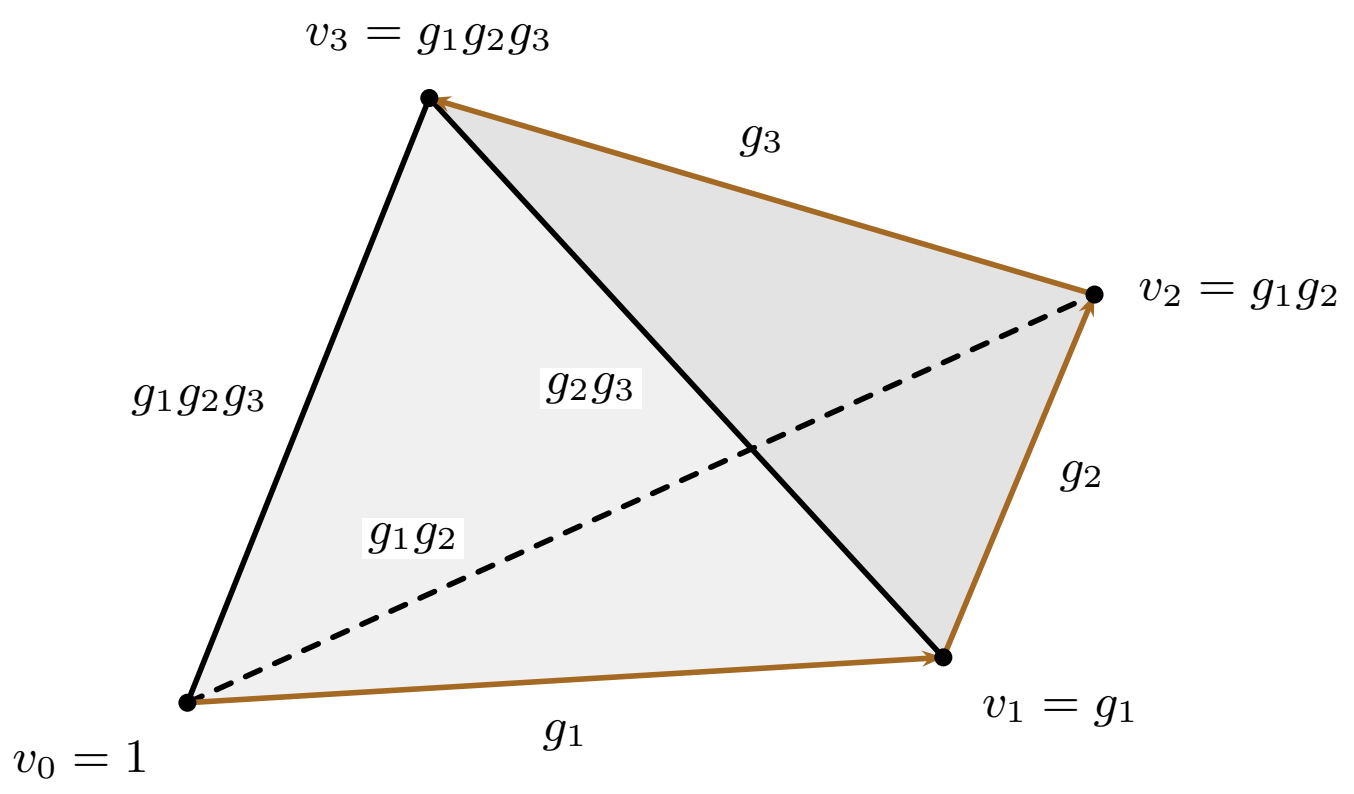
\begin{figure}[htbp]
  \centering
  \begin{tikzpicture}[x=1cm,y=1cm,font=\small,
    line cap=round,line join=round]
    \coordinate (v0) at (0,0);
    \coordinate (v1) at (5,0.3);
    \coordinate (v2) at (6,2.7);
    \coordinate (v3) at (1.6,4);
    \fill[anyonModule!6] (v0)--(v1)--(v3)--cycle;
    \fill[anyonModule!11] (v1)--(v2)--(v3)--cycle;
    \draw[anyon module,dashed] (v0)--(v2)
      node[pos=0.32,above left=2pt,fill=white,inner sep=1pt] {$g_1g_2$};
    \draw[anyon module] (v0)--(v3)
      node[midway,left=4pt] {$g_1g_2g_3$};
    \draw[anyon module] (v1)--(v3)
      node[pos=0.55,below left=3pt,fill=white,inner sep=1pt] {$g_2g_3$};
    \draw[draw=anyonJunction,-{Stealth[length=4pt]},line width=1.05pt]
      (v0)--node[below=4pt] {$g_1$}(v1);
    \draw[draw=anyonJunction,-{Stealth[length=4pt]},line width=1.05pt]
      (v1)--node[right=4pt] {$g_2$}(v2);
    \draw[draw=anyonJunction,-{Stealth[length=4pt]},line width=1.05pt]
      (v2)--node[above=4pt] {$g_3$}(v3);
    \foreach \v in {v0,v1,v2,v3} {
      \fill[anyonInk] (\v) circle (1.7pt);
    }
    \node[below left=4pt] at (v0) {$v_0=1$};
    \node[below right=4pt] at (v1) {$v_1=g_1$};
    \node[right=5pt] at (v2) {$v_2=g_1g_2$};
    \node[above=5pt] at (v3) {$v_3=g_1g_2g_3$};
  \end{tikzpicture}
  \caption{The WZ tetrahedron defining $\omega_k(g_1,g_2,g_3)$.
  The edge $v_a\to v_b$ carries $v_a^{-1}v_b$; the three highlighted
  edges carry the arguments $g_1,g_2,g_3$.}
  \label{fig:app-tetra}
\end{figure}

Throughout this work, we will use a precise representative for the class $[\omega]$ to perform explicit manipulations. Our conventions below are that $\tr$ is normalized so that long roots have squared length two (for $G=SU(N)$, this is the trace in the defining representation).
The Wess--Zumino three-form is
\be
H_k=\frac{k}{12\pi}\tr(g^{-1}\dd g)^3,
\qquad \dd H_k=0,\qquad
\int_\Sigma H_k\in2\pi\bZ
\label{eq:WZ-three-form}
\ee
for every closed integral three-cycle $\Sigma$ \cite{Witten:1983ar}. The WZ term $H_k$ is invariant under (constant) left and right translations
\be
H_k(h_L g h_R^{-1}) = H_k \, ,
\ee
exactly. 
Our ensuing construction is easier to understand using homogeneous cocycles $\Omega(v_0,v_1,v_2,v_3)$, which are related to $\omega$ via:
\bea
\omega(g_1,g_2,g_3)
  &=\Omega(1,g_1,g_1g_2,g_1g_2g_3),\\
\Omega(v_0,v_1,v_2,v_3)
  &=\omega(v_0^{-1}v_1,v_1^{-1}v_2,v_2^{-1}v_3).
\label{eq:app-dictionary}
\eea
We now consider elements $v_0, v_1, v_2, v_3 \, \in \, G$ and build an embedded tetrahedron by choosing paths $\gamma[v_i, v_j]$ between them, triangular fillings $\Delta[v_i,v_j,v_k]$ and an oriented filling $\cT[v_0,v_1,v_2,v_3]$, as in Figure \ref{fig:app-tetra}.
We claim that:
\be \label{eq: wzw-rep}
\Omega_k (v_0, v_1, v_2, v_3) \equiv  \exp\left( - i \int_{\cT[v_0,v_1,v_2,v_3]} H_k \right) \, ,
\ee
defines a nontrivial cocycle $[\omega_k] \in H^3_{\text{Borel}}(G,U(1))$.
It can be checked that:
\be
\delta \Omega_k(v_0,v_1,v_2,v_3,v_4) = \prod_{j=0}^4 \Omega_k(v_0, ... , \hat{v}_j , ..., v_4)^{(-1)^j} = \exp\left(- i \int_{\Sigma} H_k \right) = 1 \, ,
\ee
so that $\Omega_k$ is closed.  Similarly, due to the quantization on closed cycles, two different fillings $\cT$ and $\cT'$ differ by a closed cycle $\Sigma = \cT \cup \overline{\cT ' }$, which leaves the class invariant. Similar arguments can be used to prove that the class is independent of choices for $\gamma[v_i,v_j]$ and $\Delta[v_i, v_j, v_k]$.

Bi-invariance of $H_k$ gives some useful identities:
\bea
\Omega_k(av_0,av_1,av_2,av_3)&=\Omega_k(v_0,v_1,v_2,v_3),\\
\Omega_k(v_0b,v_1b,v_2b,v_3b)&=\Omega_k(v_0,v_1,v_2,v_3).
\label{eq:app-bi-invariance}
\eea
Consequently the inhomogeneous representative is invariant under
simultaneous conjugation:
\be
\omega_k(ag_1a^{-1},ag_2a^{-1},ag_3a^{-1})
  =\omega_k(g_1,g_2,g_3).
\label{eq:app-conjugation-invariance}
\ee
Permuting the ordered vertices changes the orientation of $\cT$, leading to:
\be
\Omega_k(v_{\sigma(0)},v_{\sigma(1)},v_{\sigma(2)},v_{\sigma(3)})
  =\Omega_k(v_0,v_1,v_2,v_3)^{\operatorname{sgn}\sigma}.
\label{eq:app-antisym}
\ee
Ordinary normalization follows when adjacent vertices coincide. In computations, we choose $\gamma[v_i,v_j]$ to be a geodesic edge $\gamma(t) = v_i \exp\left( t \log(v_i^{-1} v_j) \right)$, and iterated geodesic cones for $\Delta$ and $\cT$. For $G=SU(2)$ these fillings do not depend on the ordering of the vertices; in general we assume such a symmetric choice.
On the same chart, this ensures that a degenerate simplex with any repeated vertices can also be assigned phase one. This will be used in our manipulations.

If the tetrahedron further lies in a chart where $H_k = d B_k$ admits a trivialization, we can also write:
\be \label{eq:app-stokes}
\Omega_k(v_0,v_1,v_2,v_3)
  =\exp\!\left(-i\int_{\partial\mathcal T[v_0,v_1,v_2,v_3]}B_k \right) \, ,
\ee
where:
\be
\partial \cT[v_0, v_1, v_2, v_3] = \Delta[v_1,v_2,v_3] - \Delta[v_0, v_2, v_3] + \Delta[v_0, v_1, v_3] - \Delta[v_0,v_1,v_2] \, , 
\ee
is the boundary.

\vspace{2mm} \noindent \emph{Global data.}
In the previous manipulations, we have assumed that $\cT[v_0,v_1,v_2,v_3]$ lies in a single patch of our covering of $G$.
When a tetrahedron crosses several patches, the single potential $B$
in~\eqref{eq:app-stokes} must be replaced by local potentials $B_i$ and
their transition data. 
Choose a good cover $\{U_\alpha\}$ of $G$, so that
all nonempty finite intersections are contractible. The WZ gerbe
$\cG_k$ is represented by two-forms $B_\alpha$ on $U_\alpha$,
one-forms $\mathcal A_{\alpha\beta}$ on $U_\alpha\cap U_\beta \equiv U_{\alpha\beta}$, and phases
$\zeta_{\alpha \beta \gamma}$ on $U_\alpha\cap U_\beta\cap U_\gamma \equiv U_{\alpha\beta\gamma}$, satisfying \cite{Gawedzki:2002se}
\bea 
\dd B_\alpha=H_k,\qquad B_\alpha-B_\beta&=\dd\mathcal A_{\alpha\beta},\\
\mathcal A_{\beta \gamma}-\mathcal A_{\alpha \gamma}+\mathcal A_{\alpha \beta}
  &=i\zeta_{\alpha \beta \gamma}^{-1}\dd\zeta_{\alpha \beta \gamma},\\
\zeta_{\beta \gamma \delta}\zeta_{\alpha \gamma \delta}^{-1}\zeta_{\alpha \beta \delta}\zeta_{\alpha \beta \gamma}^{-1}&=1.
\label{eq:app-gerbe-descent}
\eea
These equations can be used to patch together the cocycle data.

Let $\Sigma\to G$ be a closed oriented surface, triangulated so that
each face $f$, edge $e$ and vertex $v$ lies in a chosen patch with
index $\alpha_f,\beta_e,\gamma_v$. Pullbacks along the surface map are implicit.
Choose an orientation on every edge, and let $\epsilon(f,e)$ and
$\epsilon(e,v)$ be the incidence signs in the boundaries of $f$ and
$e$. In our $\exp(-i\int H_k)$ convention the amplitude is
\bea
\mathcal W_k(\Sigma)
  ={}&\exp\!\left[-i\sum_f\int_f B_{\alpha_f}
    + i\sum_{e\subset f}\epsilon(f,e)
      \int_e\mathcal A_{\alpha_f \beta_e}\right]\\
  &\times\prod_{v\subset e\subset f}
    \zeta_{\alpha_f \beta_e \gamma_v}(v)^{\epsilon(f,e)\epsilon(e,v)}.
\label{eq:app-gerbe-holonomy}
\eea
Changing a face label changes its integral by a boundary integral of
$\mathcal A$. The edge factors cancel this change, and their endpoint
corrections are canceled by the $\zeta$ factors. The quadruple-overlap
identity makes the result independent of the order of these changes.
Thus $\mathcal W_k$ is independent of the patch labels and of
subdivision. It is the inverse of the usual gerbe holonomy, and
\be
\Omega_k(v_0,v_1,v_2,v_3)
  =\mathcal W_k(\partial\mathcal T_{0123})
  =\exp\!\left(-i\int_{\mathcal T_{0123}}H_k\right).
\ee

This form is also invariant under changes in the local trivializations. Gauge equivalent data are connected by:
\bea
B'_\alpha&=B_\alpha+\dd\Lambda_\alpha,\\
\mathcal A'_{\alpha \beta}
  &=\mathcal A_{\alpha \beta}+\Lambda_\alpha-\Lambda_\beta
    -i\chi_{\alpha \beta}^{-1}\dd\chi_{\alpha \beta},\\
\zeta'_{\alpha \beta \gamma}
  &=\zeta_{\alpha \beta \gamma}\chi_{\beta \gamma}^{-1}\chi_{\alpha \gamma}\chi_{\alpha \beta}^{-1}.
\label{eq:app-gerbe-gauge}
\eea
The face, edge and vertex changes cancel in
\eqref{eq:app-gerbe-holonomy}. 
In particular, a careful treatment shows that the vanishing of $H_k$ on an abelian subgroup does not force
the restricted anomaly to also vanish: the transition
data can remain nontrivial. In practice, it is simpler to compute the data directly from the abelianized anomaly, without using the WZW cocycle in this case.

\section{Drinfeld center and distributional quantum double}
\label{app:drinfeld-double}
In this appendix, we review the machinery behind the mathematical description of the twisted quantum double for a continuous group. We will first introduce a distributional presentation for continuous $G$, which we will use to extract local lattice terms. We then comment on the appropriate global prescription.

\subsection{The double and its representations}
An object of the center
\be
X \in \cZ(\Vec_G^\omega) \, ,
\ee
is a $G$-graded object with half-braidings
$c_{x,X}:\bde_x\otimes X\to X\otimes\bde_x$.
They take grade $q$ to $xqx^{-1}$. We use the associator
$\alpha_{a,b,c}=\omega(a,b,c)$ and write $A=D(G)^\omega$ for
the double realizing these conventions:
\be
\cZ(\Vec_G^\omega) \simeq \Rep(A),\qquad
A=D(G)^{\omega}.
\ee
For finite $G$, its basis is $(\bde_g,x)$ with $g,x\in G$, where $\bde_g$
are projectors onto grade $g$. The product is
\be
(\bde_g,x)\star(\bde_h,y)
  =\delta_{h,x^{-1}gx}\,\theta_g(x,y)^{-1}(\bde_g,xy),
\label{eq:double-product}
\ee
with
\be
\theta_g(x,y)
  =\frac{\omega(g,x,y)\,
    \omega\!\left(x,y,{}^{xy} g\right)}
  {\omega\!\left(x,{}^x g,y\right)},
\label{eq:theta-double}
\ee
the transgression. The pentagon equation implies
\be
\theta_g(x,y)\theta_g(xy,z)
  =\theta_g(x,yz)\theta_{{}^x g}(y,z),
\label{eq:double-theta-identity}
\ee
which makes the product associative.

For compact continuous $G$, we replace the Kronecker delta ($\delta_{g,h}$) by a Dirac delta ($\delta(g,h)$), and the sum over group elements by an integral over the Haar measure, so that:
\be
\int_G d \mu(g) \delta(g) = 1 \, .
\ee
To avoid singularities, we can smear the objects in the double by square integrable functions on $G \times G$:
\be
a_F \equiv \int_G d \mu(q) \int_G d \mu(x) \, F_q(x) \, (\bde_q, x) \, .
\ee
The multiplication becomes:
\be
a_{F} \star a_{G} = a_{F * G} \, ,
\ee
with the twisted convolution:
\be
(F*G)_q(z)
  =\int_Gd\mu(x)\,F_q(x)
    G_{x^{-1}qx}(x^{-1}z)\,\theta_q(x,x^{-1}z)^{-1}.
\label{eq:double-convolution}
\ee
We will connect this presentation with the sections of a twisted Fell bundle at the end of this section.
The involution is
\be
(\bde_q,x)^*
  =\theta_q(x,x^{-1})(\bde_{{}^x q},x^{-1}).
\label{eq:double-star}
\ee
When $G$ is anomaly-free, representations of $D(G)$ are described by simple objects $X_{[z],\lambda}$, where $[z]$ is a conjugacy class, and $\lambda \in \text{Rep}(Z_G(z))$ is a representation of the centralizer of $z$. This is completely analogous for discrete and continuous groups. For an anomaly-free group, the representations of the continuous quantum double have been studied in e.g. \cite{Koornwinder:1996uq, Koornwinder:1998xg}.

\subsection{Transgression and the simple objects}
We now describe how a nontrivial anomaly changes the spectrum of simple objects via transgression.

Fix a flux $z$. It is preserved by the left action of
$\widetilde{x}=\int_Gd\mu(p)(\bde_p,x)$ for $x\in Z_G(z)$.
Restricting to the centralizer gives the transgression:
\be
\theta_z(x,y)
  =\frac{\omega(z,x,y)\omega(x,y,z)}{\omega(x,z,y)},
\qquad x,y\in Z_G(z) \, , \quad [\theta_z] \in H^2(Z_G(z), U(1)) \, ,
\label{eq:center-transgression}
\ee
This appears as $\theta_z^{-1}$ in the quantum double multiplication. We label its
ordinary vector-space dual on module junctions
$\Hom_\cM(g\triangleleft X,gz)$ by $\lambda$, with multiplier
$\theta_z$. The simple objects are then labeled by
\be
X_{[z],\lambda},\qquad
\lambda\in\Irr^{\theta_z}(Z_G(z)),\qquad
\lambda(x)\lambda(y)=\theta_z(x,y)\lambda(xy),
\label{eq:center-simple-objects}
\ee
where $[z]$ is a conjugacy class and $\lambda$ an irreducible
projective representation. This construction also applies to
compact continuous groups in the Hilbert-space formulation
~\cite{Jia:2026vcr}.

\vspace{2mm} \noindent \textbf{Local triviality and spectral flow.}
Let us consider the case of a simply connected Lie group in more detail.
For generic flux $z=e^{2\pi i\tau}$, the centralizer is a maximal torus
$T$. 
At fixed $z$, $H^2_{\rm Borel}(T,U(1))=0$, so $\theta_z$ can be
trivialized. The exact expression \cite{Jia:2025uun} for the abelianized cocycle is:
\be
\theta_z(x,y) = \exp \left( \pi i k \langle \tau, n[\xi, \eta] \rangle \right) \, , \qquad  n[\xi, \eta] = \xi + \eta - [\xi + \eta] \, ,
\ee
with $x = e^{2 \pi i \xi} \equiv t_\xi$, $y= e^{2\pi i \eta}$ and $[ \, \cdot \, ]$ denoting restriction to the fixed logarithmic branch.
The trivialization is given by:
\be
\theta_z(x,y)=\frac{\nu_\tau(x)\nu_\tau(y)}{\nu_\tau(xy)},
\qquad
\nu_\tau(t_\xi)=e^{\pi i k\langle\tau,\xi\rangle}.
\label{eq:center-toral-trivialization}
\ee
An ordinary torus character has the form
$\chi_\ell(t_\xi)=e^{2\pi i\langle\ell,\xi\rangle}$, with
$\ell$ in the weight lattice $P$. The phase $\nu_\tau$ can be absorbed in a projective character:
\be
\lambda_\ell(t_\xi)
  =\nu_\tau(t_\xi)\chi_\ell(t_\xi)
  =e^{2\pi i\langle\ell+k\tau/2,\xi\rangle},
\qquad \ell\in P.
\label{eq:center-shifted-weights}
\ee
As $\tau$ crosses a wall of the Weyl alcove it jumps by a coroot $\alpha_i^\vee$, and the weight $\ell$ undergoes a nontrivial spectral flow, which for $G=SU(2)$ reads
\be
\ell \longmapsto \ell + \frac{k}{2} \alpha_i^\vee\, .
\ee
The spectral flow is the nontrivial datum which descends from the anomaly.

\subsection{Quasi-Hopf structure}
In the same convention, the coproduct and counit are
\bea
\Delta(\bde_q,x)
  &=\int_{ab=q}\gamma_x(a,b)^{-1}(\bde_a,x)\otimes(\bde_b,x),\\
\gamma_x(a,b)
  &=\frac{\omega(a,b,x)\omega(x,{}^x a,{}^x b)}
  {\omega(a,x,{}^x b)},\\
\varepsilon(\bde_q,x)&=\delta(q),\qquad
1_A=\int_Gd\mu(q)\,(\bde_q,1).
\label{eq:non-abelian-double-coproduct}
\eea

The quasi-associator is
\be
\Phi=\int_{G^3}d\mu(a)d\mu(b)d\mu(c)\,
\omega(a,b,c)
(\bde_a,1)\otimes(\bde_b,1)\otimes(\bde_c,1),
\label{eq:double-quasi-associator}
\ee
so that
\be
(\unit\otimes\Delta)\Delta(a)
  =\Phi\,(\Delta\otimes\unit)\Delta(a)\,\Phi^{-1}.
\ee
Its action is therefore the same $\omega$ as the graded
associator. 
These equations are algebraic identities for finite groups and
distributional local formulas in the continuous setting. 

\subsection{The Fell-bundle description}
We conclude with remarks on the global structure of the continuum algebraic data in the Drinfeld double. These have a groupoid interpretation, by mapping the pairs $(\bde_q,x)$ to arrows:
\be
(\bde_q,x) \equiv (q,x): {}^x q\longrightarrow q
\ee
in the conjugation groupoid $\mathcal I=G//_{\rm Ad}G$.
Two arrows $(q,x)$ and $(h,y)$ compose precisely when
$h=x^{-1}qx$, and their composite is $(q,xy)$.
Put a complex line bundle $\Sigma_{\omega,(q,x)}$ over each arrow.
In local frames $\sigma(q,x)$, its product reads
\be
\sigma(q,x)\sigma(x^{-1}qx,y)
  =\theta_q(x,y)^{-1}\sigma(q,xy).
\label{eq:fell-frame-product}
\ee
This defines a Fell line bundle over $G//_{\rm Ad}G$. The multiplication in the double is the local multiplication
coefficient of the Fell bundle. Its section algebra has exactly the
convolution~\eqref{eq:double-convolution}, and its universal
$C^*$-completion is
\be
C^*(G//_{\rm Ad}G,\Sigma_\omega).
\ee
The distributional notation $(\bde_q,x)$ is a Dirac-delta section in these
frames.

\section{Anomalies from anyon chain operators}
\label{app: EN}
This section computes the anomalies of the lattice symmetry actions
\eqref{eq:left-symmetry-action} and~\eqref{eq:right-symmetry-action}
by considering the interval restriction of symmetry operators \cite{Else:2014vma}. We write $\omega=\omega_k$ and
use the geometric representative of Appendix~\ref{app:cocycle}.

\subsection{The endpoint obstruction}
Let $I=[m,n]$ be a lattice interval. Define $U_{a,I}(h)$, with $a=L,R,V$, by transforming
its sites and internal Villain coordinates, and retaining the bond
phases for $i=m,\ldots,n-1$.
Their composition takes the form
\be
U_{a,I}(x)U_{a,I}(y)
 =\Gamma_a^l(x,y)\,\Gamma_a^r(x,y)^{-1}U_{a,I}(xy),
\label{eq:EN-interval}
\ee
where $\Gamma_a^l$ and $\Gamma_a^r$ are supported at the two ends.
Since they stand to the left
of $U_{a,I}(xy)$, their argument $s$ is the boundary value
\emph{after} the transformation.

The Else--Nayak obstruction is the phase in the associativity
relation at one endpoint~\cite{Else:2014vma}:
\bea
\Gamma_a^l(x,y)\Gamma_a^l(xy,z)
 &=\omega_a^{\rm EN}(x,y,z)\,
\left( {}_a^x\Gamma_a^l(y,z)\right)\Gamma_a^l(x,yz),\\
{}_a^x O&=U_{a,I}(x)O\,U_{a,I}(x)^{-1}.
\label{eq:EN-obstruction}
\eea
Associativity of four symmetry operators makes $\omega_a^{\rm EN}$
a three-cocycle. Multiplying the endpoints by a phase changes it by a
coboundary. For continuous $G$ its class lies in
$H^3_{\rm Borel}(G,U(1))\simeq H^4(BG,\bZ)$, as in
\eqref{eq:continuous-anomaly-class}. 

\subsection{Left symmetry}
On one bond $(g_i,g_{i+1})$, the pentagon identity gives
\be
\frac{\varphi_L(y;g_i,g_{i+1})
 \varphi_L(x;yg_i,yg_{i+1})}
 {\varphi_L(xy;g_i,g_{i+1})}=
 \frac{\omega(y^{-1},x^{-1},xyg_{i+1})}
 {\omega(y^{-1},x^{-1},xyg_i)} \, ,
\label{eq:EN-left-bond}
\ee
and the product over $I$ telescopes. In terms of the outgoing boundary
site $s=xyg_m$, the left endpoint factor is
\be
\Gamma_L^l(x,y;s)=\omega(y^{-1},x^{-1},s)^{-1}.
\ee
Conjugating a diagonal function gives
${}_L^x f(s)=f(x^{-1}s)$. Substitution into
\eqref{eq:EN-obstruction} yields
\be
\omega_L^{\rm EN}(x,y,z)
 =\frac{\omega(z^{-1},y^{-1},x^{-1}s)
         \omega(z^{-1}y^{-1},x^{-1},s)}
        {\omega(y^{-1},x^{-1},s)
         \omega(z^{-1},y^{-1}x^{-1},s)}
 =\omega(z^{-1},y^{-1},x^{-1}).
\ee
The second equality is the cocycle identity on
$(z^{-1},y^{-1},x^{-1},s)$. In homogeneous notation,
\be
\omega(z^{-1},y^{-1},x^{-1})
 =\Omega(xyz,xy,x,1)
 =\Omega(1,x,xy,xyz)=\omega(x,y,z).
\ee
Reversing the four vertices is the even permutation $(03)(12)$.
Thus the geometric representative gives
\be
\omega_L^{\rm EN}=\omega,\qquad
[\omega_L^{\rm EN}]=[\omega].
\label{eq:EN-left-anomaly}
\ee

\subsection{Right symmetry}
Write a bond as $(g,gq;v)$, where $v=x_{i+1,i}$ is its Villain
coordinate. The half-braiding factors
in $U_R(x)U_R(y)$ contribute
\be
\frac{\theta_{q^{xy}}(x,yv)^{-1}
      \theta_{q^y}(y,v)^{-1}}
     {\theta_{q^{xy}}(xy,v)^{-1}}
 =\theta_{q^{xy}}(x,y)^{-1}.
\label{eq:EN-Villain-composition}
\ee
This follows from~\eqref{eq:double-theta-identity}.

The two module-associator phases satisfy, by the pentagon identity
\be
\frac{\varphi_R(y;g,gq)
 \varphi_R(x;gy^{-1},gqy^{-1})}
 {\varphi_R(xy;g,gq)}
=\theta_{q^{xy}}(x,y)\,
 \frac{\omega(g(xy)^{-1},x,y)}
      {\omega(gq(xy)^{-1},x,y)}.
\label{eq:EN-right-associators}
\ee
The full defect
therefore telescopes to the endpoints, with
\be
\Gamma_R^l(x,y;s)=\omega(s,x,y),\qquad
s=g_m(xy)^{-1}.
\ee
Now ${}_R^x f(s)=f(sx)$, so
\be
\omega_R^{\rm EN}(x,y,z)
 =\frac{\omega(s,x,y)\omega(s,xy,z)}
        {\omega(sx,y,z)\omega(s,x,yz)}
 =\omega(x,y,z)^{-1}.
\label{eq:EN-right-anomaly}
\ee
The two chiralities hence carry the opposite level-$k$ classes
$[\omega]$ and $-[\omega]$.

\subsection{Vector symmetry}
As a final consistency check we verify that the vector symmetry is anomaly-free.
Choose $U_V(h)=U_R(h)U_L(h)$, with the left action applied first.
On one bond its action is
\bea
\cU_V(h)\ket{g,gq;v}
 &=\varphi_V(h;g,q)\,
   \theta_{hqh^{-1}}(h,v)^{-1}\\
 &\quad\times\ket{hgh^{-1},hgqh^{-1};hv},
\eea
where the three module-associator factors are
\be
\varphi_V(h;g,q)
 =\omega(h^{-1},hg,q)
  \frac{\omega(hgh^{-1},h,q)}
       {\omega(hgh^{-1},hqh^{-1},h)}.
\ee
We first reduce their composition, then include the Villain factor.
In homogeneous notation,
\be
\varphi_V(h;g,q)
 =\frac{\Omega(1,h^{-1},g,gq)
        \Omega(h^{-1},gh^{-1},g,gq)}
       {\Omega(h^{-1},gh^{-1},gqh^{-1},gq)}.
\label{eq:EN-vector-tetrahedra}
\ee
Using again the pentagon equation one finds
\be
\frac{\varphi_V(x;ygy^{-1},yqy^{-1})
        \varphi_V(y;g,q)}
       {\varphi_V(xy;g,q)}=
 \frac{\theta_{g^{xy}}(x,y)\,
       \theta_{q^{xy}}(x,y)}
      {\theta_{(gq)^{xy}}(x,y)}.
\label{eq:EN-vector-associators}
\ee
The Villain contribution is again
\eqref{eq:EN-Villain-composition}, combining it with the above leaves the full one-bond defect
\be
\frac{\theta_{g^{xy}}(x,y)}
     {\theta_{(gq)^{xy}}(x,y)}.
\ee
Multiplying along $I$ gives the endpoint operator
\be
\Gamma_V^l(x,y;s)=\theta_s(x,y),\qquad
s= g_m{}^{xy}.
\label{eq:EN-vector-endpoint}
\ee
Since ${}_V^x f(s)=f({}^x s)$, its obstruction is
\be
\omega_V^{\rm EN}(x,y,z)
 =\frac{\theta_s(x,y)\theta_s(xy,z)}
        {\theta_{{}^x s}(y,z)\theta_s(x,yz)}=1.
\label{eq:EN-vector-anomaly}
\ee
Thus the diagonal copy of $G$ is anomaly-free. 

\section{Non-abelian interaction terms}
\label{app:onetwoleg}

In this appendix we give the full derivation of the interaction terms used in the main text. Section \ref{app: seconeleg} describes the detailed derivation of the one-leg terms, Section \ref{app:two-leg-evaluation} derives the finite two-leg terms and the relevant identities to check its symmetry. Section \ref{app:lattice-B} explicitly computes the lattice B-field for our choice of representative cocycle, and computes its continuum limit.

\subsection{One-leg term} \label{app: seconeleg}

We start from the identification $\rho={}_AA$ and
\be
\End_A(\rho)=A^{\rm op} \, ,  \qquad A = D(G)^\omega \, .
\ee 
The oriented one-leg operator
is adjoint right multiplication. We evaluate its action on wavefunctions, then linearize to obtain the covariant derivative
in the Villain potential. Right multiplication in the double is
\be
(\bde_q,x)\star(\bde_p,z)
 =\delta(p,{}^x q)\theta_q(x,z)^{-1}(\bde_q,xz).
\ee
With $R_b(a)=a\star b$ and $\widetilde z=\int_Gd\mu(p)(\bde_p,z)$,
we choose $\widehat z=R_{\widetilde z}^\dagger$. It preserves $q$ and
acts on the Villain wavefunction by
\be \label{eq: apppartrap}
(\widehat z\psi_q)(x)=\theta_q(x,z)\psi_q(xz).
\ee
The groupoid identity~\eqref{eq:double-theta-identity} gives the composition law
\be
(\widehat z_1\widehat z_2\psi_q)(x)
 =\theta_{{}^x q}(z_1,z_2)
   (\widehat{z_1z_2}\psi_q)(x).
\label{eq:one-leg-composition}
\ee
The idea behind the construction is to interpret \eqref{eq: apppartrap} as a finite parallel transport on the Villain fiber. After choosing a basis of generators we can compute its covariant derivative and curvature, and finally find a ``nice'' trivializing connection leading to a simple Hamiltonian.

\vspace{2mm}\noindent\textbf{Curvature.}
We use the left- and right-invariant frames
\be
\label{frames}
(V_a\psi)(x)=\frac{d}{dt}\psi(xe^{itT^a})\Big|_{t=0}\,,\qquad
(E_a\psi)(x)=\frac{d}{dt}\psi(e^{itT^a}x)\Big|_{t=0}\,,\qquad
[V_a,V_b]=-f^{abc}V_c\,,
\ee
and define the (left) generators $\nabla_a$ of the one-leg transport by 
\be
(\widehat z\psi_q)(x)=\psi_q(x)+\epsilon^a(\nabla_a\psi_q)(x)+O(\epsilon^2) \, ,
\ee
for $z=e^{i\epsilon^aT^a}$. Their commutator takes the form
\be
[\nabla_a,\nabla_b]=-f^{abc}\nabla_c-iF_{q,ab}(x)\, \, .
\ee
The field strength $F_{q, ab}(x)$ can be extracted from the transgression by Taylor expansion:
\be
\log \left[ \frac{\theta_{{}^x q}(e^{t Z_1}, e^{s Z_2})}{\theta_{{}^x q}(e^{s Z_2}, e^{t Z_1})} \right] = -i st F_q(x)[Z_1, Z_2] + ... \, ,
\ee
notice that $F_q(x) \equiv F_{{}^x q}$ depends on the two Villain labels only via the combination ${}^x q$. Choosing a local trivialization $F = d A$ amounts to trivializing $\theta$ locally:
\be
\theta_q(x,y) = \frac{\nu_q(x) \nu_{{}^x q}(y)}{\nu_q(xy)} \, ,
\ee
and changing basis to:
\be
\widetilde{\ket{\Psi}} = \prod_{\ell} \nu_{q_\ell}(x_\ell) \ket{\Psi} \, ,
\ee
leads to a connection (in the left-basis):
\be
\nabla_a = V_a - i A_{{}^x q,a} \, ,  \qquad \nu_q(z) = 1 - i \epsilon^a A_{q, a} \, .
\ee

\begin{figure}[htbp]
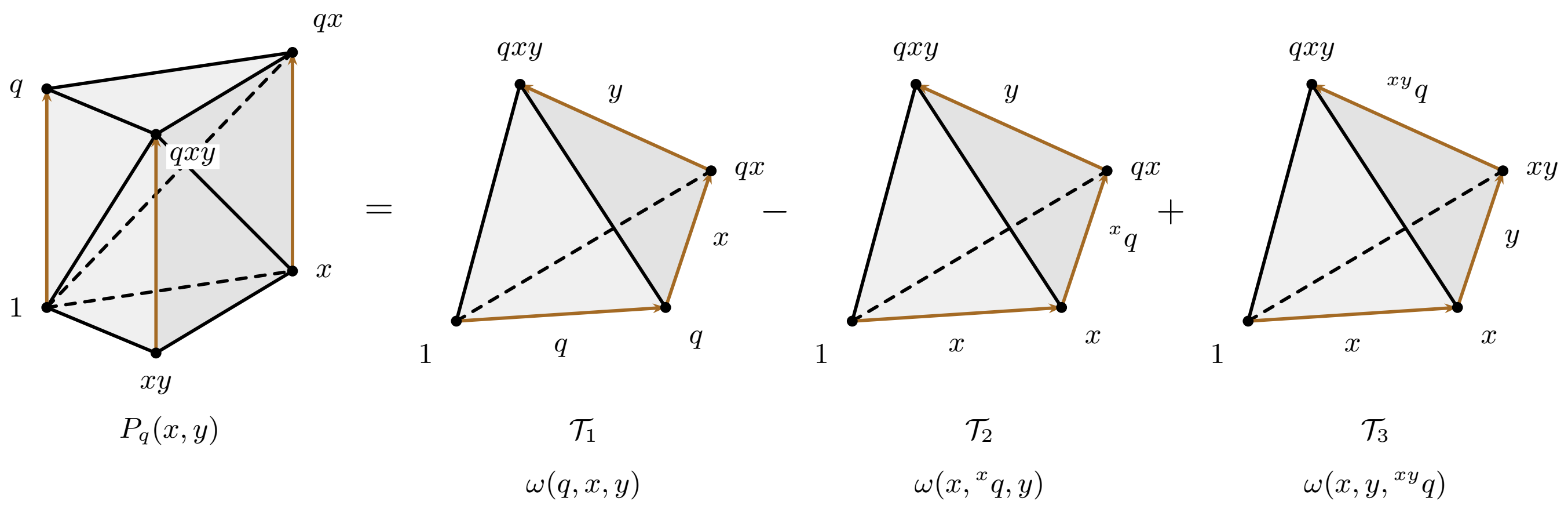

  \centering
  \TransgressionPrismDiagram
  \caption{The prism for the transgression $\theta_q(x,y)$, with bottom
  triangle $(1,x,xy)$ and its left translate $(q,qx,qxy)$.
  }
  \label{fig: prism}
\end{figure}

\begin{figure}[htbp]
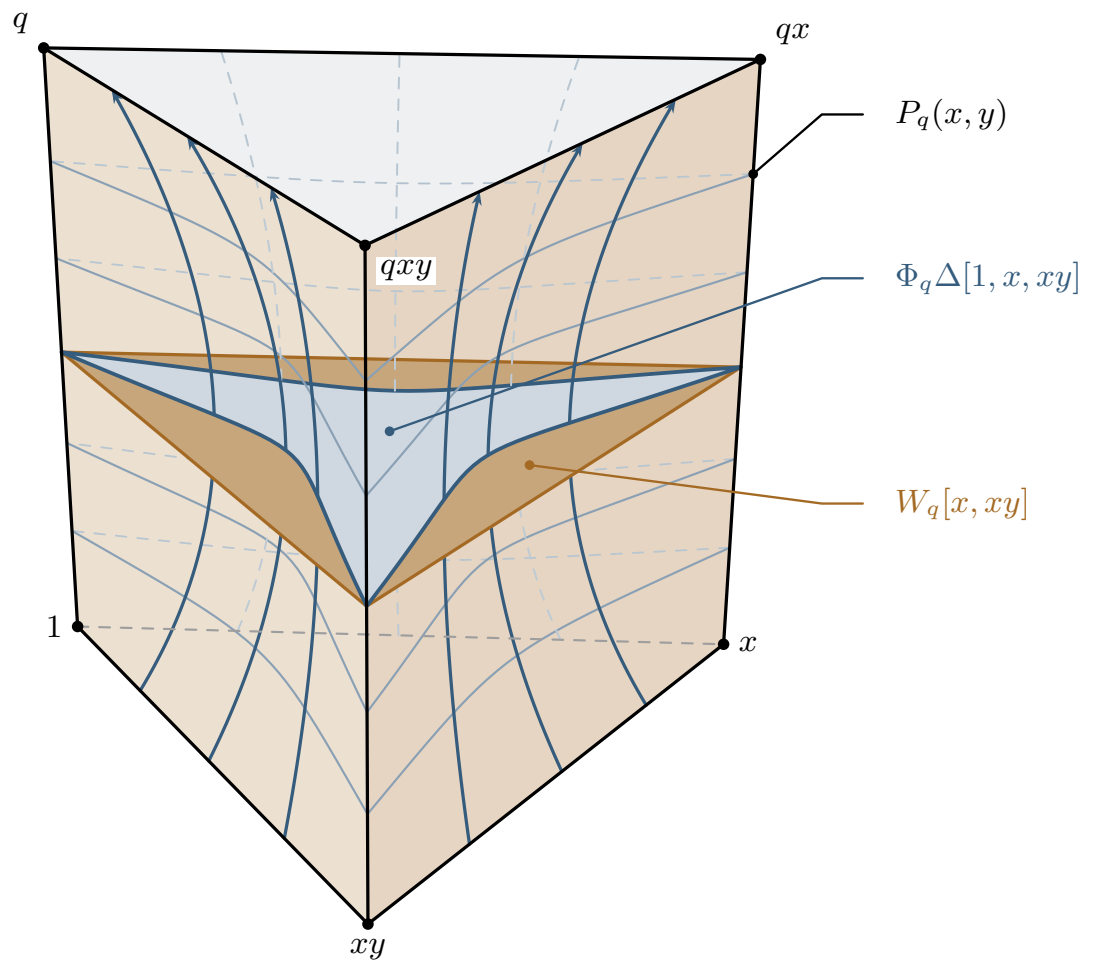

  \centering
  \TransgressionPrismSweepDiagram
  \caption{The blue sweep lies inside $P_q(x,y)$. Orange regions are the
  $W_q$ corrections, while the middle section is the parallel-transported $\Delta[1,x,xy]$.}
  \label{fig: prism2}
\end{figure}

\vspace{2mm} \noindent \textbf{Curvature from prism flux.} An instructive derivation of the form of the field strength comes from considering the transgression written using our representative WZW cocycle:
\be
\label{theta-prism}
\theta_q(x,y)=\frac{\Omega_k(1,q,qx,qxy)\,\Omega_k(1,x,xy,qxy)}{\Omega_k(1,x,qx,qxy)}
=\exp\left(-i\int_{P_q(x,y)}H_k\right)\, .
\ee
where $P_q(x,y)$ is the prism in Figure \ref{fig: prism}. 
We decompose this into two parts:
\be
P_q(x,y) = \Phi_q \Delta[1,x,xy] +  W_q[1,x] + W_q[x,xy] - W_q[1,xy] \, ,
\ee
where the first term is swept by parallel transporting $\Delta[1,x,xy]$ on the left by $q_s = e^{2\pi i s \tau}$, while $W_q[a,b]$ is the region between the swept and the triangulated wall over the edge $\gamma[a,b]$.
The transport is generated by the map $\Phi_q(s) = e^{2\pi i s \tau} \equiv q_s$. We thus have:
\be
\exp\left( -i \int_{P_q(x,y)} H_k \right) = \exp\left( -i \int_0^1 ds \int_{\Delta[1,x,xy]} \Phi_q^* H_k(\partial_s) \right) \times \left[\frac{\kappa_q(x) \kappa_{{}^x q}(y)}{\kappa_q(xy)} \right]\, ,
\ee
where:
\be
\kappa_q(x) = \exp \left( -i \int_{W_q[1,x]} H_k \right) \, .
\ee
The setup is shown in Figure \ref{fig: prism2}.

The first term can be computed explicitly: the pullback of the right-invariant Maurer--Cartan form along $\Phi_q$ is very simple:
\be
\Phi^*_q \Theta = 2 \pi \tau ds + \Ad_{q_s}\Theta \, .
\ee
Using invariance under conjugation of the inner product, the pullback of the WZW term reads:
\be
\Phi_q^* H_k = - i \frac{k}{2} ds \wedge \langle \tau \, , \Theta \wedge \Theta \rangle + ... \, ,
\ee
where the $...$ indicate terms with no $ds$ which give trivial integrals. The integral over $ds$ becomes trivial to perform, so we have:
\be
\theta_q(x,y) = \exp\left( - \frac{k}{2} \int_{\Delta[1,x,xy]} \langle \tau, \Theta \wedge \Theta \rangle  \right) \times \left[ \frac{\kappa_q(x) \kappa_{{}^x q}(y)}{\kappa_q(xy)} \right]\, .
\ee
The small $(x,y)$ expansion shows that $\kappa_q(x) = 1 + O(|\log(x)|^2)$, and we arrive at:
\be
F_{q} = - i \frac{k}{2}  \langle \tau, \, \Theta \wedge \Theta \rangle \, ,
\ee
together with:
\be \label{eq: thetaDelta}
\theta_q(x,y) = \exp\left( - i \int_{\Delta[1,x,xy]} F_q \right) \times \left[\frac{\kappa_q(x) \kappa_{{}^x q}(y)}{\kappa_q(xy)} \right] \, .
\ee

\vspace{2mm} \noindent \textbf{Connection in the right-invariant frame.} Due to the appearance of the right-invariant Maurer-Cartan form, the expression of the covariant derivative is simpler in the right-invariant frame:
\be
 D_a = (E_a - i \Gamma_{q,a}) \, .
\ee
In this basis, we can choose the unique (due to Whitehead's lemma) right-invariant connection with curvature $F_q$:
\be
\Gamma_q = - \frac{k}{2} \langle \tau, \, \Theta \rangle \, .
\ee
We finally remember that the covariant derivative must be computed at ${}^x q$ instead of $q$. In the right-invariant frame this acts by conjugation also on $\Theta$ and leads to the expansion:
\be
\widehat{z} \widetilde{\psi_q}(x) = \widetilde{\psi_q}(x) + \epsilon^a (\Ad_x D)_a \widetilde{\psi_q}(x) + ... \, .
\ee
We also have the finite formula:
\be \label{nu-explicit}
\nu_q(x) = \exp\left(- i \int_{\gamma[1,x]} \Gamma_q \right) \, \kappa_q(x) \, .
\ee
\vspace{2mm} \noindent \textbf{Existence of local trivialization.} We conclude with some comments on the existence of a local trivialization of the transgression data. Using the invariance of the WZW representative under conjugation, it can be shown that the transgression $\gamma_x(a,b)$ is a two-cocycle in its arguments (with $x$ kept fixed), thus there exists a local trivialization:
\be \label{eq:app-nu-gamma}
\gamma_x(a,b) = \frac{\nu_{ab}(x)}{\nu_a(x) \nu_b(x)} \, ,
\ee
this trivialization is unique as a semisimple $G$ has no nontrivial local characters. The coproduct compatibility $\Delta(a \star b) = \Delta(a) \star \Delta(b)$ furthermore implies that:
\be
\label{double-compatibility}
\frac{\theta_{uv}(x,y)}{\theta_u(x,y)\,\theta_v(x,y)}=\frac{\gamma_x(u,v)\,\gamma_y({}^xu,{}^xv)}{\gamma_{xy}(u,v)}\, ,
\ee
from which it follows that $\chi_q(x,y)=\theta_q(x,y)\,\nu_q(xy)/(\nu_q(x)\nu_{{}^xq}(y))$ satisfies $\chi_{uv}=\chi_u\chi_v$. It is a local character of $G$ in $q$, hence $\chi_q=\chi_e=1$, and
\be
\label{eq:app-local-nu}
\theta_q(x,y)=\frac{\nu_q(x)\,\nu_{{}^xq}(y)}{\nu_q(xy)}
\ee
holds for all $x,y$.

\subsection{Two-leg term}
\label{app:two-leg-evaluation}

We compute the matrix elements of $K_{X; (f,g)}$ for
$X=X_{[z],\mathrm{Reg}}$ and arbitrary junction vectors $f,g$.
We first construct the $\cI_f,\cJ_g$ junctions explicitly, then apply the four maps in
Figure~\ref{fig: mod-ass} to the input anyon chain state. 

\vspace{2mm}\noindent\textbf{Junction vectors.}
Since $\rho={}_AA$ is the left regular module, an $A$-linear map $T$
from $\rho$ is determined by its value at the unit. For any left
$A$-module $M$,
\be
\Hom_A(A,M)\simeq\underline M,\qquad
T\longmapsto T(1_A) \, ,
\label{eq:app-regular-Hom}
\ee
via  $T(a)=a\cdot T(1_A)$. $\underline{M}$ denotes the
underlying vector space. In particular
\bea
\Hom_A(A,A\otimes X)&\simeq\underline A\otimes \underline X \, ,\\
\Hom_A(A,X\otimes A)&\simeq \underline X\otimes\underline A.
\eea
We choose the vectors $1_A\otimes|f\rangle_X$ and
$|g\rangle_X\otimes1_A$. Thus we fix the $\underline A$ entry
to the unit:
\be
1_A = \int_G d\mu(q)  (\bde_q, 1) \, .
\ee
Changing this datum amounts to composition by a one-leg interaction while $f,g$ are the remaining free data. We can define inner products in these vector spaces:
\bea
\langle g | f \rangle_X  &= \int_G d \mu(r) \overline{g}(r) f(r) \, ,  \\
\langle G , F \rangle_A  &= \int_G d\mu(q) \int_G d\mu(x) \overline{G}_q(x) F_q(x) \, ,
\eea
where we have expanded $F = \int d\mu(q) \int d \mu(x) F_q(x)  \, (\bde_q, x)$. These can be used to define the adjoints of the junction maps.
Their $A$-linear extensions are
\bea
\cI_f(a)&=\Delta(a)(1_A\otimes|f\rangle_X),\\
\cJ_g(a)&=\Delta(a)(|g\rangle_X\otimes1_A).
\label{eq:app-junction-extension}
\eea
The coproduct acts by left multiplication on the $A$ factor and
by the appropriate $A$ representation $\pi_X$ on $V_X$. For $X=X_{[z],\mathrm{Reg}}$, the left module is $A(\bde_z,1)$.
Right multiplication by $(\bde_z,1)$ gives the following elements in the double
\be
(\bde_q,x)\star(\bde_z,1)=\delta(q,z^{x})(\bde_q,x),
\ee
so its image has basis
\be
|r\rangle_X \equiv (\bde_{z^r},r) \, ,
\ee
with grading $z^r$ which is left implicit. The grade-$z$ subspace has
$r\in Z_G(z)$ and carries the twisted regular action of the
centralizer; this identifies the object with
$X_{[z],\mathrm{Reg}}$. The left action of $D(G)^\omega$ on this vector space is:
\be
\pi_X(\bde_q,x)|r\rangle_X
 =\delta(q,z^{xr})\theta_q(x,r)^{-1}|xr\rangle_X \, .
\label{eq:app-auxiliary-action}
\ee
We expand a vector $\ket{f}_X \in L^2(G)$ in this basis as
\be
|f\rangle_X=\int_Gd\mu(r)f(r)|r\rangle_X \, ,
\ee
and similarly
\be
\ket{g}_X = \int_G d \mu(r') g(r') \ket{r'}_X \, .
\ee

\vspace{2mm}\noindent\textbf{Computing the two-leg term.}
Let $q_-=g_{i-1}^{-1}g_i$ and
$q_+=g_i^{-1}g_{i+1}$, similar shorthands are used for the other link variables. The input anyon chain state in
Figure~\ref{fig: mod-ass} corresponds to the vector:
\be
|\Psi\rangle =
\bigl(|g_{i-1}\rangle\otimes(\bde_{q_-},x_-)\bigr)
 \otimes(\bde_{q_+},x_+),
\label{eq:app-two-leg-input-vector}
\ee
in the double notation, where $|g_{i-1}\rangle$ is the unit vector of the one-dimensional
object of grade $g_{i-1}$. The right module action is
\be
g_{i-1}\triangleleft U=g_{i-1}\otimes F(U),\qquad
\alpha^\cM_{g_{i-1};U,V}=\alpha_{g_{i-1},F(U),F(V)}.
\label{eq:app-module-associator}
\ee
Thus on grades $p,q$, the module associator multiplies by
$\omega(g_{i-1},p,q)$.
We now perform the explicit steps (1) (2) (3) (4) of Figure~\ref{fig: mod-ass} and choose a local basis of the Hilbert space to simplify the final expression. Since there will be many labels we provide a table for the reader:

\begin{center}
\small
\begin{minipage}[c]{0.40\linewidth}
\centering
\TwoLegCoordinateGuide
\end{minipage}\hspace{0.05\linewidth}
\begin{minipage}[c]{0.53\linewidth}
\renewcommand{\arraystretch}{1.4}
\setlength{\arrayrulewidth}{0.25pt}
\arrayrulecolor{black!25}
\begin{tabular}{@{}cc@{}}
\hline
Label & Definition \\[3pt]
\hline\noalign{\vskip3pt}
$q_\pm$ & $q_-=g_{i-1}^{-1}g_i,\quad q_+=g_i^{-1}g_{i+1}$. \\[3pt]
\hline\noalign{\vskip3pt}
$r$ & $\displaystyle |f\rangle_X=\int_Gd\mu(r)\,f(r)|r\rangle_X$. \\[3pt]
\hline\noalign{\vskip3pt}
$s$ & $\deg|x_- r\rangle_X=z^{x_- r}=s^{-1},\quad g_i'=g_i s$. \\[3pt]
\hline\noalign{\vskip3pt}
$w$ & $w=x_+^{-1}x_-$. \\[3pt]
\hline\noalign{\vskip3pt}
$r'$ & $\begin{gathered}
  \displaystyle |g\rangle_X=\int_Gd\mu(r')\,g(r')|r'\rangle_X,\\
  r'=wr.
  \end{gathered}$ \\[3pt]
\hline
\end{tabular}
\arrayrulecolor{black}
\end{minipage}
\end{center}
In the final formulas, $r$ is the only integrated variable, while the auxiliary $s, w$ are functions of $r$ and the input data of the anyon chain.
On the position eigenstates the two-leg term implements a map:
\be
g_i \mapsto g_i s  \, , \qquad q_- \mapsto q_- s \, , \qquad q_+ \mapsto s^{-1} q_+ \, , \qquad x_\pm \mapsto x_{\pm} \, ,
\ee
and the moves (1) (2) (3) (4) compute the coefficient of said map.

\vspace{2mm} \noindent \emph{(1) Emission.}
\[
\TwoLegLocalMove{1}
\]
The initial map $\cI_f$ is applied to the leftmost junction in the double $(\bde_{q_-}, x_-)$. The coproduct gives:
\be
\Delta(\bde_{q_-}, x_- ) = \int_{uv = q_-} \gamma_{x_-}(u,v)^{-1} (\bde_u , x_-) \otimes (\bde_v, x_-) \, ,
\ee
and is applied to:
\be
1_A \otimes \ket{f}_X = 1_A \otimes \int_G d\mu(r) f(r) \ket{r}_X \, .
\ee
The action on $\ket{r}_X$ gives 
\be
(\bde_v, x_-) \ket{r}_X = \delta(v, z^{x_- r}) \, \theta_{v}(x_-, r)^{-1} \ket{x_- r}_X  \, , 
\ee
and we introduce $s^{-1} = z^{x_- r}$.
 Solving the constraint $uv = q_-$ in the coproduct integral gives $u = q_- s$ and 
\be
\cI_f(\bde_{q_-},x_-)
 =\int_Gd\mu(r)\,
 \frac{f(r)}{\gamma_{x_-}(q_-s,s^{-1})\theta_{s^{-1}}(x_-,r)}
 (\bde_{q_-s},x_-)\otimes| x_- r\rangle_X.
\label{eq:app-two-leg-emission}
\ee
The degree of the propagating object in $\underline{X}$ is $z^{x_- r} = s^{-1}$.
The intermediate object is now
$\bigl(g_{i-1}\triangleleft(\rho\otimes X)\bigr)\triangleleft\rho$.

\vspace{2mm} \noindent \emph{(2) The first module associator.}
\[
\TwoLegLocalMove{2}
\]
We apply $(\alpha^\cM_{g_{i-1};\rho,X})^{-1}\triangleleft1_\rho$.
This separates the emitted $X$ from the left $\rho$ and contributes
\be
\omega(g_{i-1},q_-s,s^{-1})^{-1} \, ,
\label{eq:app-two-leg-left-move}
\ee
where the last two entries are the grades of \eqref{eq:app-two-leg-emission}.
The component of $g_{i-1}\triangleleft\rho$ now has grade
$g_i'=g_{i-1}(q_-s)=g_i s$ and the middle site gets right-translated. The configuration is $\bigl((g_{i-1}\triangleleft \rho )\triangleleft X\bigr)\triangleleft\rho$.

\vspace{2mm} \noindent \emph{(3) The second module associator.}
\[
\TwoLegLocalMove{3}
\]
We now apply $\alpha^\cM_{g_{i-1}\triangleleft\rho;X,\rho}$ to group
$X$ with the right $\rho$. Its first argument is the component
of grade $g_i s$, so this map contributes
\be
\omega(g_i s,s^{-1},q_+) \, ,
\label{eq:app-two-leg-right-move}
\ee
where the last two arguments are the grades of $\ket{x_- r}_X \otimes (\bde_{q_+}, x_+) $, which are the outgoing elements in the operation.
The resulting $X\otimes\rho$ factor has grade $s^{-1}q_+$ and we are in the configuration $(g_{i-1}\triangleleft \rho )\triangleleft (X \otimes \rho)$.

\vspace{2mm} \noindent \emph{(4) Absorption.}
\[
\TwoLegLocalMove{4}
\]
Taking the adjoint of the junction formula gives:
\be
\cJ^\dagger_g( \ket{r}_X \otimes (\bde_q, x)) = \overline{g(x^{-1}r)} \gamma_x(z^r, q) \theta_{z^r}(x,x^{-1}r) (\bde_{z^r q}, x)  \, .
\ee
It can be verified for example that $\cJ^\dagger_g \cJ_g = 1_A$ for unit-normalized $\ket{g}_X$.
This map must be applied to the incoming vector $\ket{x_- r} \otimes (\bde_{q_+}, x_+)$ giving
\be
\cJ_g^\dagger\bigl(|x_- r\rangle_X\otimes(\bde_{q_+},x_+)\bigr)
 =\overline{g( x_+^{-1} x_- r)}\,\gamma_{x_+}(s^{-1},q_+)
   \theta_{s^{-1}}(x_+,r')\,(\bde_{s^{-1}q_+},x_+) \, .
\label{eq:app-two-leg-junction-J}
\ee
We introduce the shorthand $w = x_+^{-1} x_-$. When grade $s$ is exchanged through $X$, the outgoing graded components in $\rho \otimes \rho$ after the action of the two-leg term are:
\be
(\bde_{q_- s}, x_-) \otimes (\bde_{s^{-1} q_+}, x_+) \, .
\ee

\vspace{2mm} \noindent \emph{The full two-leg operator.} Multiplying the four steps (1) (2) (3) (4) and computing the scalar product of vectors in $\underline X$ we are left with an integral over $r$, which spans the conjugacy class of $[z]$, dressing the following operator:
\bea
K_{[z];(f,g)}|\Psi\rangle
 &=\int_Gd\mu(r)\,f(r)\overline{g(wr)}\,
   T_s L_s Z_s|\Psi\rangle,\\
L_s
 &=\frac{\gamma_{x_+}(s^{-1},q_+)}
         {\gamma_{x_-}(q_-s,s^{-1})}
   \frac{\theta_{s^{-1}}(x_+,wr)}
        {\theta_{s^{-1}}(x_-,r)} \, , \qquad  s \equiv s(r) =  (z^{x_- r})^{-1} \, , \qquad w = x_+^{-1} x_- \, .
\label{eq:app-two-leg-raw-kernel}
\eea
Here $T_s$ translates $g_i$ to $g_i s$, and the two module associators give
\be
Z_s
 =\frac{\omega(g_i s,s^{-1},q_+)}
         {\omega(g_{i-1},q_-s,s^{-1})}
 =\omega(g_{i-1},q_-,s)\omega(g_i s,s^{-1},q_+).
\label{eq:app-two-leg-Z}
\ee
The second line uses the orientation identity for our representative cocycle. 

\vspace{2mm}\noindent\textbf{Change of basis.}
We now simplify the expression by locally trivializing the transgression $\theta_q$ with the function $\nu$ of \eqref{nu-explicit}. Using~\eqref{eq:app-local-nu} and \eqref{eq:app-nu-gamma}, this implies
\be
L_s=
\frac{\nu_{q_-s}(x_-)\nu_{s^{-1}q_+}(x_+)}
     {\nu_{q_-}(x_-)\nu_{q_+}(x_+)}
\frac{\nu_{z^{w r}}(w r)}{\nu_{z^r}(r)} \, ,
\label{eq:app-two-leg-phase-cancellation}
\ee
as the intermediate factors $\nu_{s^{-1}}(x_\pm)$
cancel between $\gamma$ and $\theta$. We now work in the auxiliary (tilde) basis
\bea
&\widetilde{|\Psi\rangle}
 &&=\nu_{q_-}(x_-)\nu_{q_+}(x_+)|\Psi\rangle,\\
&\widetilde{|r\rangle_X}
 &&=\nu_{z^r}(r)|r\rangle_X \, .
\eea
Recall that the initial state is graded by elements $(\bde_{q_-}, x_-) \otimes (\bde_{q_+}, x_+)$, while the final state by $(\bde_{q_- s}, x_-) \otimes (\bde_{s^{-1} q_+}, x_+)$. These give rise to a factor $\left(\frac{\nu_{q_-s}(x_-)\nu_{s^{-1}q_+}(x_+)}
     {\nu_{q_-}(x_-)\nu_{q_+}(x_+)}\right)^{-1}$, which cancels four of the $\nu$s. Changing the junction vector basis to the tilded basis $\widetilde{\ket{r}}_X$ cancels the remaining two. Substitution in~\eqref{eq:app-two-leg-raw-kernel} thus cancels the whole $L_s$ and gives
\be
K_{[z];(f,g)}\widetilde{|\Psi\rangle}
 =\int_Gd\mu(r)\,f(r)\overline{g(wr)}\,
   T_{s(r)}Z_{s(r)}\widetilde{|\Psi\rangle},
\label{eq:app-two-leg-kernel}
\ee
where we now use $f,g$ to denote the coefficients of $\widetilde{\ket{f}}_X , \, \widetilde{\ket{g}}_X$ in the tilded frame, with a slight abuse of notation.
For $f=g=1$, changing variables to $u=x_-r$ gives the class average
with $s=(z^u)^{-1}$.

\vspace{2mm} \noindent \textbf{Symmetry of the two-leg term.} The form of the junction operators $\cI_f$ and $\cJ_g$ allows us to check that the two-leg terms commute with the right symmetry action (the left one can be checked by using the pentagon equation). The key property, which we derive here, is:
\be
\cI_f \, c_{h,\rho} = c_{h, \rho \otimes X} \, \cI_f \, , \quad \forall h \in G , \, f \in L^2(G)
\ee
where $c$ is the half braiding on $\cZ(\Vec_G^\omega)$, and, in the double notation, is implemented by left multiplication by the integrated element:
\be
\widetilde{h} = \int d\mu(p) (\bde_p, h) \, .
\ee
First, we must compute $c_{h, \rho \otimes X}$, which is given by:
\be
c_{h, \rho \otimes X} = \alpha^{-1}_{\rho, X, h}(1_\rho \otimes c_{h, X}) \alpha_{\rho, h, X} (c_{h, \rho} \otimes 1_X) \alpha^{-1}_{h,\rho,X} \, ,
\ee
where the associators are used to change the order of the brackets $h \otimes (\rho \otimes X) \to (h \otimes \rho) \otimes X$,  $(\rho \otimes h ) \otimes X \to \rho \otimes (h \otimes X)$,  $\rho \otimes (X \otimes h) \to (\rho \otimes X) \otimes h$. Decomposing $X$ and $\rho$ in graded components $q$, $p$ respectively the associators evaluate to:
\be
\frac{\omega(p^h,  h ,q)}{\omega(h,p,q) \omega(p^h , q^h, h)} = \gamma_h(p^h, q^h)^{-1} \, .
\ee
This can be conveniently represented as an algebra element:
\be
c_{h, \rho \otimes X} = \Delta(\widetilde{h}) = \int_{G \times G} d \mu(u) d \mu(v) \gamma_h(u,v)^{-1} \left( (\bde_u, h)_\rho  \otimes (\bde_v, h)_X \right) \, .
\ee
We now have:
\be
c_{h, \rho \otimes X} \, \cI_f(a) =  \Delta(\widetilde{h})  \Delta(a) ( 1 \otimes \ket{f}_X ) = \Delta( \widetilde{h} \star a)  ( 1 \otimes \ket{f}_X ) = \cI_f( \widetilde{h} \star a) = \cI_f \, c_{h, \rho}(a) \, ,
\ee
where we have recognized the half braiding $\widetilde{h} \star a$. By taking adjoints (and redefining $h$) we can similarly prove:
\be
\cJ_g^\dagger \, c_{h, (X \otimes \rho) } = c_{h , \rho} \, \cJ_g^\dagger \, ,
\ee
which can be used to complete the proof in the main text.

\subsection{Transition functions and gerbe data} \label{app: chartwise}
After having derived the expression for the two-leg term in a local patch, let us discuss how to transport the operators and wavefunctions over patches, and explain how this connects to the gerbe structure.
We consider patches $U_\alpha$, $U_\beta$, $U_\gamma$ and let $U_{\alpha \beta}$, $U_{\alpha \beta \gamma}$ denote double and triple overlaps, respectively. 
We will use a different expression for the module associator, namely:
\be \label{eq: modulegerbe}
\psi^\alpha (g; p, q) = \exp\left(-i \int_{\Delta[g,gp,gpq]} B_k^\alpha \right) \, ,
\ee
where we use $g \equiv g_{i-1}$ and $p \equiv q_-, \,q \equiv q_+$ to lighten the notation. It is simple to check that in the radial gauge this reduces to $\omega_k$, and is related to this choice by a gauge transformation.
This form will make the gerby data most transparent.
On a double overlap $U_{\alpha \beta}$ the 3-cocycle changes by a coboundary
\be
\omega_k^\beta  = \omega_k^\alpha \delta c^{\alpha \beta} \, .
\ee
while
\be
\psi^\beta(g;p,q) = \psi^\alpha(g;p,q) \, c^{\alpha \beta}(p,q) \frac{\mu^{\alpha\beta}(g;pq)}{\mu^{\alpha \beta}(gp;q) \mu^{\alpha \beta}(g;p)} \, ,
\ee
where the factor $c^{\alpha \beta}(p,q)$ encodes a change of basis in the junction space $\Hom(\rho \otimes \rho, \rho)$ (and thus is related to the change in representative of $\omega$), and $\mu^{\alpha\beta}$ follows from the change of basis in the module junctions' vector space, instead. 
Using \eqref{eq: modulegerbe} and \eqref{eq:app-gerbe-descent} we identify:
\be
\mu^{\alpha \beta}(g;p) = \exp\left(-i \int_{\gamma[g,gp]} \cA^{\alpha\beta}_k \right) \, .
\ee
The basis elements of the double transform as $(\bde_q,x)^\beta = b^{\alpha\beta}(q,x)^{-1} (\bde_q,x)^\alpha$, where we have defined the auxiliary function:
\be
b^{\alpha\beta}(q,x) = \frac{c^{\alpha\beta}(x, {}^x q)}{c^{\alpha\beta}(q,x)} \, .
\ee
Thus their product keeps its form with $\theta^\beta_q$. This leads to the following transformation law for basis vectors $|\Psi\rangle$:
\be \label{eq: patch-basis}
|\Psi\rangle_\beta  = \prod_{\ell =[i+1, i]} b^{\alpha\beta}(q_\ell,x_\ell)^{-1}  \cU^{\alpha \beta} |\Psi \rangle_\alpha = \prod_{\ell} b^{\alpha\beta}(q_\ell,x_\ell)^{-1}  \exp\left( i \int_{\gamma[g_0,..., g_N]} \cA_k^{\alpha\beta} \right) |\Psi\rangle_\alpha \, .
\ee
Due to our choice of representative, $c^{\alpha \beta}$ can be chosen to be conjugation invariant. 
One further has:
\bea
\theta_q^{\beta}(x,y) &= \theta^{\alpha}_q (x,y) \frac{b^{\alpha\beta}(q,x) b^{\alpha\beta}({}^x q,y)}{b^{\alpha\beta}(q,xy)} \, , \\
\gamma_x^\beta(u,v) &= \gamma_x^\alpha(u,v) \frac{b^{\alpha\beta}(uv,x)}{b^{\alpha\beta}(u,x) b^{\alpha\beta}(v,x)} \, ,
\eea
and
\be
\nu^\beta_q(x) = \nu^\alpha_q(x) \, b^{\alpha\beta}(q,x) \, .
\ee
Thus the tilded basis elements $\nu_q(x) (\bde_q,x)$ are the same in the two patches. 
This leads to the following transformation law:
\be
\widetilde{|\Psi\rangle}_\beta =   \cU^{\alpha \beta} \widetilde{|\Psi\rangle}_\alpha \, ,
\ee
and to the fact that the auxiliary vectors $\widetilde{\ket{r}}_X$ transform as:
\bea
\widetilde{\ket{r}}_X^{\beta, +} &= c^{\alpha \beta}(s^{-1}, q)^{-1} \  \,\widetilde{\ket{r}}_X^{\alpha, +} \, , \\
\widetilde{\ket{r}}_X^{\beta, -} &= c^{\alpha \beta}(p s, s^{-1})^{-1} \widetilde{\ket{r}}_X^{\alpha, -} \, ,
\eea
where $\pm$ refer to the right and left leg, respectively. Let us now consider a triple overlap $U_{\alpha\beta\gamma}$ and the composition of changes of patch $\alpha \to \beta \circ \beta \to \gamma$ and $\alpha \to \gamma$. Denoting $M^{\alpha\beta\gamma} = \frac{c^{\alpha\beta} c^{\beta \gamma}}{ c^{\alpha \gamma}}$ and $N^{\alpha \beta \gamma} = \frac{\mu^{\alpha \beta} \mu^{\beta \gamma}}{\mu^{\alpha \gamma}}$ we find, from the module map consistency:
\be
M^{\alpha\beta \gamma}(p,q) = \frac{N^{\alpha \beta \gamma}(gp;q) N^{\alpha \beta \gamma}(g;p)}{N^{\alpha\beta\gamma}(g;pq)} \, .
\ee
while the cocycle $\omega$ gives:
\be
M^{\alpha\beta\gamma}(p,q) = \delta \lambda^{\alpha\beta\gamma}(p,q) \, .
\ee
Substituting back into the first equation we solve it by:
\be
N^{\alpha\beta \gamma}(g;q) = \zeta^{\alpha\beta \gamma}(gq)/\zeta^{\alpha\beta\gamma}(g) \lambda(q) \, .
\ee 
This finally leads to the following composition laws for transition functions on the Hilbert space:
\be
\cU^{\alpha \beta} \cU^{\beta \gamma} \left[\cU^{\alpha \gamma}\right]^{-1} = \frac{\zeta^{\alpha \beta \gamma}(g_0)}{\zeta^{\alpha \beta \gamma}(g_N)}  \, \prod_{\ell} \lambda(q_{\ell})^{-1} .
\ee
Assuming that the class is unchanged among links ($\lambda^{\alpha\beta\gamma}=1)$, comparing with \eqref{eq:app-gerbe-descent} we identify $\zeta^{\alpha\beta\gamma}$ with the corresponding gerbe data. It can be checked that all of the gerbe gauge transformations now act appropriately on the Hilbert space.
Lastly, consider a quadruple overlap $U_{\alpha\beta\gamma\delta}$. Compatibility of the associativity condition (a.k.a. the pentagon equation) implies that $N^{\beta \gamma \delta} \left[ N^{\alpha \gamma \delta}\right]^{-1} N^{\alpha \beta \delta} \left[ N^{\alpha \beta \gamma} \right]^{-1}$ must be trivial. This leads to the closedness condition on $\zeta^{\alpha \beta \gamma}$:
\be
\zeta_{\beta \gamma \delta}\zeta_{\alpha \gamma \delta}^{-1}\zeta_{\alpha \beta \delta}\zeta_{\alpha \beta \gamma}^{-1}=1.
\ee

\vspace{2mm} \noindent \textbf{Transformation of one- and two-leg operators.}
Using the above the transformation laws of one- and two-leg operators are simple to extract in both the standard and local (tilded) basis.

\vspace{2mm} \noindent \emph{One-leg terms.} The integrated element $\widetilde{z}$ has unit components in the basis $(\bde_p, z)$ of a given patch, and thus defines different elements of $\End(\rho)$ in different patches. The corresponding operators are related by
\be
\widehat{z}^\beta = b^{\alpha\beta}({}^x q, z) \, \widehat{z}^\alpha \, ,
\ee
Evaluating both sides with \eqref{eq: apppartrap}, or with its tilded form $\widehat{z} \widetilde{\psi_q}(x) = \nu_{{}^x q}(z) \widetilde{\psi_q}(xz)$ gives:
\bea
\widehat{z}^\beta  \widetilde{\psi_q}(x)  &= b^{\alpha\beta}({}^x q, z) \, \widehat{z}^\alpha \widetilde{\psi_q}(x) \, , \\
\widehat{z}^\beta  \psi_q(x)  &= \frac{b^{\alpha\beta}({}^x q, z) b^{\alpha \beta}(q,x)}{b^{\alpha\beta}(q,xz)} \widehat{z}^\alpha \psi_q(x) \, .
\eea

\vspace{2mm} \noindent \emph{Two-leg terms.} 
We use that, with the current choices:
\be
Z_s = \psi(g_i s; s^{-1}, q_+)  \left[ \psi(g_{i-1}; q_- s, s^{-1}) \right]^{-1} \, .
\ee
From the transformation of $\psi^\alpha$ the phase $Z_s$ changes as
\be
Z_s^\beta = Z_s^\alpha \, \frac{c^{\alpha\beta}(s^{-1}, q_+)}{c^{\alpha\beta}(q_- s, s^{-1})} \, \frac{\mu^{\alpha\beta}(g_{i-1}; q_- s) \, \mu^{\alpha\beta}(g_i s; s^{-1} q_+)}{\mu^{\alpha\beta}(g_{i-1}; q_-) \, \mu^{\alpha\beta}(g_i; q_+)} \, .
\ee
The first factor is compensated by the transformation of the auxiliary vectors $\widetilde{\ket{r}}_X^{\pm}$, while the second is the ratio of the transition functions $\cU^{\alpha\beta}$ of the initial and final states. Thus, for fixed junction vectors, the kernel \eqref{eq:app-two-leg-kernel} of the two-leg operator in the tilded basis transforms as
\be
K_{[z]; (f,g)}^\beta = [\cU^{\alpha \beta}]^{-1} \,  K_{[z];(f,g)}^\alpha  \left[\cU^{\alpha \beta} \right] \, .
\ee
In the standard basis the same holds with $\cU^{\alpha\beta}$ replaced by $\cU^{\alpha\beta} \prod_\ell b^{\alpha\beta}(q_\ell,x_\ell)^{-1}$, the additional factors being provided by the transformation of $L_s$ and of the vectors $\ket{r}_X$. Finally, for a change of the potential $B_k^\alpha$ at fixed representative $\omega_k$ one has $c^{\alpha\beta} = b^{\alpha\beta} = 1$: the one- and two-leg operators are then the same in all patches, and only the basis vectors transform.

\subsection{Lattice B-field from the WZW cocycle}
\label{app:lattice-B}
In this appendix we evaluate explicitly the lattice B-field by differentiating the WZW cocycle \eqref{eq: wzw-rep}. 
In homogeneous notation, the two factors in~\eqref{eq:app-two-leg-Z}
are the tetrahedra
\be
Z_s
 =\Omega(1,g_{i-1},g_i,g_i s)\,
   \Omega(1,g_i s,g_i,g_{i+1}).
\ee
We will now take $s \to 1$ and expand this term explicitly.

\vspace{2mm}\noindent\emph{Radial gauge.}
We choose the local WZ potential $B_k$ so that it vanishes whenever
one of its arguments points along a radial path from the identity.
On a logarithm chart, these paths are $t\mapsto e^{tY}$, with
$g=e^Y$. If $R$ is their radial tangent field, the gauge-fixing condition is
\be
\iota_R B_k=0 \, .
\ee
Let $g\in G$ and $U,V\in T_gG$; the notation $(\eta)_g(U,V)$
means the two-form $\eta$ at $g$ evaluated on the tangent vectors $U,V$.

\begin{figure}[tbp]
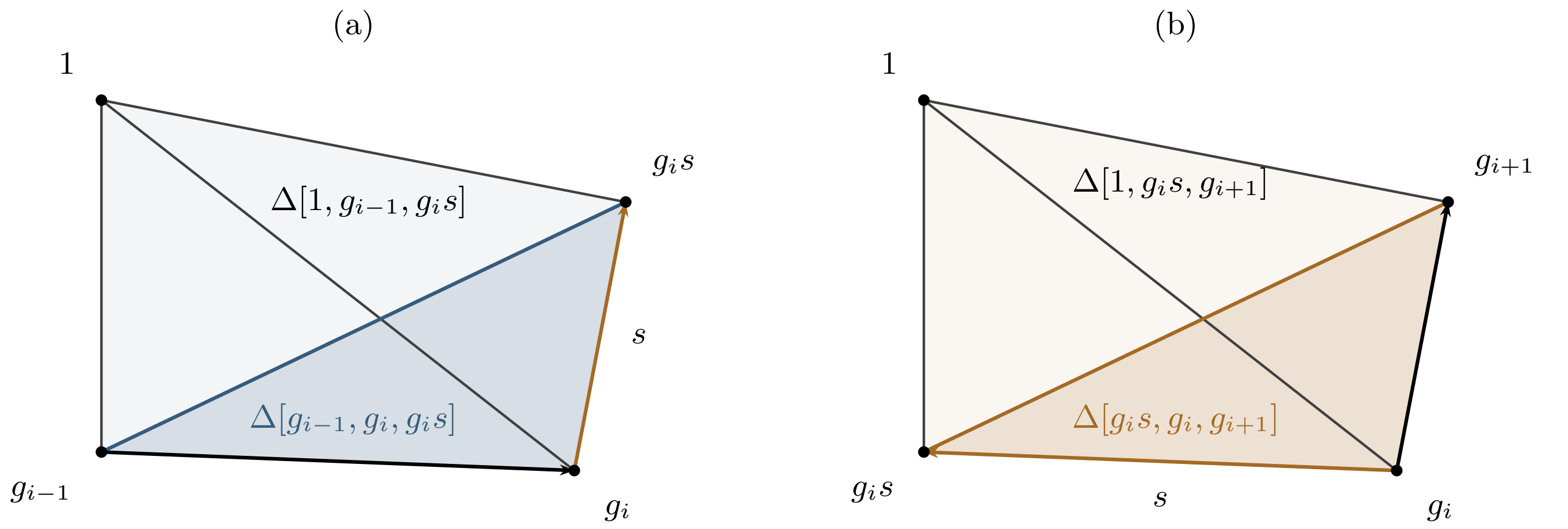

  \centering
  \RadialGaugeSUtwo
  \caption{Separate views of the two tetrahedra defining $Z_s$.
  The front radial triangles are shaded more lightly; their
  pullback of $B_k$ vanishes in radial gauge.}
  \label{fig:radial-gauge-su2}
\end{figure}

\vspace{2mm}\noindent\emph{From tetrahedra to faces.}
Choose the faces containing $1$ as radial cones, as in
Figure~\ref{fig:radial-gauge-su2}.
Concretely, the cone over an edge $\gamma(t)$ is parametrized by
$\exp[u\log\gamma(t)]$, with $0\leq u\leq1$.
Its tangent is radial, so the pullback of $B_k$ onto this
two-dimensional face vanishes in our gauge.
Stokes' theorem therefore leaves
only two terms:
\be
\log Z_s
 =-i\int_{\Delta[g_{i-1},g_i,g_i s]}B_k
  -i\int_{\Delta[g_i s,g_i,g_{i+1}]}B_k.
\label{eq:app-two-leg-B-faces}
\ee

\vspace{2mm}\noindent\emph{Moving one endpoint.}
Parametrize $s=e^{i\epsilon T^a}$. 
The two $s$-dependent geodesics $\gamma_\pm$ connecting $g_{i-1} \to g_i s$ and $g_i s \to g_{i+1}$ are parametrized as:
\bea
\gamma_-(t;\epsilon)
 &=g_{i-1}\exp\!\left[t\log(q_-e^{i\epsilon T^a})\right],\\
\gamma_+(t;\epsilon)
 &=g_i e^{i\epsilon T^a}
   \exp\!\left[t\log(e^{-i\epsilon T^a}q_+)\right],
\qquad 0\leq t\leq1 .
\label{eq:app-moving-links}
\eea
At $\epsilon=0$ these coincide with $\gamma[g_{i-1},g_i]$ and
$\gamma[g_i,g_{i+1}]$, respectively.
Their displacement vector fields are
\be
\xi_\pm^a(t)=\left.\partial_\epsilon
\gamma_\pm(t;\epsilon)\right|_0 \, ,
\ee
which vanish at the fixed points $g_{i-1}$, $g_{i+1}$ and equal $g_i iT^a$
at the common endpoint $g_i s$.

To parametrize the thin faces $\Delta[g_{i-1}, g_i, g_i s]$ and $\Delta[g_is, g_i, g_{i+1}]$ use the strips swept out by these
geodesics. The one-form $\iota_{\xi_\pm^a}B_k$ measures the
strip flux per unit displacement, since
$(\iota_{\xi_\pm^a}B_k)(\dot\gamma_\pm)
 =B_k(\xi_\pm^a,\dot\gamma_\pm)$.
Differentiating~\eqref{eq:app-two-leg-B-faces} gives
\be
k\cB_{a,i} = -i \partial_\epsilon \log Z_{s = e^{i \epsilon T^a}} \vert_{\epsilon = 0} 
 =\int_{\gamma[g_{i-1},g_i]}\iota_{\xi_-^a}B_k
   +\int_{\gamma[g_i,g_{i+1}]}\iota_{\xi_+^a}B_k \, .
\label{eq:app-lattice-B}
\ee

\vspace{2mm}\noindent\emph{The continuum limit.}
Let $g_j=g(x_j)$ with $x_j = a j$, and lattice spacing $a$. We evaluate the continuum limit of the lattice B-field.
In this limit, the links $\gamma[g_i, g_{i+1}]$ become short and using the vanishing of the displacement vector at the fixed endpoints we find:
\bea \label{eq: pullback}
\partial_\epsilon\gamma_-(t;\epsilon)|_0
 &=t\,g_i iT^a+O(a),\\
\partial_\epsilon\gamma_+(t;\epsilon)|_0
 &=(1-t)\,g_i iT^a+O(a),
\eea
while
$\partial_t\gamma_\pm=a\,\partial_xg_i+O(a^2)$.
Integrating over $t \in [0,1]$, each of the two terms contributes half of the continuum B-field, with \eqref{eq: pullback} providing the map to the physical WZW fields:
\be
k\cB_{a,i}
 =a\,\sum_b (B_k)_{ab}(g_i)\cJ^b(x_i)+O(a^2),
\qquad g^{-1}\partial_xg=i \sum_b T_b\cJ^b .
\label{eq:app-lattice-B-continuum}
\ee

\vspace{2mm} \noindent \textbf{The $SU(2)$ lattice B-field.}
Let us evaluate the lattice B-field for $G = SU(2)$.
We parametrize a group element using quaternionic coordinates
\be 
g=c_g1+i\mathbf v_g\cdot\boldsymbol\sigma \, ,
\ee
where
$c_g=\cos\chi_g$, $|\mathbf v_g|=\sin\chi_g$ and $0\leq\chi_g<\pi$ denotes the polar angle from the identity on the sphere. With the shorthand $c_i=c_{g_i}$ and $\mathbf v_i=\mathbf v_{g_i}$ we have
\be
\cB_{a,i}
 =\frac{c_i\mathbf e_a-\mathbf v_i\times\mathbf e_a}{\sqrt2}
 \cdot\left\{\mathbf v_i\times
 \left[b(g_{i-1},g_i)\mathbf v_{i-1}
       -b(g_{i+1},g_i)\mathbf v_{i+1}\right]\right\},
\label{eq:app-SU2-lattice-B}
\ee
with $\mathbf e_a$ the Cartesian unit vector and
\be
b(g,h)=\frac{
 \chi_g\sin\chi_g
 -\dfrac{\chi_h}{\sin\chi_h}\,\mathbf v_g\cdot\mathbf v_h
 -\dfrac{d}{\sin d}(c_h-c_g\cos d)
 }{2\pi|\mathbf v_g\times\mathbf v_h|^2},
\qquad d=\arccos(c_gc_h+\mathbf v_g\cdot\mathbf v_h).
\label{eq:app-SU2-endpoint-kernel}
\ee
Degenerate pairs are evaluated by continuity of the line integrals
on the chosen chart. The continuum limit can be checked directly from
expression~\eqref{eq:app-SU2-lattice-B}. The coincident kernel is
\be
b(g,g)=\frac{\chi-\sin\chi\cos\chi}{4\pi\sin^3\chi}.
\ee
Expanding $\mathbf v_{i\pm1}=\mathbf v_i\pm a\,\partial_x\mathbf v_i+O(a^2)$,
the first-order variations of $b$ multiply $\mathbf v_i$ and
drop out of the cross product in~\eqref{eq:app-SU2-lattice-B}.
The remaining terms give~\eqref{eq:app-lattice-B-continuum}, with
\be
(B_k)_{ab}(g)
 =\frac{k(2\chi-\sin2\chi)}{8\pi\sin^2\chi}
   \epsilon_{ab}{}^c n_c \, ,
\ee
with $n^a$ the unit vector on the conjugacy sphere of $g$. These are precisely the components of the WZW B-field for $SU(2)$ in the radial gauge.

\end{document}